%% file: RSD_Accepted.tex
\documentclass[fleqn,usenatbib]{mnras}

\usepackage{newtxtext,newtxmath}

\usepackage[T1]{fontenc}
\usepackage{ae,aecompl}

\usepackage{graphicx}	
\usepackage{amsmath}	
\usepackage{amssymb}	
\usepackage{subfigure}

\usepackage{colortbl}

 \newcommand{\todo}[1]{\relax}

\title[Structure growth rate measurement from the anisotropic LRG correlation function]{The clustering of the SDSS-IV extended Baryon Oscillation Spectroscopic Survey DR14 LRG sample: structure growth rate measurement from the anisotropic LRG correlation function in the redshift range 0.6 < z < 1.0}

\author[ M. Icaza-Lizaola et al.]{\parbox{\textwidth}{
\Large
M. Icaza-Lizaola$^{1,2,3}$\thanks{E-mail: miguel.a.de-icaza-lizaola@durham.ac.uk},
M. Vargas-Maga\~na$^{1}$\thanks{E-mail: mmaganav@fisica.unam.mx},
S. Fromenteau$^{4}$\thanks{E-mail: sfroment@icf.unam.mx},
S. Alam$^{5}$,
B. Camacho$^{1}$,
H. Gil-Marin$^{6}$,
R. Paviot$^{7}$,
Ashley Ross$^{8}$,
Donald P. Schneider$^{9,10}$,
Jeremy Tinker$^{11}$,
Yuting Wang$^{12}$,
Cheng Zhao$^{11}$,
Abhishek Prakash$^{13}$,
G.Rossi$^{14}$,
Gong-Bo Zhao$^{15,16}$,
Irene Cruz-Gonzalez$^{2}$,
Axel de la Macorra$^{1}$.
}\vspace*{4pt} \\
\scriptsize $^{1}$Instituto de F\'isica, Universidad Nacional'Aut\'onoma de M\'exico, Apdo. Postal 20-364, Ciudad de M\'exico, Mexico.\vspace*{-2pt} \\
\scriptsize $^{2}$Instituto de Astronom\'ia, Universidad Nacional Aut\'onoma de M\'exico, Apdo. Postal 70-264, Ciudad de M\'exico, Mexico.\vspace*{-2pt} \\
\scriptsize $^{3}$Institute for Computational Cosmology, Department of Physics, University of Durham, South Road, Durham DH1 3LE, UK.\vspace*{-2pt} \\
\scriptsize $^{4}$Instituto de Ciencias F\'isicas, Universidad Nacional Aut\'onoma de M\'exico, Av. Universidad s/n, 62210 Cuernavaca, Mor., Mexico. \vspace*{-2pt} \\
\scriptsize $^{5}$Institute for Astronomy, University of Edinburgh, Royal Observatory, Blackford Hill, Edinburgh, EH9 3HJ , UK. \vspace*{-2pt} \\
\scriptsize $^{6}$Institut de Ciencies del Cosmos, Universitat de Barcelona ICCUB. \vspace*{-2pt} \\
\scriptsize $^{7}$Aix Marseille Université, CNRS, LAM, Laboratoire d'Astrophysique de Marseille, Marseille, France. \vspace*{-2pt} \\
\scriptsize $^{8}$Center for Cosmology and AstroParticle Physics, The Ohio State University, Columbus, OH 43210, USA. \vspace*{-2pt}.\\
\scriptsize $^{9}$Department of Astronomy and Astrophysics, The Pennsylvania State University, University Park, PA 16802. \vspace*{-2pt} \\
\scriptsize $^{10}$Institute for Gravitation and the Cosmos, The Pennsylvania State University, University Park, PA 16802. \vspace*{-2pt} \\
\scriptsize $^{11}$New York University, Center for Cosmology and Particle Physics, Department of Physics, 726 Broadway, Room 1005, New York, NY 10003, USA. \vspace*{-2pt}.\\
\scriptsize $^{12}$National Astronomy Observatories, Chinese Academy of Science, Beijing, 100101, P.R.China. \vspace*{-2pt} \\
\scriptsize $^{13}$IPAC, California Institute of Technology, MC 100-22, 1200 E California Boulevard, Pasadena, CA 91125, USA. \vspace*{-2pt} \\
\scriptsize $^{14}${Department of Physics and Astronomy, Sejong University, Seoul, 143-747, Korea}. \vspace*{-2pt} \\
\scriptsize $^{15}$National Astronomical Observatories of China, Chinese Academy of Sciences, 20A Datun Road, Chaoyang District, Beijing 100012, P.R. China. \vspace*{-2pt} \\
\scriptsize $^{16}$University of Chinese Academy of Sciences, Beijing, 100049, P.R.China. \vspace*{-2pt} \\
}

\date{Accepted 2019 December 5. Received 2019 December 4; in original form 2019 September 16}

\begin{document}
\label{firstpage}
\pagerange{\pageref{firstpage}--\pageref{lastpage}}
\maketitle

\begin{abstract}
 We analyze the anisotropic clustering of the Sloan Digital Sky Survey-IV Extended Baryon Oscillation Spectroscopic Survey (eBOSS) Luminous Red Galaxy Data Release 14 (DR14) sample combined with  Baryon Oscillation Spectroscopic Survey (BOSS) CMASS sample of galaxies in the redshift range 0.6$<z<$1.0, which consists of 80,118 galaxies from eBOSS and 46,439 galaxies from the BOSS-CMASS sample. The eBOSS-CMASS Luminous Red Galaxy sample has a sky coverage of 1,844 deg$^2$, with an effective volume of 0.9 Gpc$^3$. The analysis was made in configuration space using a Legendre multipole expansion. The Redshift Space Distortion signal is modeled as a combination of the Convolution Lagrangian Perturbation Model and the Gaussian Streaming Model.
  We constrain the logarithmic growth of structure times the amplitude of dark matter density fluctuations, $f (z_{\rm eff})\sigma_8(z_{\rm eff})=0.454 \pm0.134 $, and the Alcock-Paczynski dilation scales which constraints the angular diameter distance $D_A(z_{eff})=1466.5  \pm 133.2  (r_s/r_s^{\rm fid})$ and $H(z_{\rm eff})=105.8 \pm  15.7  (r_s^{\rm fid}/r_s)  \mathrm{km\,s^{-1}\,Mpc^{-1}}$, where $r_s$ is the sound horizon at the end of the baryon drag epoch and $r_s^{\rm fid}$ is its value in the fiducial cosmology at an effective redshift $z_{\rm eff}=0.72$.  These results are in full agreement with the current $\Lambda$-Cold Dark Matter ($\Lambda$-CDM) cosmological model inferred from Planck measurements. This study is the first eBOSS LRG full-shape analysis i.e. including Redshift-Space Distortions (RSD) simultaneously with the Alcock-Paczynski (AP) effect and the Baryon Acoustic Oscillation (BAO) scale.
\end{abstract}




\section{Introduction}

\label{section:intro}

\input{intro_Accepted}

\section{Data}
\label{section:data}
\input{data_Accepted}

\section{Mocks}
\label{section:mocks}
\input{mocks_Accepted}

\section{Modeling Redshift Space Distortions}
\label{section:model}
\input{model_Accepted}

\subsection{Including the Alcock-Paczynski Effect}
\label{section:AP}
\input{AP_Accepted}

\section{Methodology}
\label{section:metho}
\input{methodology_Accepted}

\section{Testing for systematic uncertainties}
\label{section:systematics}

\input{Nseries_Accepted}

\input{QPM_EZ_Accepted}

\section{Results on the LRG DR14 sample}
\label{sec:results}
\input{results_Accepted}

\section{Conclusions}

\input{Conclusions_Accepted}

\section*{Acknowledgements}

The  authors  of  this  paper  would  like  to  thank  Peder  Norberg,Shaun Cole, and Carlton Baugh from Durham University for thediscussions about the methodology and for the help given with thedesign of the methods for testing the accuracy of our results.

The authors thank Alejandro \'Aviles for the useful discussionsover the model generation using CLPT-GSRSD.

MI is supported by a PhD Studentship from the Durham Centrefor Doctoral Training in Data Intensive Science, funded by the UKScience and Technology Facilities Council (STFC, ST/P006744/1)and Durham University. MDL also acknowledges support from theSTFC through ST/P000541/1.

MV is partially supported by Programa de Apoyo a Proyectos de Investigaci\'on e Innovaci\'on Tecnol\'ogica (PAPIIT) No IA102516, NoIA101518, Proyecto Conacyt Fronteras No 281 and from ProyectoLANCAD-UNAM-DGTIC-319.

SF is supported by Programa de Apoyo a Proyectos de Investigaci\'on e Innovaci\'on Tecnol\'ogica (PAPIIT) No IA101619.

GR acknowledges support from the National Research Founda-tion of Korea (NRF) through Grant No. 2017077508 funded by theKorean Ministry of Education, Science and Technology (MoEST),and from the faculty research fund of Sejong University in 2018.

This work used the DiRAC@Durham facility managed by the In-stitute for Computational Cosmology on behalf of the STFC DiRACHPC  Facility  (www.dirac.ac.uk).  The  equipment  was  funded  byBEIS  capital  funding  via  STFC  capital  grants  ST/K00042X/1,ST/P002293/1 and ST/R002371/1, Durham University and STFCoperations grant ST/R000832/1. DiRAC is part of the National e-Infrastructure

This research used resources of the National Energy Research Scientific Computing Center, a DOE Office of Science User Facility supported by the Office of Science of the U.S. Department of Energy under Contract No. DE-AC02-05CH11231.

Funding for the Sloan Digital Sky Survey IV has been provided by the Alfred P. Sloan Foundation, the U.S. Department of Energy Office of Science, and the Participating Institutions. SDSS acknowledges support and resources from the Center for High-Performance Computing at the University of Utah. The SDSS web site is www.sdss.org.
SDSS is managed by the Astrophysical Research Consortium for the Participating Institutions of the SDSS Collaboration including the Brazilian Participation Group, the Carnegie Institution for Science, Carnegie Mellon University, the Chilean Participation Group, the French Participation Group, Harvard-Smithsonian Center for Astrophysics, Instituto de Astrof\'isica de Canarias, The Johns Hopkins University, Kavli Institute for the Physics and Mathematics of the Universe (IPMU) / University of Tokyo, the Korean Participation Group, Lawrence Berkeley National Laboratory, Leibniz Institut for Astrophysik Potsdam (AIP), Max-Planck-Institut for Astronomie (MPIA Heidelberg), Max-Planck-Institut for Astrophysik (MPA Garching), Max-Planck-Institut for Extraterrestrische Physik (MPE), National Astronomical Observatories of China, New Mexico State University, New York University, University of Notre Dame, Observatorio Nacional / MCTI, The Ohio State University, Pennsylvania State University, Shanghai Astronomical Observatory, United Kingdom Participation Group, Universidad Nacional Aut\'onoma de M\'exico, University of Arizona, University of Colorado Boulder, University of Oxford, University of Portsmouth, University of Utah, University of Virginia, University of Washington, University of Wisconsin, Vanderbilt University, and Yale University.





\bibliographystyle{mnras}
\bibliography{biblio_cmu}




\appendix

\include{Appendices_Accepted}


\end{document}

%% file: intro_Accepted.tex
The standard cosmological model ($\Lambda$-CDM) accurately describes most observations. However, the acceleration of the expansion of our Universe requires the existence of a dominating source of exotic energy, i.e., the Dark Energy. This energy remains undetected to this day, which has led to many searches for an alternative explanation. One possibility is to modify the geometric part of Einstein's equations, which corresponds to changing the General Relativity (hereafter GR) equations rather than invoking a new component in the stress-energy tensor. A common modification to GR is the addition of a cosmological constant,  $\Lambda$, coupled to the metric. However, it is not possible to distinguish between $\Lambda$ and a specific case of Dark Energy with a constant equation of state $w=-1$.

Another way to reproduce cosmological observations is to modify the gravity model. Various alternative gravitational models have been studied during the past 50 years which can be grouped in different families. Extra-field theories, such as $f(R)$  \citep{Sotiriou2010}, Tensor-Scalar theories, extra-dimension theories, such as DGP \citep{Fang2008}, braneworld, and string gravity models, and higher-order theories such as the Galileons model \citep{2015_galileon_rev} are some of them.

All modified gravity models must recover the GR results at the local scale (i.e., for high density) where GR has been strongly tested; this is generally solved by invoking screening mechanisms. Therefore, any modification has to appear in the context of weak gravity and large scales; this is the reason why cosmology, and more particularly Large-Scale-Structures (LSS) observations, is the appropriate framework for this study.

Cosmological constraints on the theory of gravity are primarily produced from LSS observations, the most important of these being Supernovae \citep{Riess1998, Perlmutter1999}, Baryon Acoustic Oscillations \citep{Eisenstein2005, Alam2017} and weak lensing \citep{Sheldon2004}, and from the early Universe through Cosmic Microwave Background observations, when the density contrast was of the order of $\sim 10^{-5}$  \citep{Planck2015_Mod_Grav}.

Large-scale peculiar velocities, combined with standard clustering, are a unique framework to distinguish between the various models of gravity. However, obtain precise relative velocity measurements at large scales ((>10 $h^{-1}$ Mpc)) is challenging. The Kinetic Sunyaev-Zel'dovich effect is a possibility \citep{Mueller2014} but requires measurements of massive galaxy clusters with high precision on the SZ signal estimation. Conversely, we can directly use the imprint of these velocities on the redshift measurement through the Redshift Space Distortions (RSD) in the anisotropic correlation function of galaxies (or other tracers of the dark matter) \citep{Kaiser1987,Hamilton1992,Cole1995,Peacock2001,Cabre2009,Alam2015,Satpathy2017,Zarrouk2018}.
The measured redshift is the sum of the Hubble flow, the Doppler effect due to the peculiar velocities of the observer and the observed galaxies, and a small contribution from gravitational redshift. If the peculiar velocities are randomly distributed (i.e. from satellite galaxies inside clusters), then they only contribute as a noise.  They are, however, correlated with the density field, revealing cosmological information, in particular allowing us to distinguish between dark energy models or deviations from GR. The Redshift Distortion introduces anisotropies in the  galaxy-galaxy two-point correlation function, particularly if we stack the information around over-densities, where these tracers live. Performing an anisotropic study, i.e., using the angle with respect to the line-of-sight as a statistical breakdown, we can detect the coherent deformations of the 3D two-point correlation function predicted by the \cite{Kaiser1987} effect.

BAO and supernova measurements are constraints on the expansion history of the Universe. However, it has been shown that an appropriate choice of the equation of state $w(a)$ can allow different cosmological models to have the same expansion history \citep{Linder2005}.
In order to break this degeneracy  one can complement expansion history observations with the clustering history of the structures through the measurement of the linear growth rate:

\begin{equation}
\label{growth_rate}
f(a)=\frac{d \ln D(a)}{d \ln a},
\end{equation}
where $D(a)$ is the linear growth factor as a function of  the scale factor $a$, and it quantifies the degree of structure at that time. In this paper we extend the growth rate $f$ measurement from previous surveys to an effective redshift of $z=0.72$ using the Luminous Red Galaxies (LRGs) sample from the extended Baryon Oscillation Spectroscopic Survey (eBOSS; \citealt{Dawson}).

The paper is organized as follows:  Section \ref{section:data} presents the data, Section \ref{section:mocks} describes the mock catalogs used for the estimation of the covariance matrix and for our systematics checks. Section \ref{section:model} presents the modeling of the RSD signal as well as the parametrization used for the Alcock-Paczynski test.
 Section \ref{section:metho} describes the methodology followed in our analysis.
  Section \ref{section:systematics} presents our analysis, using mock catalogs, of the systematic effects associated with our methodology. The results for the eBOSS-CMASS sample are presented in Section \ref{section:results}. Finally, the cosmological implications of this work are reviewed in Section \ref{section:discussion}.

%% file: data_Accepted.tex
Our sample of spectroscopic data was collected during the first two years of eBOSS \citep{Dawson}, which is the cosmological component of the fourth generation of the Sloan Digital Sky Survey (SDSS-IV; \citealt{Blanton}).
All of our spectra were obtained by the Sloan 2.5m telescope using two multi-object spectrographs \citep{Smee} at Apache Point Observatory in New Mexico, USA \citep{Gunn}.
All of these data belong to the SDSS Data Release 14 \citep{Abolfathi}, of which we analyze the Luminous Red Galaxies (LRG) Sample.
 The LRG targets were selected based on updated photometric data from SDSS I/II/III imaging \citep{Fukugita,Gunn1998}
  for which the calibration of the photometric data was updated following the procedure presented in \cite{Schlafly}.
 The target selection process also used infrared photometry data from the Wide-Field Infrared Survey Explorer (WISE; \citealt{Wright}). The WISE satellite observed the entire sky using four infrared channels respectively centered at 3.4 $\mathrm{\mu}$m (W1), 4.6 $\mathrm{\mu}$m (W2), 12 $\mathrm{\mu}$m (W3), and 22 $\mathrm{\mu}$m (W4). The eBOSS LRG sample uses the W1 and W2 bands. Given that stars have different properties than galaxies in infrared (particularly due to the galactic dust), the WISE data allow us to reduce the stellar contamination, it is also useful for extending the redshift range with respect to BOSS. The target selection follows the algorithm described in \cite{Prakash}.

\subsection{eBOSS-CMASS Sample}
 Our eBOSS DR14 LRG Sample includes data of the first two years of the eBOSS program combined with the BOSS  CMASS data \citep{Alam2017} which overlaps with the eBOSS footprint in a redshift range of $0.6<z<1.0$. This approach allows construction of a more complete sample without decreasing the median redshift.

The eBOSS-CMASS sample is composed of 80118 galaxies from eBOSS  and 46439 from CMASS, yielding a total of 126,557 galaxies. The numbers separated by Galactic hemisphere are listed in table \ref{tab:data_stats}. The sky coverage in the North Galactic Cap (hereafter NGC) is 1011.15 $\mathrm{deg}^2$ and 788.09 $\mathrm{deg}^2$ in the South Galactic Cap (hereafter SGC), giving a total solid angle of 1844.0 $\mathrm{deg}^2$.
The effective volume of eBOSS is 0.618 Gpc$^3$ which increases up to  0.9 Gpc$^3$ when considering the eBOSS-CMASS sample.

\begin{figure}

\subfigure{\includegraphics[width=80mm,height=60mm]{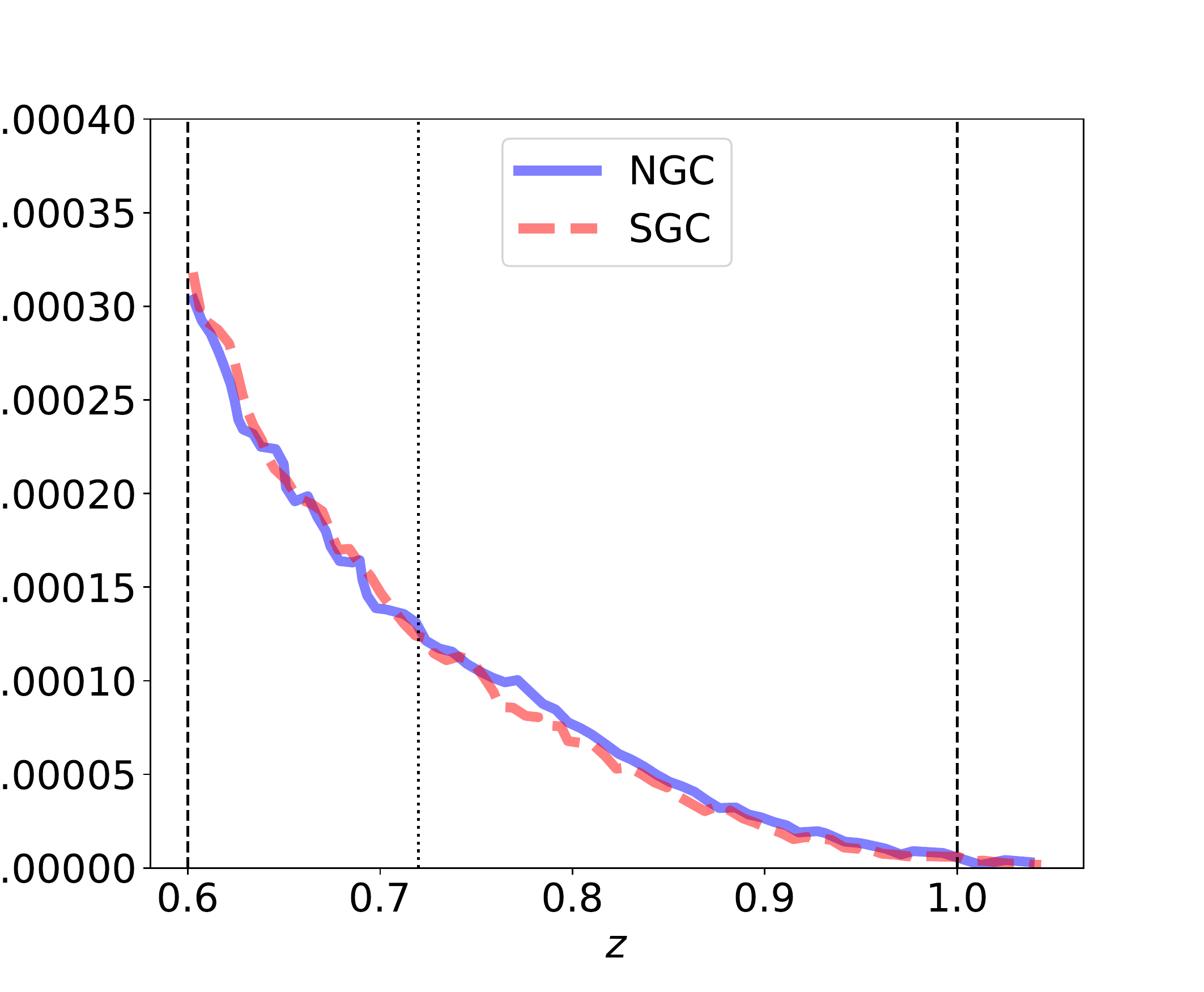}}

\caption{Number density of the LRG sample as a function of the redshift for both hemispheres, the solid blue lines correspond to the North Galactic Cap (NGC) and the dashed red lines to the South Galactic Cap (SGC); the dashed vertical lines indicate the redshift cuts applied. The median redshift of the sample is $z=0.72$ and is represented by the vertical dotted line.}
\label{fig:nz}
\end{figure}

\begin{table*}
\caption{Characteristics of the LRG data catalogs used. The left panel corresponds to the BOSS CMASS sample from DR12, the right to the eBOSS LRG DR14 sample. $N_{\rm star}$ and $N_{\rm qso}$ are the number of objects whose spectra were determined to be stars or quasars instead of LRGs. $N_{\rm zfail}$ is the number of objects whose redshift measurement was not reliable, and $N_{\rm cp}$ the number of objects without spectra due to close pair effects. The last line reports the number of galaxies and the effective volume of our final sample, which is a combination of the CMASS and eBOSS samples.\todo{Moificate the table. The last column $N_{\rm noz}$ is never filled.}}
\label{tab:data_stats}
\begin{tabular}{@{}lccccccc}
\hline
\multicolumn{8}{c}{CMASS LRG Sample DR12 }\\
\hline
Catalogue &Area (deg$^2$)&	Total redshifts \\
\hline
CMASS-BOSS NGC	&1011.15 &	26149	&-	&	-&-&-	&-\\
CMASS-BOSS SGC	&788.09 &	20290	&-	&	-&-&-&-\\
\hline
\multicolumn{8}{c}{ eBOSS LRG DR14 Sample }\\
\hline
Catalogue &	$N_{\rm gal} $&	$N_{\rm star}$ &$N_{\rm qso}$& $N_{\rm cp}$&$N_{\rm zfail}$ &$A_{\rm eff}[\rm deg^2]$& $V_{\rm eff}[\rm Gpc^3]$\\
\hline
eBOSS NGC	& 45826&			2897&	18&2263&4957 & 1033.4&0.356\\
eBOSS SGC	&34292&			4273&	18&1687& 4366 &811.6&0.262 \\
\hline
Total &80118 &7170& 36 &3950& 9323 &1844.0 & 0.618\\
\hline
eBOSS-CMASS&126557 &&&&&&0.900\\
\hline

\end{tabular}
\end{table*}
Figure \ref{fig:nz} shows the number density of the sample as a function of redshift for both hemispheres, the solid blue lines correspond to the NGC and the dashed red lines to the SGC; the dashed vertical lines indicate the redshift cuts applied for our analysis.
 The median redshift of the sample is $z=0.72$, which is represented by the vertical dotted line.

\subsection{Footprint  and Masks}
\label{sec:masks}
\begin{figure}
\includegraphics[width=90mm,trim = 3cm 2.0cm 3cm 0cm,]{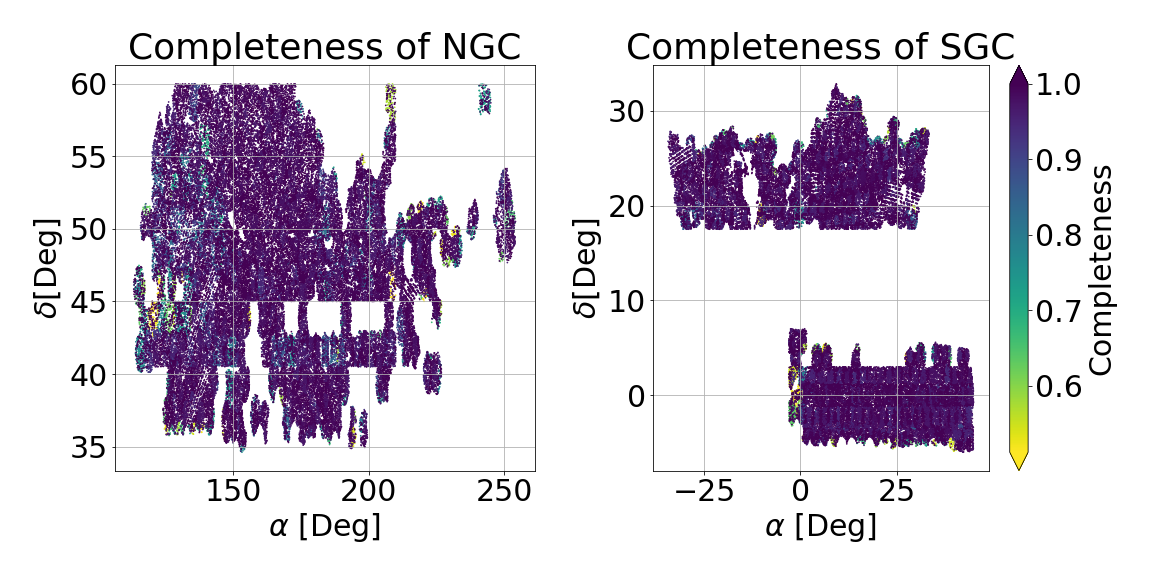}
\caption{Angular mask of the North Galactic Cap NGC (left) and the South Galactic Cap SGC (right). The color indicates the targeting completeness of the DR14 LRG sample in a given area of the sky, which is computed using equation \ref{Comp_Equation}. Regions of low targeting completeness (where $C<0.5$) were not included in the final sample.}
\label{Completeness}
\end{figure}

The left and right panels of Figure \ref{Completeness} display the sky coverage of the galaxy sample for the NGC and SGC respectively, where the color scale indicates the targeting completeness defined as:
\begin{equation}
\label{Comp_Equation}
C=\frac{N_{\rm gal}+N_{\rm qso}+N_{\rm star}+N_{\rm cp}+N_{\rm zfail}}{N_{\rm target}},
\end{equation}
where
\begin{itemize}
\item $N_{\rm gal}$ is the number of  galaxies with good quality eBOSS spectra.
\item $N_{\rm cp}$ is the number of galaxies without spectra due to the fiber collision effect. Two fibers cannot be closer than 62$''$ on a given plate.
\item $N_{\rm star}$ denotes the number of observed objects which are spectroscopically confirmed to be stars.
\item $N_{\rm zfail}$ denotes the  number of objects for whom the measured spectra lacks sufficient qualities to provide a confident redshift measurement.
\end{itemize}
The targeting completeness is computed by sector, and the mean completeness is $96.3\%$ (where the NGC has an average completeness of $95.9\%$ and the SGC  $96.9\%$). We only use data from regions with a completeness higher than 0.5 (this value is smaller than the completeness used in BOSS).

Certain areas in the sky have to be excluded from the final data sample. The maps of these excluded regions are known as veto masks and have to be removed from our random catalogs as well. The veto masks used in eBOSS were:

\begin{itemize}
\item The Collision priority veto mask that excludes regions that are closer than 62$''$  from an already observed target, as any object inside this radius would not be observed due to fiber collision.
\item  The Bright veto mask which excludes regions around stars that are part of the Tycho catalog \citep{Hog2000} with Tycho BT magnitudes larger than 6 and lower than 11.5. The excluded radius is magnitude-dependent and it goes from 0.8$'$ to 3.4$'$.  An additional mask excludes regions around bright galaxies and other objects \citep{Rykoff2014}; it is also magnitude-dependent and goes from a radius of 0.1$'$ to 1.5$'$.
\item The Bad fields veto mask excludes regions of the sky with bad photometry. If the local sky is badly determined (as occasionally happens in regions with complex backgrounds), the core of an object can be strongly negative.
\item The Extinction mask excludes regions where the Galactic extinction is such that $E(B -V ) > 0.15$ or where the seeing FWHM is larger than 2.3$''$, 2.1$''$, and 2.0$''$ in $g$, $r$, and $i$ bands, respectively.
\item The Center Focal Plane mask excludes LRG targets that lie within 92$''$ of the center of the telescope focal plane, where a center post holds the plate and prevents fibers from being assigned.
 \end{itemize}
 The total masked area is 12.3\% for the NGC and 18.2\% for the SGC.

 \subsection{Catalogue for LSS analysis}
Two data catalogs that differ in their treatment of the photometric systematics and of spectroscopic incompleteness were used to create the sample for our study.
The first is a BOSS-like catalog where traditional weighting schemes are applied, described in \cite{Ross2017}, to the data. The second is denoted as the "official catalog", and it was used in \cite{Bautista2017} for performing the  BAO analysis. Here some improvements with respect to previous analysis were implemented: the forward modeling of the randoms for the spectroscopic incompleteness and the multilinear regression and subsampling of the randoms for the photometric systematics.

In this section, we briefly review both methodologies, first describing the different treatments of the photometric systematics, and then the procedures used for dealing with the redshift incompleteness. Finally, we summarize the weights applied to the data for both cases and also the subsampling techniques used in the random catalogs in each case.

 \subsubsection{Correcting for Photometric Systematics}
Here we will give a brief description of the two methodologies for correcting photometric systematics:
\begin{itemize}
\item Iterative method ("BOSS-like") was developed in \cite{Ross2017}.
The basic idea is to include the systematics in an iterative way and estimate at each step the associated weights. For the eBOSS LRG sample, we studied the correlation
of the mean density as a function of seven potential observational systematics related with SDSS photometry: stellar density, $i$-band depth, $z$-band sky flux, $z$-FWHM, and $r$-band extinction\footnote{Additionally we explored two additional maps derived from WISE photometry:
one for the median number of single-exposure frames per pixel in the WISE W1 band (denoted as WISE W1 Cov Med) and the median of accumulated flux per pixel in the WISE W1 band (denoted by WISE W1 Med).}. We followed the iterative method starting with the main systematics reported in previous works. Figure \ref{fig:photosys} displays the mean density of data, $N_{\rm gal}$, normalized by the random number density, $N_{\rm ran}$, as a function of six of the seven systematics considered in the analysis before and after corrections. The most significant weights are those due to stellar density ($w_{\rm star}$), followed by the $r$-band extinction ($w_{\rm ext}$), airmass ($w_{\rm air}$), and $z$-band sky flux ($w_{\rm sky}$).
The systematics related with the WISE maps did not have any strong correlation requiring correction, thus we decided not to include them in the weight estimation.
\begin{figure}

\subfigure{\includegraphics[width=90mm,height=110mm,trim = 1.5cm 2cm 2cm 2cm, clip]{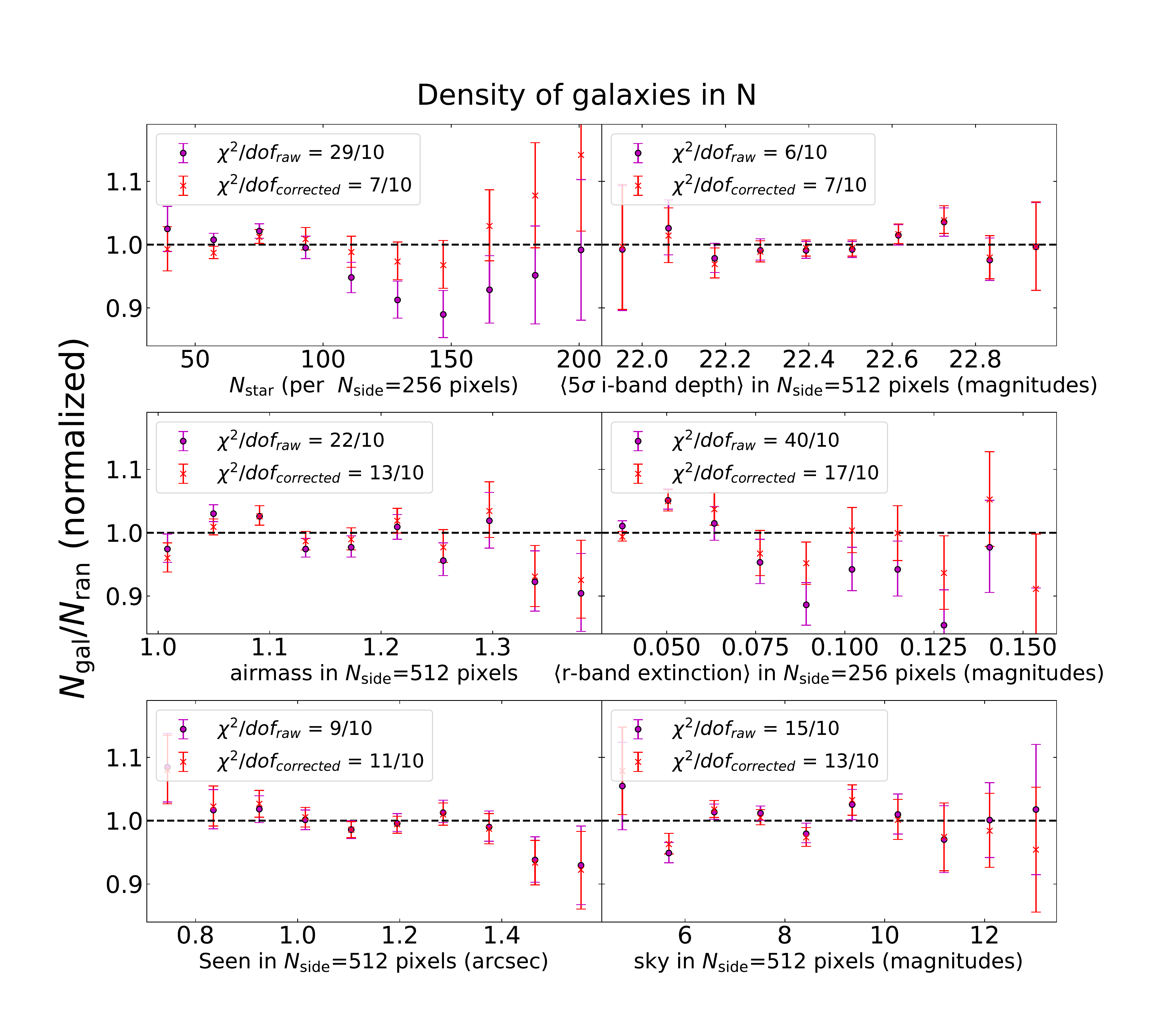}}
\subfigure{\includegraphics[width=90mm,height=110mm,trim = 1.5cm 2cm 2cm 2cm, clip]{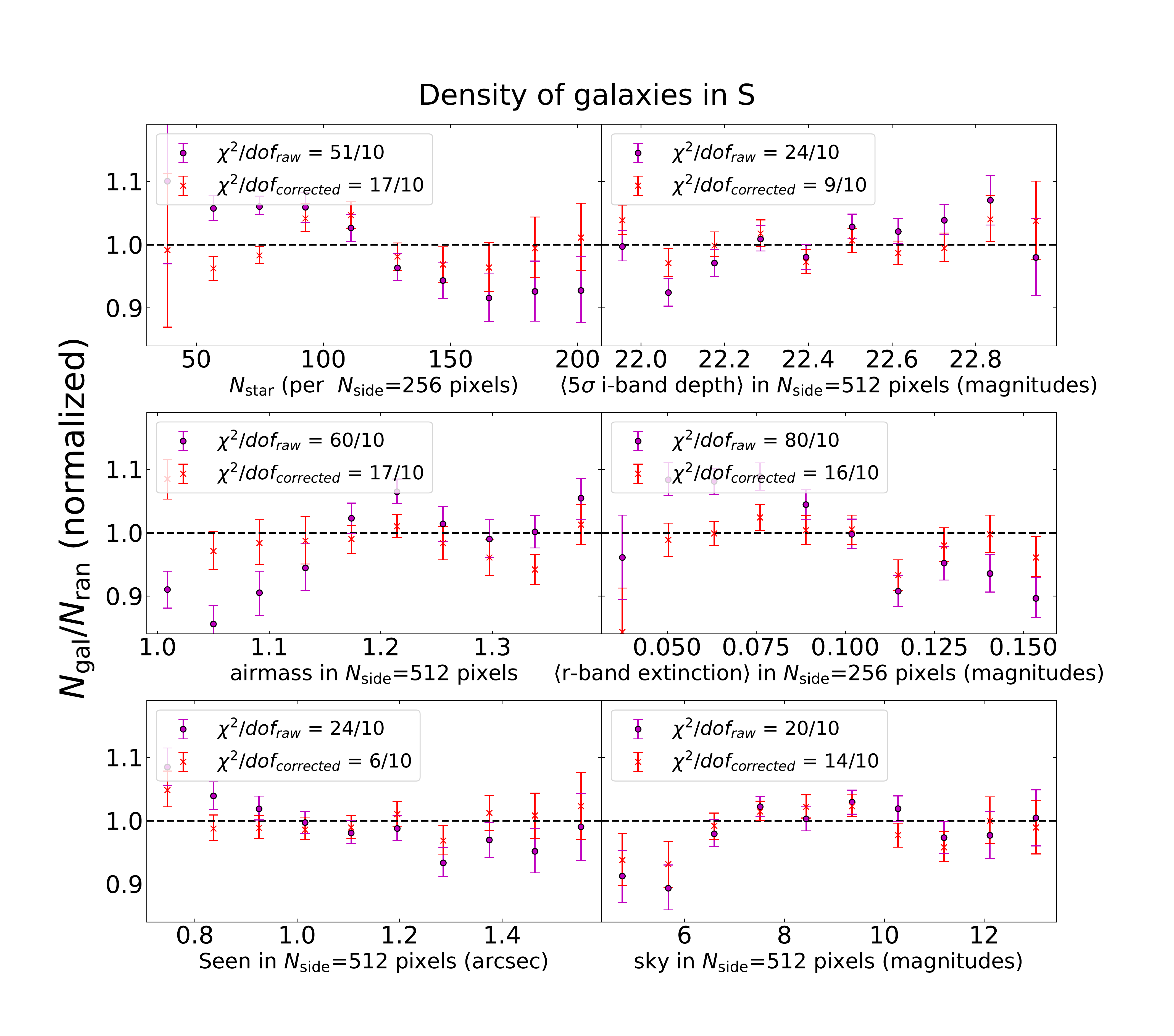}}
\caption{We show the mean density of data, $N_{\rm gal}$, normalized by the random number density, $N_{\rm ran}$ as a function of six of the nine systematics considered in the analysis. The most significant weights were those due to stellar density ($w_{\rm star}$), r-band extinction ($w_{\rm ext}$), airmass ($w_{\rm air}$), and z-band sky flux ($w_{\rm sky}$).}
\label{fig:photosys}
\end{figure}
We calculate a weight for each galaxy  that takes in account a linear relationship for each potential systematic.
\begin{equation}
w(\rm sys)=\frac{1}{mx+b}.
\end{equation}
The total systematic weight $w_{\rm systot}$, is defined as
\begin{equation}
\label{weight_Equation}
w_{\rm SYSTOT}=w_{\rm star} \, w_{\rm ext} \, w_{\rm air} \, w_{\rm sky}
\end{equation}
\item Multi-regression Method: We followed the same methodology presented in \cite{Bautista2017}, where the correlation between systematic maps and density were computed using a multilinear regression of the seven systematic maps instead of the iterative method. The advantage of this method is that it does not assume the systematics are independent, as does the iterative method. Additionally, in the official catalogs, instead of using weights associated with galaxies, the randoms are subsampled following the correlation found with the multi-regression method; the subsampling of the randoms or the weighting scheme of the galaxies should yield the same results; the main differences observed in the catalogs should be derived uniquely from the Iterative/Multi-regression methodologies.
\end{itemize}

\begin{figure}
\subfigure{\includegraphics[width=100mm,height=100mm,trim = 1.5cm 1.5cm 2.5cm 2cm, clip]{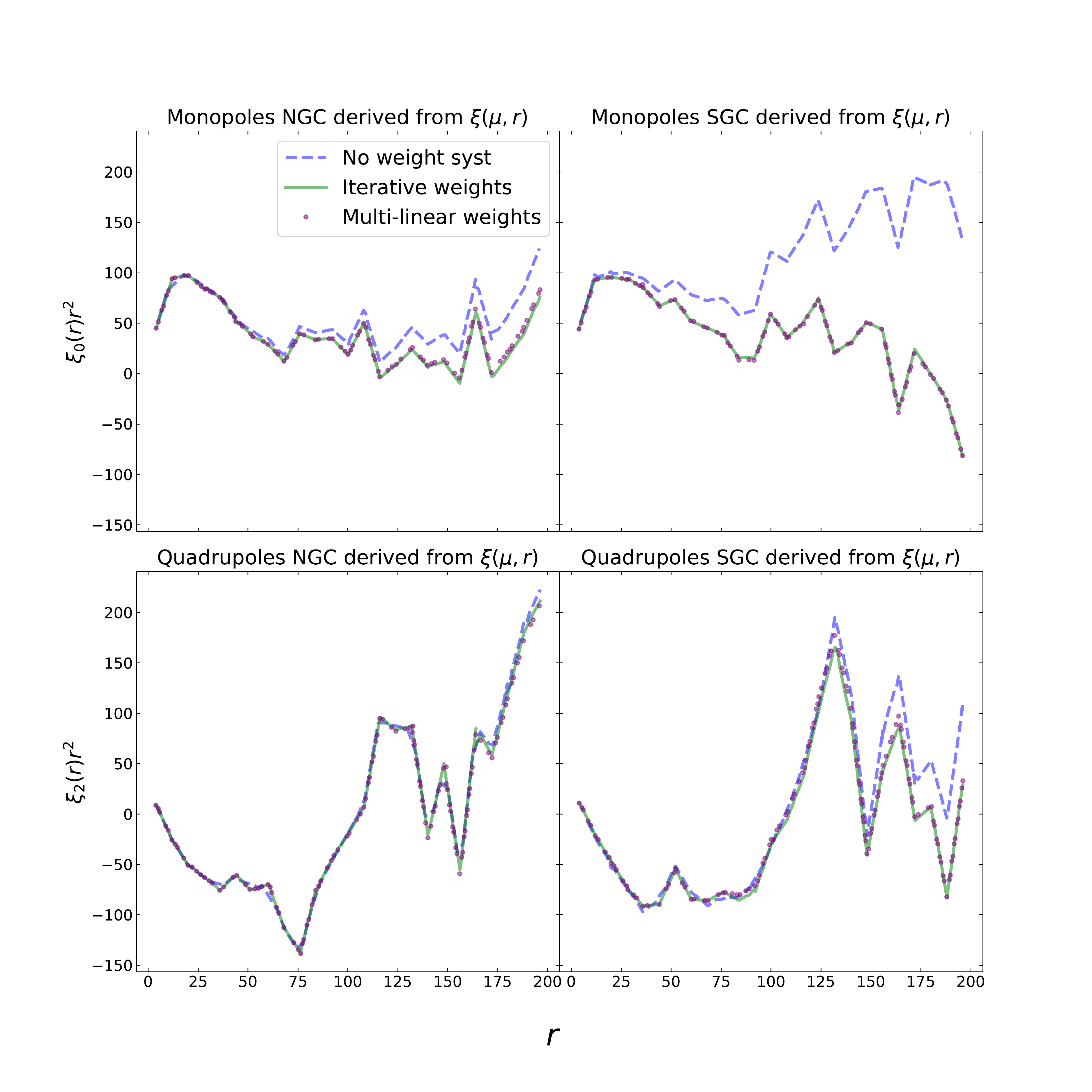}}
\caption{Multipoles for eBOSS sample (NGC left and SGC right) comparing the iterative and multilinear regression methods. The monopole from both hemispheres without corrections shows a large spurious correlation at large scales in monopole that is reduced when either of the methods for correcting the observational systematics is applied.}
\label{fig:sys}
\end{figure}

Figure \ref{fig:sys} presents the multipoles for the eBOSS sample (NGC and SGC separated), comparing the iterative and multilinear regression methods. The monopole from both hemispheres without corrections shows a large spurious correlation at large scales that is reduced when either of the methods for correcting the observational systematics is applied. There is an excellent agreement in both methods for correcting photometric systematics. The SGC does show slightly better performance using the multi-regression method.

 \subsubsection{Correcting for Spectroscopic Completeness}
 \label{sec:Corr_Spec_Comp}
Previous analyses on the eBOSS LRG sample reported that fluctuations in the (S/N) significantly affect the probability of obtaining a confident redshift (See Figure 5 of \citealt{Bautista2017}).
Additionally, the probability of obtaining a confident redshift varies across the
focal plane, decreasing near the edges (See Figure 6 of \citealt{Bautista2017}).
We define the failure rate  as:
\begin{equation}
\eta=\frac{N_{\rm gal}}{N_{\rm zfail}+ N_{\rm gal}}
\end{equation}

where the failure rate in eBOSS LRGs sample is 10$\%$, which is significantly higher than previous surveys; for example, in BOSS the failure rate was only 1.8$\%$ (This is due to eBOSS targeting fainter galaxies than BOSS).

The variations of the failure rate across the focal plane could bias the clustering measurements. In order to account for the effect of this redshift incompleteness, we applied two methods to mitigate the effect on the clustering measurements; in particular, we studied how the two techniques affect the RSD analysis.
\begin{itemize}
\item Nearest-neighbor up-weighting. The procedure followed in BOSS \citep{Reid2012} was to upweight the nearest neighbor with a good redshift and spectroscopic classification in its target class, within a sector. It has been shown that this method  introduces structure into the monopole at small scales, and also modifies the quadrupole amplitude, which could potentially affect the growth factor measurements.

\item Forward-Modeling. This approach uses a probabilistic model that depends on the the position of its fiber in the focal plane and the overall signal-to-noise ratio of the plate. The model for failures is then applied to the random sample by subsampling, mimicking the patterns retrieved in the model. For more details about this modeling we refer the reader to \cite{Bautista2017}.
\end{itemize}

\subsubsection{Data Weights}
We now specify the weights applied for each catalog and the randoms treatment.

\begin{itemize}
\item $w_{\rm SYSTOT}$. As described previously, these weights account for the fluctuations of the observational conditions that can impact the clustering signal.  For the BOSS-like method these weights are computed as described in the Iterative Method.

\item $w_{\rm FKP}$. These weights are used for both set of catalogs. They serve to optimize clustering signal-to-noise ratio for a survey with density varying with respect to the redshift. Also known as FKP weights \citep{Feldman1994}, they  are defined as:
\begin{equation}
w_{\rm FKP}=\frac{1}{1+\bar{n}(z) P_0},
\end{equation}
where $\bar{n}(z)$ is the average comoving density of galaxies as a function of redshift and $P_0$ is the value of the power spectrum at scales relevant for our study ($k = 0.14 h$.Mpc$^{-1}$).
For the eBOSS LRG sample we adopt a value of $P_0 = 10^4 h^{-3}$.Mpc$^3$, which is the same value used in the final BOSS CMASS clustering measurements. \item $w_{\rm CP}$. This weight accounts for the fiber collisions and is used for both catalogs. Targets missed due to fiber collisions do not happen randomly on the sky; they are more likely to occur in overdense regions. For mitigating this effect we followed the up-weighting technique described previously.
\item $w_{\rm NOZ}$. This weight accounts for the redshift failures. For the BOSS-like method this weight is computed for each galaxy following the up-weighting technique described in the previous section.
\end{itemize}

  For the official catalogs these weights are set to 1, as the spectroscopic incompleteness is modeled to subsample the randoms as described in the previous section.

%% file: mocks_Accepted.tex
We use three different sets of mock catalogs in our analysis. The first is a collection of 1000 Quick Particle Mesh (QPM) mocks \citep{2014MNRAS.437.2594W}, which will be used for computing the covariance matrices and for doing several systematic tests. The second one is a set of 1000 Effective Zeldovich approximation method (EZ) mocks \cite{Chuang2015}, that are used to test variance of our fitting methodology.
The third catalog is a set of 84 high-fidelity mocks called CutSky-Mocks \cite{Alam2017}. These catalogs will  be necessary for testing the accuracy of the model used.

\subsection{QPM Mock Catalogs}

We use 1000  realizations of QPM mocks using the following cosmology $\Omega_{\rm M} = 0.29$, $h=0.7$, and $\Omega_{\rm b} h^2 = 0.02247$. A Halo Occupation Distribution (HOD) framework is adopted for populating halos with galaxies following the 5-parameter method described in \cite{2012ApJ...745...16T} but taking into account the HOD parameters tuning to the DR14 eBOSS LRG sample in \cite{2017ApJ...848...76Z}.

 The same boxes were used for generating NGC and SGC mocks, thus there should be a small correlation between them (particularly in the large modes). In order to mitigate this effect, we combined mocks produced by different realizations of the NGC and the SGC.
The mask that we applied to the mocks will be described in Section \ref{sec:masks}.

\begin{figure*}
\subfigure{\includegraphics[width=55mm,height=45mm]{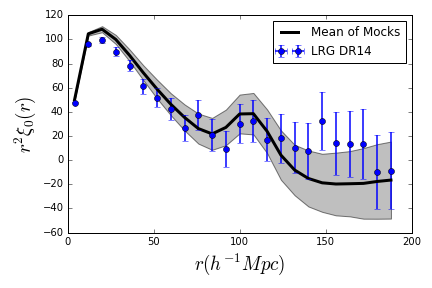}}
\subfigure{\includegraphics[width=55mm,height=45mm]{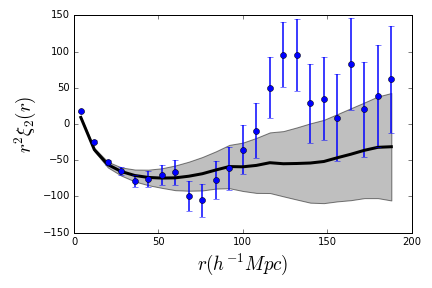}}
\subfigure{\includegraphics[width=55mm,height=45mm]{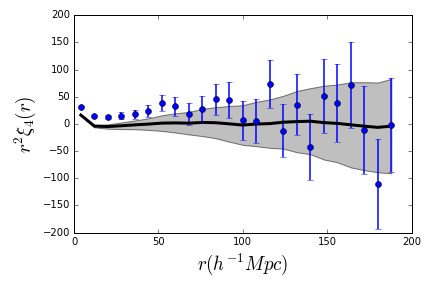}}
\caption{The Black solid lines are the mean of our 1000 QPM mocks for the Monopole (left), Quadrupole (center), and Hexadecapole (right); the shaded regions are the 1-$\sigma$ variations. The blue dots represent the data points and the associated error bars and are equal to the 1-$\sigma$ variation shown in the shaded contours.}
\label{fig:meanmocks}
\end{figure*}

Our QPM mocks are needed for two reasons: to compute an estimate of the covariance matrix and to test our methodology. Figure \ref{fig:meanmocks} shows the mean of the mocks compared with the data; the solid lines represent the mean of the mocks correlation function and the blue dots the data correlation function multipoles with their associated error bars. There is a good agreement between the data  and the mocks for scales larger that 30 $h^{-1}$Mpc; at smaller radii a mismatch appears, which might be related to the resolution of the mocks.

\subsection{EZ Mock Catalogs}
EZ simulations are light-cone mock catalogs created following the Effective Zeldovich methodology described in \cite{Chuang2015}. In order to construct the eBOSS+CMASS sample, the CMASS and eBOSS mocks are calibrated and generated separately and then combined. The CMASS mocks are constructed in four redshift bins: (0.55, 0.65), (0.65, 0.7), (0.7, 0.8), and (0.8, 1.0025), while the eBOSS mocks are constructed at five redshift bins: (0.55, 0.65), (0.65, 0.7), (0.7, 0.8), (0.8, 0.9), and (0.9, 1.05). The fiducial cosmology is a flat $\Lambda$CDM model with $\Omega_{\rm M}$ = 0.307115, $h = 0.6777$, $\sigma_8$ = 0.8225, $\Omega_{\rm b}$ = 0.048206 and $n_{\rm s} = 0.9611$. We will use these mocks to test the variance of the fitting methodology.

\subsection{N-Series Cut Sky Mocks}\label{sec:nseries}
Our N-Series Cut Sky Mock library contains 84 mocks generated with N-body simulations that where done using GADGET2 \citep{2005MNRAS.364.1105S}.
Our mocks have the angular and radial mask of BOSS NGC DR12 based on simulations with $2048^3$ particles in a volume of $(2.6 \,h^{-1}\mathrm{Gpc})^3$ corresponding to resolution particle mass about $1.5\times10^{11} \mathrm{M_{\odot}} h^{-1}$.
We used these mock catalogs to test the theoretical systematics related to our modeling methodology. The N-Series cosmology is $\Omega_{\rm M} =0.286$ , $h=0.7$, $\Omega_{\rm b} =0.047$, $\sigma_8=0.820$, and $n_{\rm s}=0.96$.

%% file: model_Accepted.tex
In order to model the different multipoles of the two-point correlation function, we use the combined Convolutional Lagrangian Perturbation Theory (CLPT) and Gaussian Streaming RSD (CLPT-GSRSD) formalism, developed by \cite{Wang2014}, \cite{Reid2011}, and \cite{Carlson2013}. In this section we briefly describe this theoretical framework.

\subsection{CLPT}

CLPT provides a non-perturbative resummation of Lagrangian perturbation to the two-point statistic in real space for biased tracers. The starting point for the Lagrangian framework is the relation between
the Lagrangian coordinates $\vec q$ that are related to the Eulerian coordinates $\vec x$  as:

\begin{equation}
\vec x (\vec q,t)=\vec q+\vec \Psi(\vec q,t),
\end{equation}
where $ \Psi(\vec q,t)$ is the displacement field at each time $t$.
 The two-point correlation function  is expanded in its Lagrangian coordinates considering the tracer $X$, in our case the Luminous Red Galaxies, to be locally biased with respect to the initially Cold Dark Matter overdensity $\delta (\vec q)$.  The expansion is performed over different orders of the Lagrangian bias function $F[\delta (\vec q) ]$, defined as:

\begin{equation}
1+\delta_X(\vec q,t)=F[\delta(\vec q)].
\end{equation}
The Eulerian contrast density field is computed convolving with the displacements:
\begin{equation}
\label{delta_model}
1+\delta_X (\vec x)=\int d^3 F\left[ \delta(\vec q) \right]\int \frac{d^3 k}{(2 \pi)^3} e^{i \vec k (\vec x - \vec q - \vec \psi(\vec q))}.
\end{equation}
Assuming that the expectation value of the $n^{th}$ derivative of the Lagrangian bias function $F$ is given by:
\begin{equation}
\label{F_der}
\langle F^n \rangle=\int \frac{d\delta}{\sqrt{2\pi} \sigma}e^{-\delta^2/2\sigma^2}\frac{d^n F}{d \delta^n},
\end{equation}
the two-point correlation function is obtained by evaluating the expression $\xi_X(\vec r) = \left< \delta_X (\vec x) \delta_X (\vec x + \vec r)\right>$ corresponding to Eq 19 of  \cite{Carlson2013} and which can be simplified using the bias expansion as in their Eq. 43:
\begin{equation}
\label{xi_model}
1+\xi_X(\vec r)=\int d^3 q M(\vec r, \vec q),
\end{equation}
where $ M(\vec r, \vec q)$ is the kernel of convolution taking into account the displacements and bias expansion up to its second derivative term. The bias derivative terms are computed using  a linear power spectrum (LPS). The LPS that we used was computed using the code CAMB \citep{2000ApJ...538..473L} for a fixed cosmology described as the fiducial cosmology of our analysis.

As we are interested in studying RSD, we also must model the peculiar velocity's effect on the clustering statistic. CLPT can compute the pairwise velocity distribution $\vec{v}_{12}$  and the pairwise velocity dispersion $\sigma_{12}$. This calculation is done following the formalism of \cite{Wang2014} which is similar to the one describe above but modifying the kernel to take into account the velocities rather than the density:

\begin{equation}
\vec{v}_{12}(r)=(1+\xi(\vec{r}))^{-1}\int {M_1}(\vec{r}, \vec{q}) d^3 q,
\end{equation}

and
\begin{equation}
\sigma_{12}(r)=(1+\xi(\vec r))^{-1}\int M_2(\vec r, \vec q)d^3q.
\end{equation}
The kernels $M_{1,2}(\vec{r}, \vec{q})$ also depend on the first two derivatives of the Lagrangian bias $\langle F' \rangle$ and $\langle F'' \rangle$, which are free parameters, in addition to the growth factor $f$, for our model. Hereafter we eliminate the brackets for the Lagrangian bias terms to have a less cumbersome notation in the following sections. 

\subsection{CLPT-GSRSD}

While CLPT generates more accurate multipoles than the Lagrangian Resummation Theory (LRT) from \cite{2008PhRvD..77f3530M} and the linear theory, we still require better performance to study the smaller scales of our quadrupoles. This represents an issue that is particularly important when doing RSD measurements as the peculiar velocities are generated by interactions that occur on the scales of clusters of galaxies.

In order to achieve the required precision, we map the real space CLPT models of the two-point statistics into redshift space following the Gaussian Streaming Model (GSM). This formalism was proposed by \cite{2011MNRAS.417.1913R}. Here, the pairwise velocity distribution of tracers is  assumed to have a Gaussian distribution that is dependent on both the separation of tracers $r$ and the angle between their separation vector and the line-of-sight  $\mu$.

The methodology of using CLPT to model the necessary inputs of a GSM was implemented by \cite{Wang2014}. Its predictions are computed via the following integral:

\begin{equation}
\label{gsrd_integral}
\begin{split}
1+\xi(r_\perp,r_\parallel)= & \int \frac{1}{\sqrt{2\pi (\sigma_{12}^2(r,\mu)+\sigma^2_{\rm FoG})}}[1+\xi(r)]\\
& \times \exp{-\frac{[r_\parallel-y-\mu v_{12}(r,\mu)]^2}{2(\sigma_{12}^2(r,\mu)+\sigma^2_{\rm FoG})}} dy,
\end{split}
\end{equation}
where, as stated , $\xi(r)$, $v_{12}(r)$, and $\sigma_{12}(r)$ are computed from CLPT.

\cite{2011MNRAS.417.1913R} demonstrated that GSM can predict accuracies of $\approx 2\%$ when DM halos are used as tracers. However, not all LRGs are central halo galaxies; approximately $20\%$ of them are satellite galaxies with a peculiar velocity respect to their host halo. Therefore, we need to consider a contribution to the velocity dispersion due to the {\it Fingers of God} (FoG) effects on non-linear scales. We have addressed this point by adding the $\sigma_{\rm FoG}$ parameter to Eq. \ref{gsrd_integral}.

To summarize, given a fiducial cosmology, our model has four free parameters $[f, F',F'', \sigma_{\rm FoG}]$. The cosmology determines the LPS used in the model. The following subsection  describes how we include variations of the cosmological parameters around the fiducial values using the Alcock-Paczynski Test.

%% file: AP_Accepted.tex
We described above the model for the RSD signal given a fixed fiducial cosmology that determines the LPS to be used. However, we can extract additional information by measuring the galaxy clustering along the line-of-sight and perpendicular to the line-of-sight, and we can extract geometrical information via the Alcock-Paczynski (AP) test  \citep{AP}. 
In this work, for extracting AP information, we use the parametrization described in \cite{Xu2012}, \cite{Vargas2014}, and \cite{Anderson2014}, which derives measurements
of the isotropic dilation of the coordinates parametrized by $\alpha$ and the anisotropic warping of the coordinates parametrized by $\epsilon$ \footnote{Note that $\alpha =  1$ and $\epsilon = 0$ for the mocks, if we use their natural cosmology as the fiducial cosmology for the analysis.}. We remind the connection with the other parametrization, that we will further use for comparison with previous works, is given by :
\begin{eqnarray}\label{eq:ae1}
\alpha = \alpha_{\perp}^{2/3} \alpha_{||}^{1/3} \nonumber \,, \\
1+ \epsilon = \left( \frac{\alpha_{||}}{\alpha_{\perp}} \right)^{1/3} \,.
\end{eqnarray}
where $\alpha_\perp$ and $\alpha_{||}$ are defined  in terms of dilation in the transverse and line-of-sight directions.

%% file: methodology_Accepted.tex
\subsection{Fiducial Cosmology}
Our analysis  is performed using the following fiducial cosmology:
\begin{eqnarray*}
 \Omega_{\rm M}=0.31,\\
  \Omega_{\rm \Lambda}=0.69, \\
  \Omega_{\rm k}=0, \\
  \Omega_{\rm b}h^2=0.022, \\
  \Omega_{\rm \nu}h^2=0.00064, \\
  w=-1,\\
  w_{\rm a}=0, \\
  h=0.676, \\
  n_{\rm s}=0.97, \\
   \sigma_8=0.8.
\end{eqnarray*}
This fiducial cosmology is different from the ones used to compute the mocks; this additional bias in our methodology has to be considered. This extra bias will be defined in Section \ref{section:systematics}.

\begin{table*}
\caption{Expected values of cosmological parameters for the QPM mocks and Fiducial Cosmology  at different redshift ranges/model. The units for H(z) are $\mathrm{km\:s^{-1}Mpc^{-1}})$ and ($\mathrm{Mpc})$ for $D_A(z)$.}
\label{tab:expectedvalues}
\begin{tabular}{@{}lccccccc}
\hline
Model &
$z$-range&
$z_{\rm eff}$&
$f(z)$&
$\sigma_8(z)$&
$f(z)\sigma_8(z)$&
$H(z)$&
$D_{\rm A}(z)$\\
\hline
\\[-1.5ex]
QPM&$[0.6,1.0]$&0.72&0.806& 0.557&0.449&-&-\\
Fiducial&$[0.6,1.0]$&0.72&0.819&0.550&0.450&101.94&1535\\
Nseries &$[0.43,0.7]$&0.5&0.740& 0.637&0.471&-&-\\
\\[-1.5ex]
\hline
\end{tabular}
\end{table*}

\subsection{2PCF Estimator}

The following section will describe the methodology used to compute the two-point clustering statistics of the DR14 LRG sample described in Section \ref{section:data}.

We are interested in constraining RSD parameters. Therefore, we must study the clustering of galaxies in two directions: the one parallel to the LOS, where peculiar velocities of infalling galaxies generate RSD, and its perpendicular direction, where no distortion occurs. We decompose the vector $\vec r$, which represents the distance between two galaxies, into two components:  $r_{||}$ parallel to the LOS  and $r_{\perp}$ that is perpendicular to it:
\begin{equation}
r^2=r^2_{||}+r^2_{\perp}.
\end{equation}

Let $\theta$ denote the angle between the galaxy pair separation and the LOS direction, and let $\mu$ be defined as  $\mu=\cos \theta$. We then have the relation:
\begin{equation}
\mu^2=\cos^2 \theta = \frac{r_{||}^2}{r^2},
\end{equation}
 and our two direction parameters will be $[r,\mu]$.

The 2D-correlation function $\xi(r,\mu)$
is  computed using the Landy-Szalay estimator \citep{LZ}:
\begin{equation}
\label{LS_estimator}
\xi(r, \mu)=\frac{DD(r, \mu)-2DR(r, \mu) +RR(r, \mu)}{RR(r, \mu)},
\end{equation}
where $DD(r,\mu), RR(r,\mu)$, and $DR(r, \mu)$ are the number of pairs of galaxies which are separated by a radial separation $r$ and angular separation $\mu$. The three symbols represent the data-data, random-random, and data-random pairs, respectively.

The multipoles are Legendre moments of the 2D-correlation function $\xi(r, \mu)$, and can be computed using the following equation:
\begin{equation}
\xi_{\ell}(r) = \frac{2\ell +1}{2} \int_{-1}^{+1} d\mu \; \xi(r,\mu) \; L_{\ell}(\mu),
\end{equation}
where $L_{\ell}(\mu)$ is the $\ell$-th order Legendre polynomial.

We will focus primarily on the monopole, the quadrupole, and the hexadecapole ($\ell=0$, $\ell=2$, and $\ell=4$).

The pair-counts were computed using the public code CUTE \citep{2012arXiv1210.1833A}. However, there are three corrections to be considered when using the LS equation (\ref{LS_estimator}):

\begin{itemize}
\item The number of galaxies in the Data catalogs ($N_D$) is approximately 50 times smaller than the ones in our random catalogs ($N_R$). Therefore the Random and Data pairs should be compared as $$\frac{DD(r,\mu)}{RR(r,\mu)}\times \frac{N_R(N_R-1)}{2}\times \frac{2}{N_D(N_D-1)}.$$
\item The number of galaxies in the SGC ($N_{D,S}$) is smaller than those in the NGC ($N_{D,N}$). Therefore the total number of pairs should be added as: $$DD(r,\mu)=\frac{2(DD_N(r,\mu)+DD_S(r,\mu))}{(N_{D,N}(N_{D,N}-1)+N_{D,S}(N_{D,S}-1))}.$$
\item Each galaxy has a particular weight $w_i$ as described in Section \ref{section:data}. Hence, the total number of galaxies in any catalog is weighted as $$N^w=\sum w_i.$$
\end{itemize}

\subsection{Fitting}
\label{Fit_metho}
Unless stated otherwise, we will be using 13 bins of $8h^{-1}$Mpc in width, in the interval between [$28h^{-1}$Mpc, $124h^{-1}$Mpc]. Given that we will be working with either the first two non-zero multipoles or the first three (depending on the test), the analysis will have a total of either 26 or 39 bins.

We will now compare our measured two-point statistics with those predicted by our model and try to find the best-fit model parameters.

In order to find identify best-fit  parameters, we
minimize the $\chi^2$ function,
\begin{equation}
\label{chi2}
\chi^2 = (\vec{m} - \vec{d})^T C^{-1} (\vec{m}-\vec{d})
\end{equation}
where $\vec{m}$ is the vector formed by the model predictions, and $\vec{d}$ is the equivalent vector observed from our data. Examining Eq. \ref{chi2} reveals that the smaller the value of $\chi^2$, the more similar $\vec{m}$ is to $\vec{d}$.

The sample covariance is defined as:
 \begin{equation}
\label{cov_mat_nonorm}
C_{ij}=\sum_{m=1}^{N_{\rm mocks}} (\xi^m_i-\bar\xi_i)(\xi^m_j-\bar\xi_j),
\end{equation}
where $N_{\rm mocks}$ is the number of mocks, and $\bar\xi_i$ is the average of the $i^{th}$ bin.

We scale the
inverse sample covariance matrix, $C^{-1}_s$, using Eq. 17 of \cite{2007A&A...464..399H}:
\begin{equation}
C^{-1} = C^{-1}_s \frac{N_{\rm mocks} - N_{\rm bins} - 2}{N_{\rm mocks} - 1}.
\end{equation}
This procedure corrects for the fact that the matrix in Eq. \ref{cov_mat_nonorm} is a biased estimate of the true inverse covariance matrix $C^{-1}$.

\begin{figure*}
\includegraphics[width=80mm]{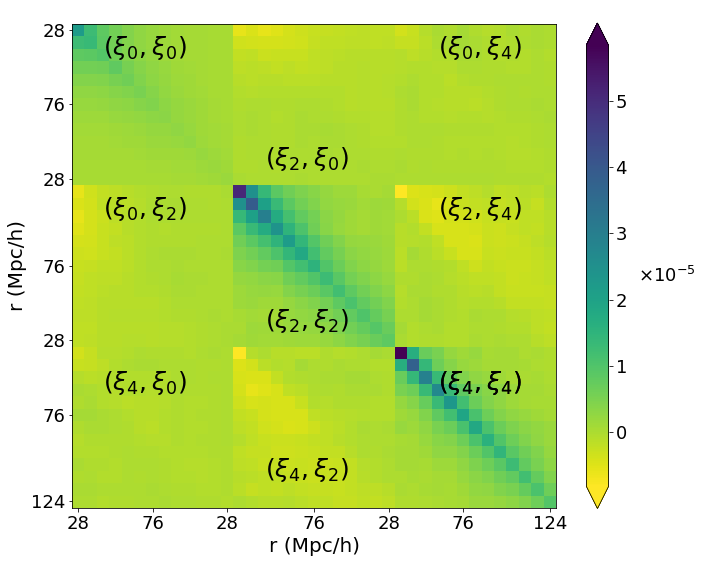}\includegraphics[width=80mm]{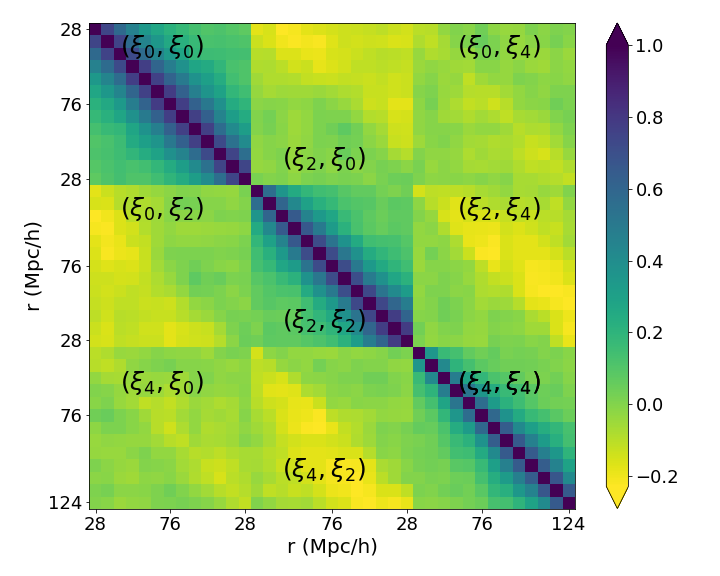}
\caption{Density map of the Covariance matrix (left) computed from our 1000 QPM Mocks simulations. The matrix has 13 bins in $r$ and 3 multipoles. The right panel presents the correlation matrix defined as $Corr_{ij}=\frac{C_{ij}}{\sqrt{C_{ii}C_{jj}}}$. The normalization is done such that the diagonal is always unity, and it shows how much covariance (off-diagonal) there is compared to our variance (diagonal). }
\label{fig:covariance}
\end{figure*}

Figure \ref{fig:covariance} shows the covariance and correlation matrix computed from 1000 QPM mocks. Most of our error arises from the elements on the diagonal (variance of a given bin), but there is also a significant contribution coming from elements outside of the diagonal (covariance between different bins).

In order to identify the best-fit parameters, we minimize the $\chi^2$ function.The minimization of the $\chi^2$ is done using the Powell algorithm \citep{2002nrca.book.....P}. This algorithm will find a unique solution if the parameter space is gaussian, which should be a fair assumption when fitting our mocks. This method is adequate for our work as it does not require us to compute the gradient of the CLPT-GSRD model with respect to the model parameters, which would be challenging. Due to the nature of the algorithm it is not necessary to specify any prior, just some starting points, if our assumption about the parameter space being somewhat gaussian is correct then any starting point should work fine and one that is close to the Best Fit should reduce the running time.

The estimate of the errors on our fits will be computed using MCMC chains, but we will only do this analysis for our data sets (Section \ref{section:results}) and not for the mocks.

%% file: Nseries_Accepted.tex
\subsection{Testing Accuracy of GSRSD Hexadecapole Model with High-Resolution Simulations}
\label{High_resolution_sys}
This section is dedicated to testing the performance of the methodology developed in Section \ref{section:metho}. Here, we will use the N-Series CutSky mocks described in Section \ref{section:mocks} to check the reliability of the CLPT model with regards to recovering the cosmological parameters. These high resolution mocks are built with the BOSS-CMASS properties that allow us to study the accuracy of the model.
We will run our fitting methodology on these high-fidelity mocks in order to test if their fiducial parameters can be recovered. The N-Series CutSky mocks have been used previously in the literature for testing the monopole- and quadrupole-only methodologies.

We fit our N-series CutSky mocks twice, the first using only the monopole and the quadrupole, and the second including the hexadecapole.  The fits are done following the methodology described in Section \ref{Fit_metho}, but here we will be using 21 bins of $5 h^{-1}$ Mpc in width, in the interval between [$27.5 h^-1$ Mpc, $127.5 h^-1$ Mpc]. We decided to choose a smaller bin-size to facilitate comparisons with other previous results. The sample covariance matrix used to perform these fits is computed using the QPM-BOSS CMASS sample re-scaled to match the mocks volume, and provides our error estimate (the covariance matrix obtained from N-Series would be quite noisy given the limited number of realizations available). The pair-counts of our mocks were computed using the mocks cosmology to transform angular positions and redshifts into comoving coordinates. To be consistent, the CLPT-GSRSD input template was also computed using the cosmology of the mocks.

The expected values of the linear growth rate of $f$ ($f_{\rm exp}$) are reported in Table \ref{tab:expectedvalues} for the natural cosmology of the mocks.
We define the bias of the growth factor estimation $b_f$ as:
 \begin{equation}
    b_f=\left< f_{\rm measured} \right >-f_{\rm exp}  \label{eq:biasf}
 \end{equation}
 We use the measurement of the Eulerian bias $b=2.3$ performed by \cite{2017ApJ...848...76Z} as reference. This estimate was computed using our same sample with the addition of an HOD model.

\begin{figure*}
\subfigure{\includegraphics[width=85mm,height=90mm,trim = 1.1cm 2cm 2cm 2cm, clip]{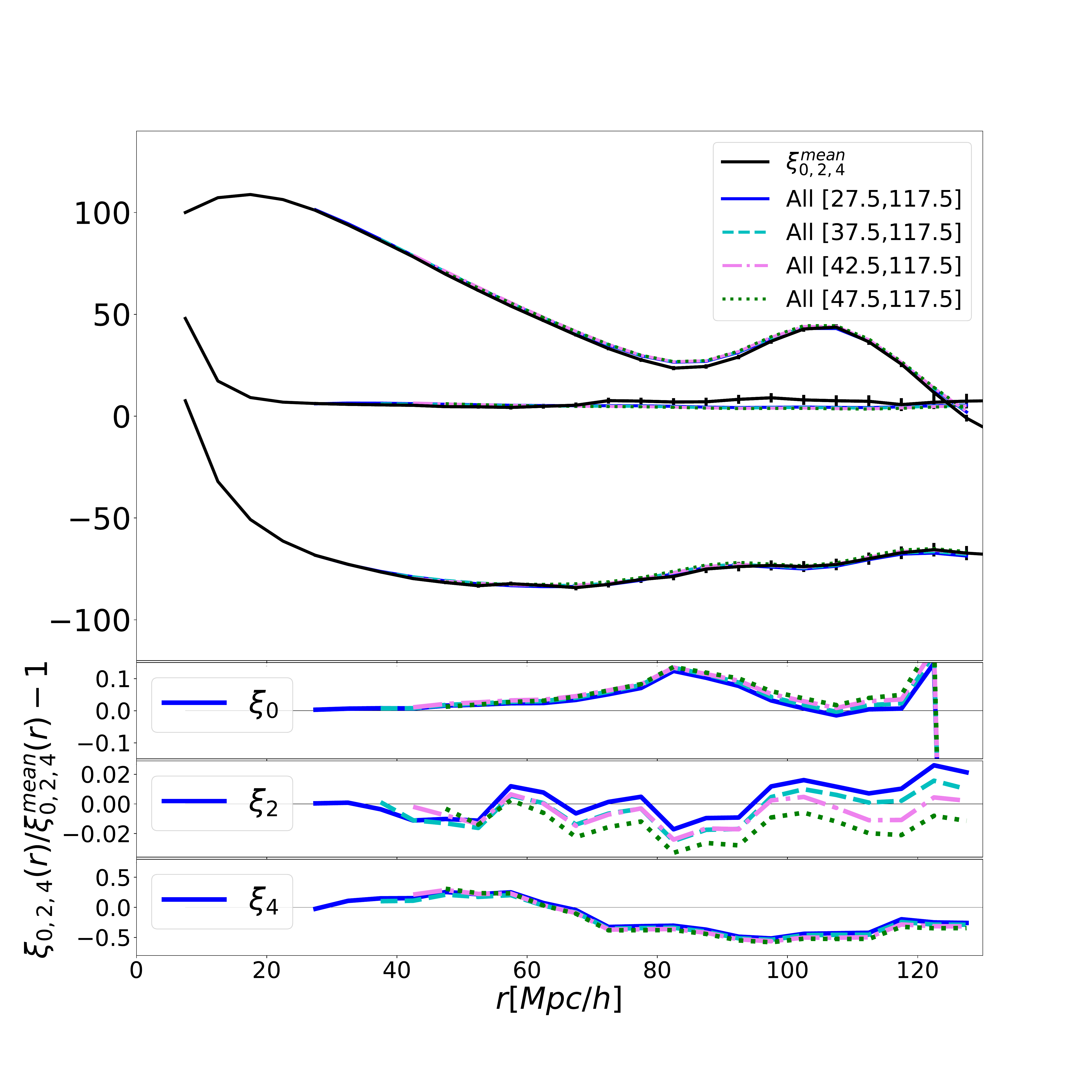}}
\subfigure{\includegraphics[width=85mm,height=90mm,trim = 1.1cm 2cm 3.5cm 2cm, clip]{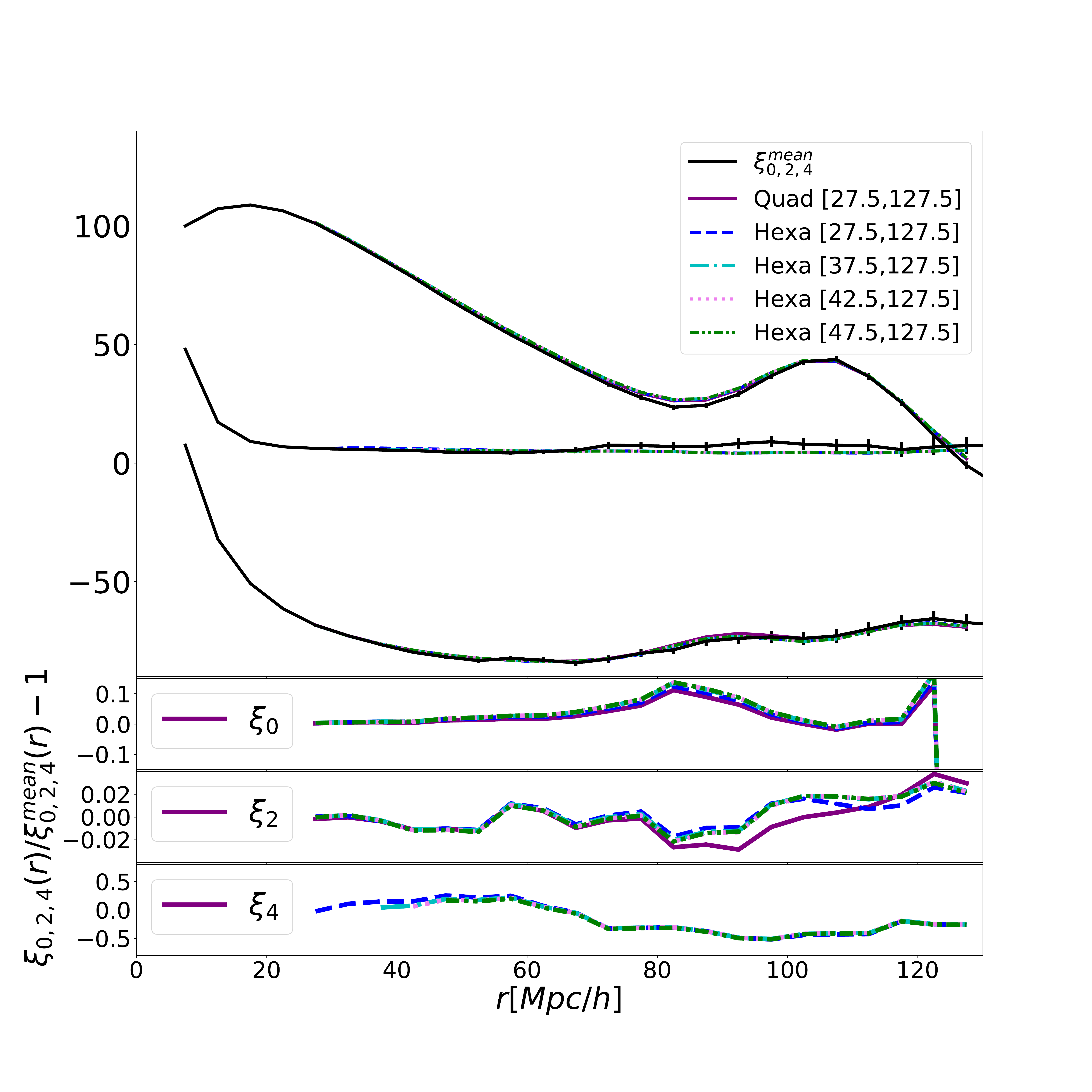}}
\caption{The mean of the mocks is shown as the black line in both plots. The error bars are computed from the re-scaled QPM mocks covariance. The left panel shows the best fit models from different lower ranges of the multipoles. In the right panel only the lower range of the hexadecapole is varied. The error plots show the quotient between the best fit model and the mean of the mocks. For all cases are residuals in the hexadecapole, the smaller residuals are obtained by the monopole + quadrupole.}
\label{fig:mean_nseries}
\end{figure*}

The left panel of Figure \ref{fig:mean_nseries} shows the mean of the multipoles; the error bars are the diagonal terms of the covariance matrix divided by $1/\sqrt{N_{\rm Mock}}$. The different colors (and line-styles) represent the best fit model for the mean of the mocks using the fiducial range at different minimum scales of the fits when the cuts are applied to all multipoles.

 The model using the hexadecapole fitted at the full range (blue lines) does not match the hexadecapole of the mean of the mocks accurately at any scale, this can be seen in the corresponding residual plot (bottom panel of the figure) where the value of the residuals is close to 50\% of the value of the model, this is very  large when compared to the residuals of the monopole and quadrupole that are around 10\% and 2\% respectively (second and third panels of the figure).
Increasing the minimum range of the fit  mostly affects the quadrupole at large scales and has little effect on the monopole and hexadecapole at any scale. By comparing the residuals of the quadrupole(third panel of the figure) of the full-range fit and the reduced-range fit we can tell that the full range fits adjusts the quadrupole  better (i.e. the blue solid line is a better fit than the green dotted one).

The right panel of Figure \ref{fig:mean_nseries} displays a similar exercise to the one in the left panel, except this time we only cut the minimum scale of the hexadecapole while leaving the other two multipoles in the full range. The changes on the quadrupole are now less severe than when varying all multipoles. By looking at the residual plot of the quadrupole (third panel of the figure) we see that the model considering the hexadecapole in the full range (green dash-dotted line) matches the quadrupole slightly better than the model using only the monopole and quadrupole (purple solid line), but it does not improve the other multipoles significantly. It is not clear that including the hexadecapole improves the fits significantly when compared to the monopole and quadrupole only case.

Table \ref{tab:MeanFitsNSERIES} reveals that the bias in $f$ is slightly larger when we include the hexadecapole in the full range than when we only use the monopole and the quadrupole ($b_f=0.005$ compared to $b_f=0.004$). However, the bias in $\epsilon$ is smaller when the hexadecapole is left out of the fits ($b_\epsilon=0.002$ compared  to $b_\epsilon=0.0005$). The bias in alpha is the same for both cases  ($b_\alpha=0.001$). The right panel of Figure \ref{fig:mean_nseries} shows that the best fit model for both cases are very similar, only showing small differences in the quadrupole at the scales in the range [80,110]  $h^{-1}$Mpc.

Reducing the range for all multipoles (second block of the table) increases the biases in $f$ and $\epsilon$. If by contrast we constrain the range only for the hexadecapole (third block), we reduce the bias in $f$ and $\alpha$  to $b_f=0.001$  and $b_\alpha<0.001$, respectively, leaving the bias value for $\epsilon$ unchanged.

In summary, there is no clear preference between the case with the 3 multipoles and just considering monopole and quadrupole. There is a trade-off between the biases in $\epsilon$ and $f$: the smaller bias in $f$ is obtained when using the hexadecapole while the smaller bias in $\epsilon$ comes from using only the monopole and quadrupole. As there is not a clear trend we will explore the hexadecapoles impact on the LRG sample analysis further.

\begin{figure*}
\subfigure{\includegraphics[width=80mm,height=80mm]{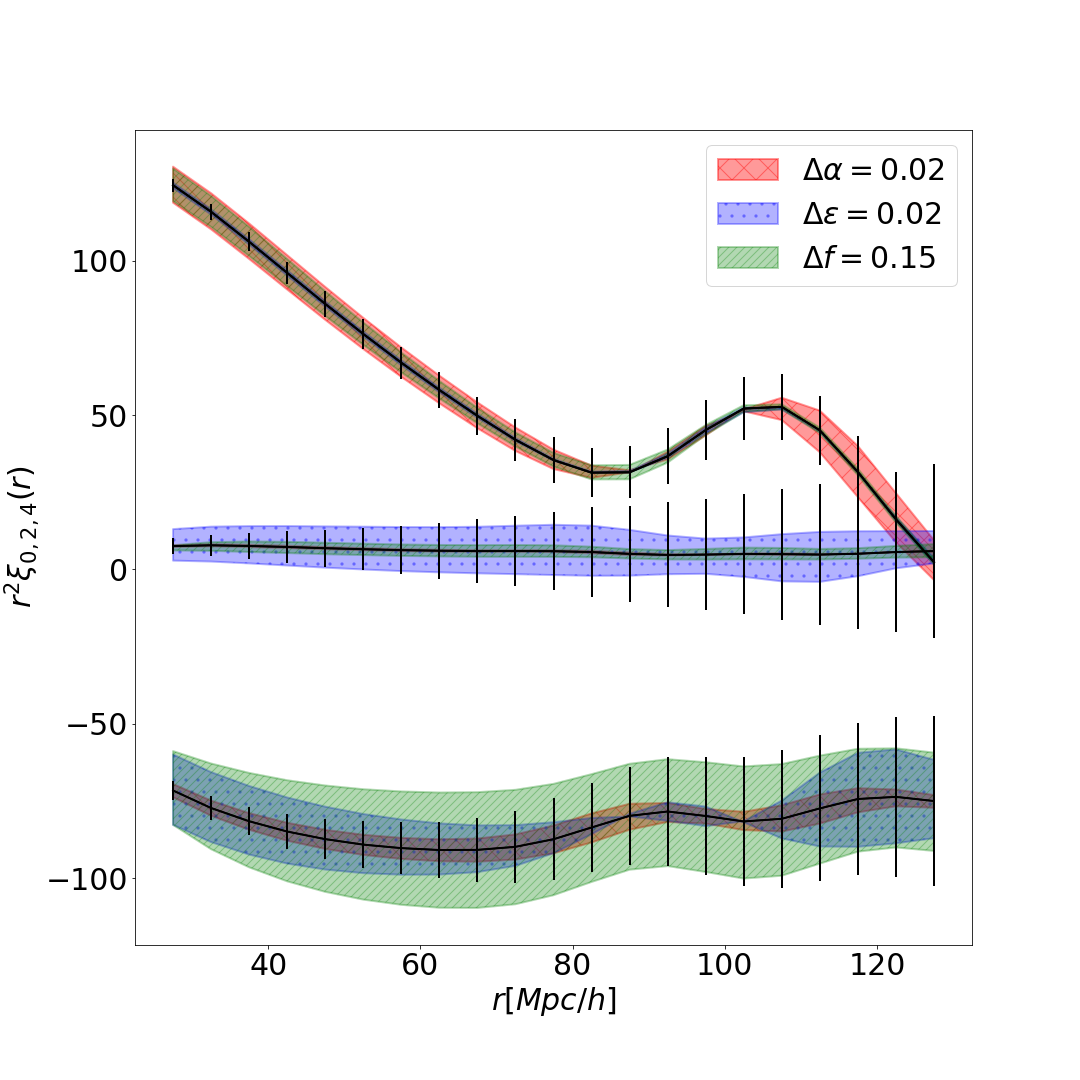}}
\subfigure{\includegraphics[width=80mm,height=80mm]{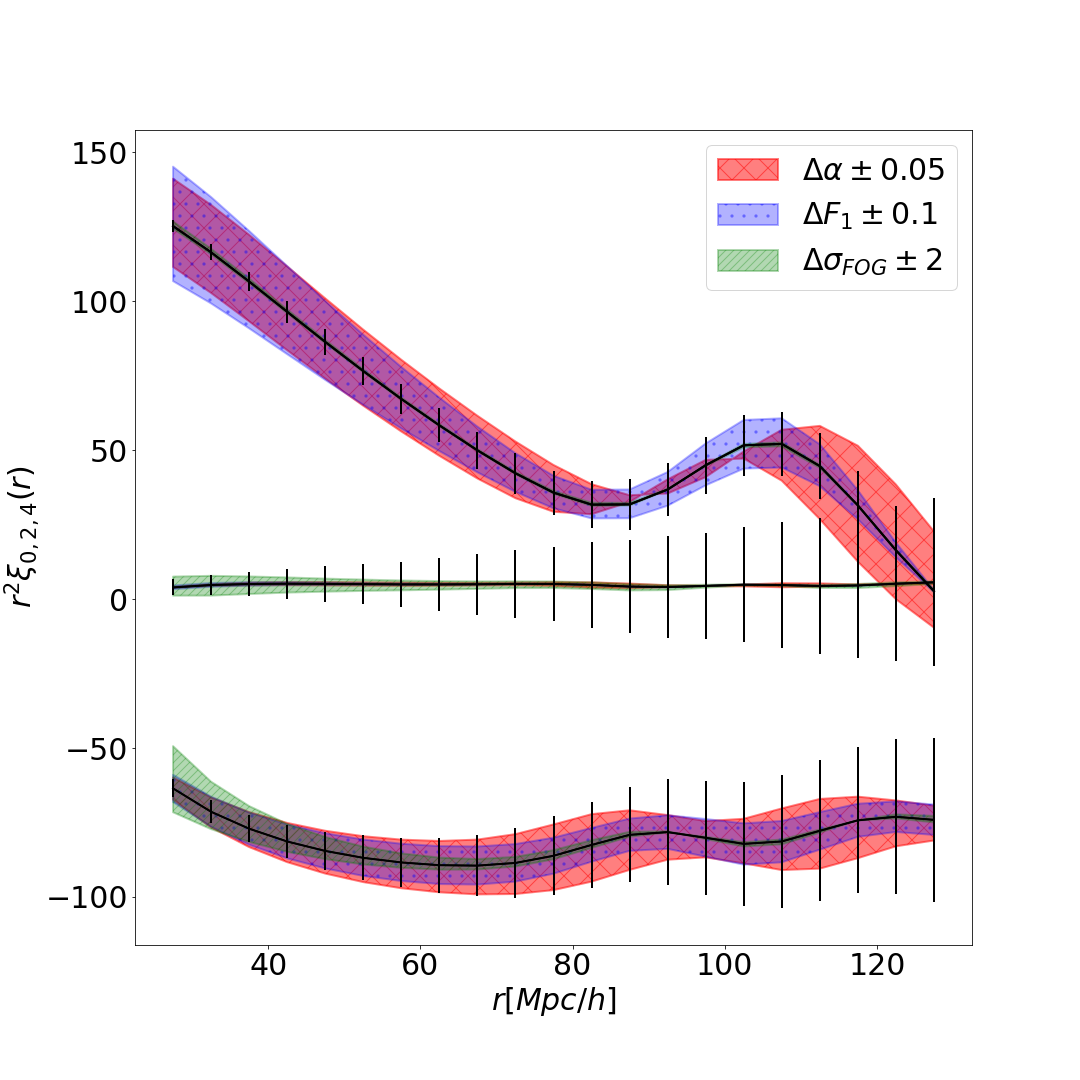}}
\caption{Left panel: Model variations for $\Delta f$, $\Delta \alpha$, and $\Delta \epsilon$ compared to the error bars coming from covariance: the behavior for the monopole (top), the hexadecapole (middle) and the quadrupole (bottom). The variations in $\epsilon$ have a large impact on the hexadecapole, while the variations in $\alpha$ and growth factor are of the same order of magnitude for the hexadecapole. This   behavior explains why fits are driven by $\epsilon$ when the hexadecapole is taken into account. Right panel: Model variations for $\Delta F'$, $\Delta \alpha$, and $\Delta \sigma_{\rm FoG}$ compared to the errors produced by covariance: the behavior for the monopole (top), the hexadecapole (middle) and the quadrupole (bottom).}
\label{fig:modelclpt}
\end{figure*}

Figure \ref{fig:modelclpt} displays the model behavior for variations of the parameters, and is included to explain the different trends observed with mocks when using multipoles up to order $\ell=2$ compared to $\ell=4$. We also indicate the variations in our model predicted by changes of $\sim20\%$ in the input parameters, that correspond to deviations of $\Delta f=0.15$, $\Delta \alpha=0.2$, and $\Delta \epsilon=0.2$ around the fiducial cosmology expected value. The error bars were obtained from the diagonal of the mocks covariance matrix.
The variations in $\epsilon$ have a large impact on the predicted hexadecapole at all scales (middle curve), while the variations of the hexadecapole due to variations on $\alpha$ and $f$ are significantly smaller and of a comparable order of magnitude. This behavior explains why the fits are driven by $\epsilon$ when the hexadecapole is included. Considering that the error bars between 20 and 60 $h^{-1}$Mpc are smaller, their constraining power is significantly larger.

As stated before, even if our results using the hexadecapole do not show significant biases, figure \ref{fig:mean_nseries} shows that the model obtained using the cosmology of the mocks does not accurately match the mean of the hexadecapole mocks at any scale, in particular at the lower scales that have more weight in the likelihood.
This mismatch in the hexadecapole is pushing $\epsilon$ to higher values and as a consequence the correlated parameters follow. Therefore, the accuracy of the model at all scales is critical for not biasing the fitted parameters.

\begin{table*}
\caption{Results from fitting the mean of N-series Mocks. The expected values for the N-series mocks are f (z = 0.5) = 0.740, $\alpha$ = 1.0 and  $\epsilon$= 0.0. The fits are done over bins of $5 h^{-1}$ Mpc each so that the full range of each multipole (27.5h $-1$ Mpc, 127.5h $-1$ Mpc) will have 21 bins.}

\label{tab:MeanFitsNSERIES}
\begin{tabular}{@{}lccccccccc}
\hline
\multicolumn{9}{c}{$\xi_0+\xi_2 $ with cuts in all multipoles}\\
\hline
Model&Range&$F'$&$F''$&$ f$&$\alpha$&$\epsilon$&$\sigma_{\rm FoG}$&$\chi^2/$d.o.f\\
\hline

$\xi_0+\xi_2$&27.5-127.5&0.999& 0.637& 0.736& 1.001& 5e-4 & 1.076&68.5/36=1.9\\
\hline
\multicolumn{9}{c}{$\xi_0+\xi_2 +\xi_4$ with cuts in all multipoles}\\
\hline

$\xi_0+\xi_2+\xi_4$&27.5-127.5&1.003& 1.034& 0.745& 1.001& -0.002&1.770& 91.2/57=1.60\\

$\xi_0+\xi_2+\xi_4$&37.5-127.5&1.014&1.708&0.735& 1.001& -0.003& 2.239& 84.0/51=1.65\\

$\xi_0+\xi_2+\xi_4$&42.5-127.5&1.022&1.870& 0.731& 0.999&-0.004& 0.530& 78.7/48=1.64 \\

$\xi_0+\xi_2+\xi_4$&47.5-127.5&1.027& 3.149& 0.721& 0.997& -0.004& 1.018&70.8/45=1.57\\
\hline
\multicolumn{9}{c}{$\xi_0+\xi_2+\xi_4$ with a cut in hexadecapole only}\\
\hline

$\xi_0+\xi_2+\xi_4$&37.5-127.5&1.010&1.543& 0.742&1.000& -0.002&2.793&86.15/55=1.57\\
$\xi_0+\xi_2+\xi_4$&42.5-127.5&1.011& 1.649& 0.741& 1.000& -0.002& 2.938& 86.18/54=1.60\\

$\xi_0+\xi_2+\xi_4$&47.5-127.5&1.012&1.697&0.741& 1.000& -0.002&2.984& 86.28/53=1.62\\
\hline
\end{tabular}
\end{table*}

\begin{figure*}
\includegraphics[width=160mm,trim = 4.5cm 5cm 6.5cm 2cm, clip]{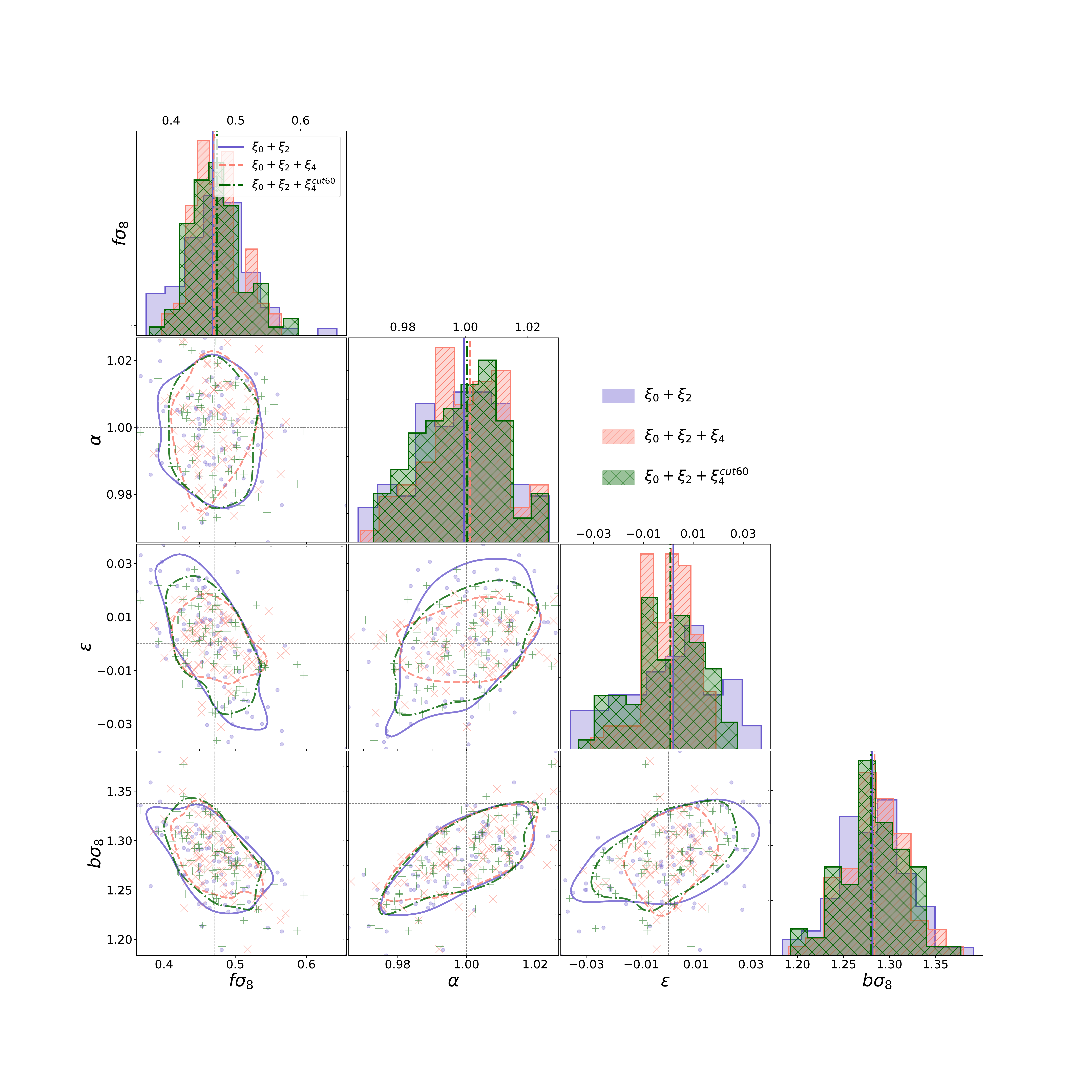}
\caption{Results from the best fits of all of the individual mocks for the four parameters of interest [$f$, $b$, $\alpha$, $\epsilon$]. Also shown are their respective best fits distributions histograms and the $1\sigma$ confidence region. The dotted black lines represent the expected value of each parameter. The colored lines in each histogram indicate the mean value of that parameter found by our fits. We present three cases: 1) fitting the complete range [27.5,127.5] $h^{-1}$Mpc using monopole and quadrupole (blue dots), 2) fitting the complete range [27.5,127.5] $h^{-1}$Mpc using monopole, quadrupole, and hexadecapole (red x's), and 3) fitting the complete range [27.5,127.5] $h^{-1}$Mpc for monopole and quadrupole and reducing the range to [47.5,127.5] $h^{-1}$Mpc for the hexadecapole (green crosses).}
\label{fig:triangular_nseries}
\end{figure*}

We now analyze the individual mocks for three cases: 1) fitting the complete range [27.5,127.5] $h^{-1}$Mpc using monopole and quadrupole, 2) fitting the complete range [27.5,127.5] $h^{-1}$Mpc using monopole, quadrupole, and hexadecapole, and 3) fitting the complete range [27.5,127.5] $h^{-1}$Mpc for monopole and quadrupole and reducing the range to [47.5,127.5] $h^{-1}$Mpc for the hexadecapole. Figure \ref{fig:triangular_nseries} show the results of the individual fits in all three cases and for the four parameters of interest [$f\sigma_8$, $b\sigma_8$, $\alpha$, $\epsilon$], as well as their respective best fit distributions histograms. The colored dashed lines indicate the mean of the best fits, and the dotted line represents the expected value of the parameters.
Table \ref{tab:nseries} presents the results of the individual fits for the parameters of interest.

The monopole- and quadrupole-only fits show a bias in the estimation of the three parameters of $|b_{f\sigma_8}|=0.003$, $|b_\alpha|=0.002$, and $|b_\epsilon|=0.0004$. The standard deviation of the distributions are $S_{f}=0.051$, $S_\alpha= 0.014$, and $S_\epsilon=0.019$ respectively; the expected values are within the dispersion.
Thus the significance of the biases are $0.5\sigma, 1.1\sigma$ and $0.2\sigma$. These numbers are in agreement with the test performed for the BOSS sample and these numbers are comparable with the results obtained in \cite{Alam2017} \footnote{The BOSS analysis only reported the result of $\Delta f\sigma_8$ for the N-Series Mocks Challenge.}.
The full range hexadecapole fits show a lower  bias in the $f$ parameters, with a value of $|b_{f\sigma_8}|=0.0007$,
$|b_\alpha|=0.00008$ and $|b_\epsilon|=0.0002$ respectively. The standard deviation of the distributions decreases for $f$, $\alpha$ and $\epsilon$, with $S_{f}=0.037$, $\alpha=0.013$ and $S_\epsilon=0.010$.
The significance of the biases decreases significantly to $<0.1\sigma$, $0.1\sigma$, and $0.2\sigma$ respectively.
Constraining the range of hexadecapole fits, produces biases of $|b_{f\sigma_8}|=5e-4$, $|b_\alpha|=0.001$, and $|b_\epsilon|=5e-4 $, while also decreasing the standard deviation of the distributions compared with the monopole and quadrupole fits, $S_{f\sigma_8}=0.042, S_\alpha= 0.013$ and $S_\epsilon=0.014$, giving a significance of the biases of $0.1\sigma, 0.9\sigma$, and $0.4\sigma$, respectively.

\begin{figure*}
\subfigure{\includegraphics[width=55mm,height=50mm]{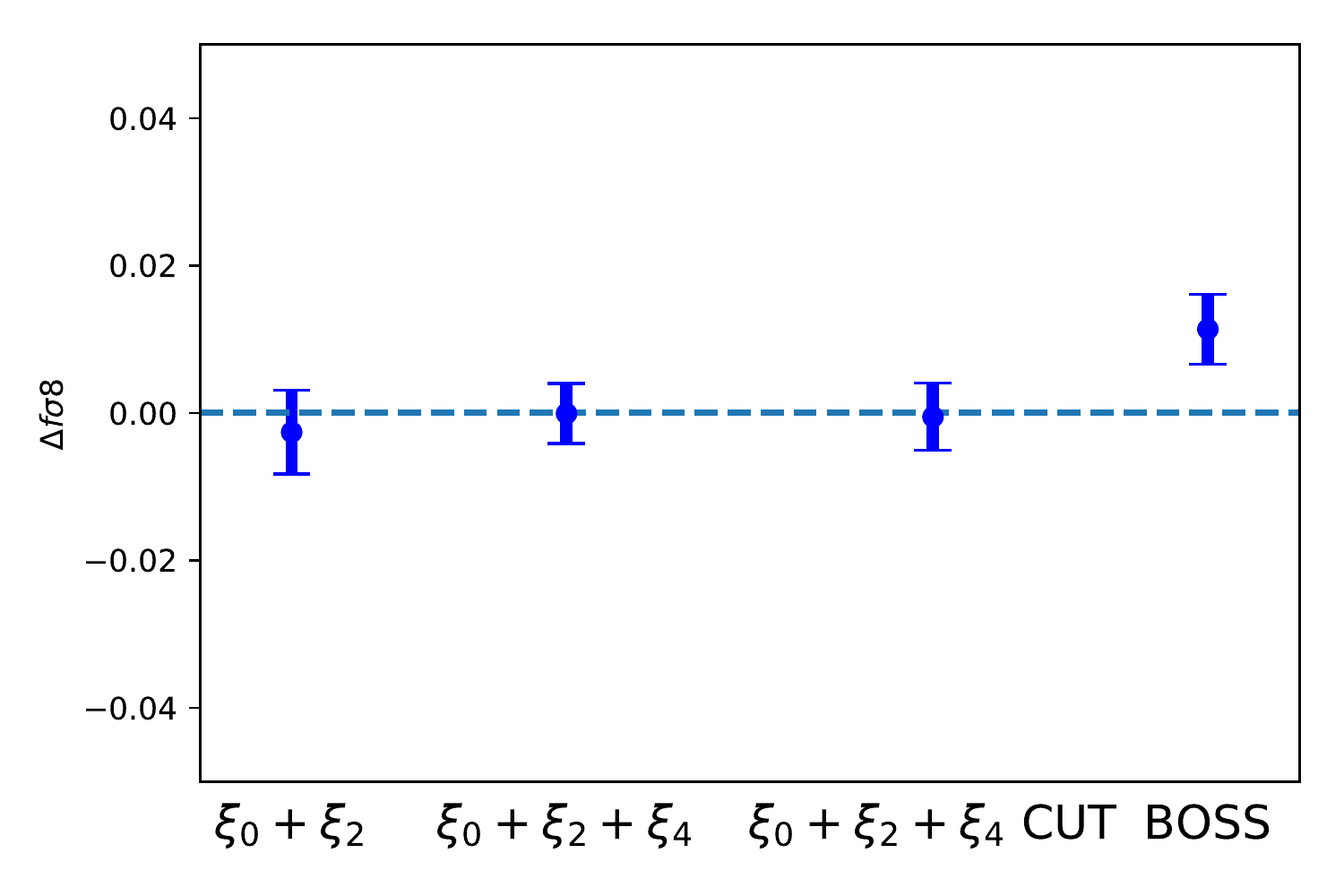}}
\subfigure{\includegraphics[width=55mm,height=50mm]{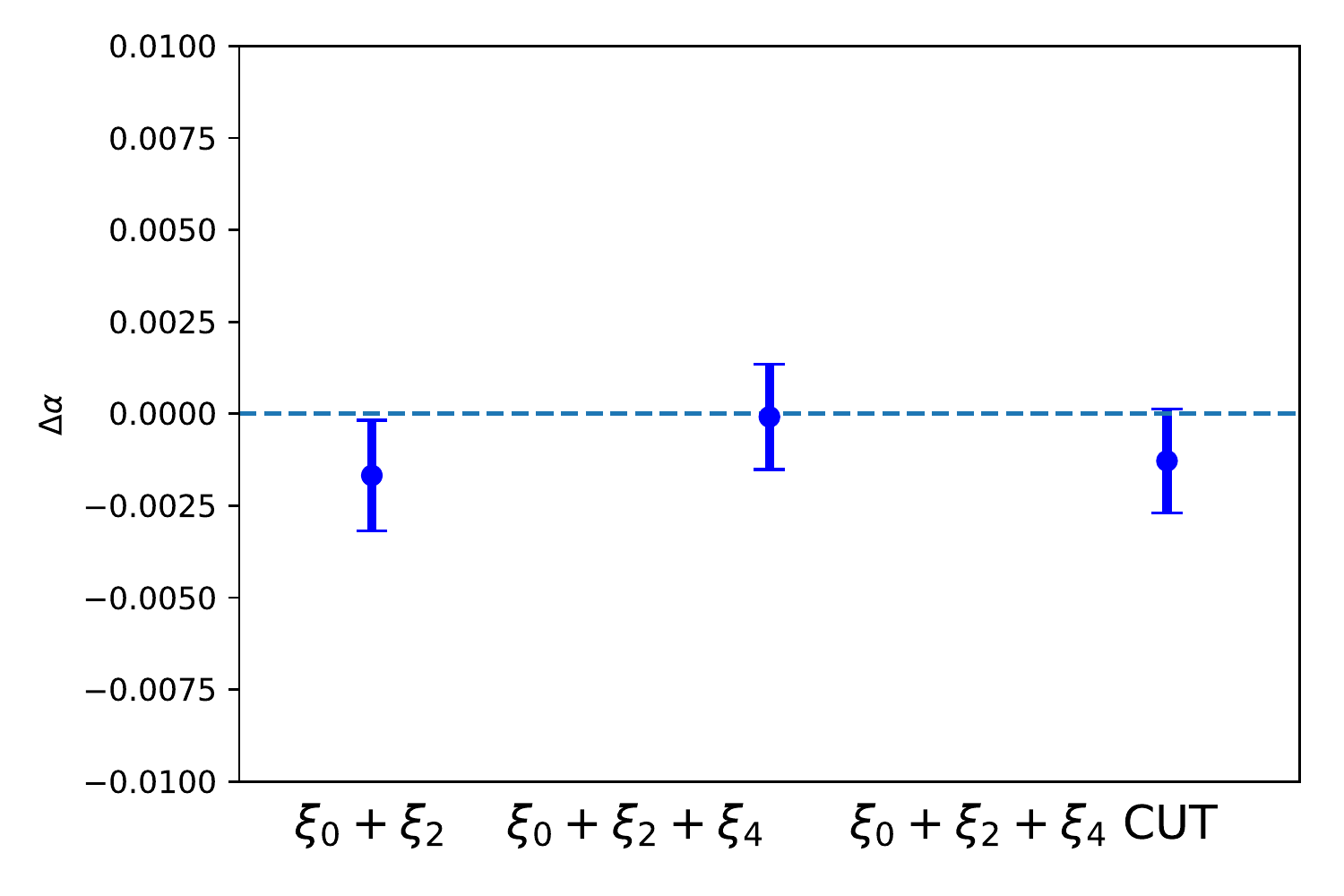}}
\subfigure{\includegraphics[width=55mm,height=50mm]{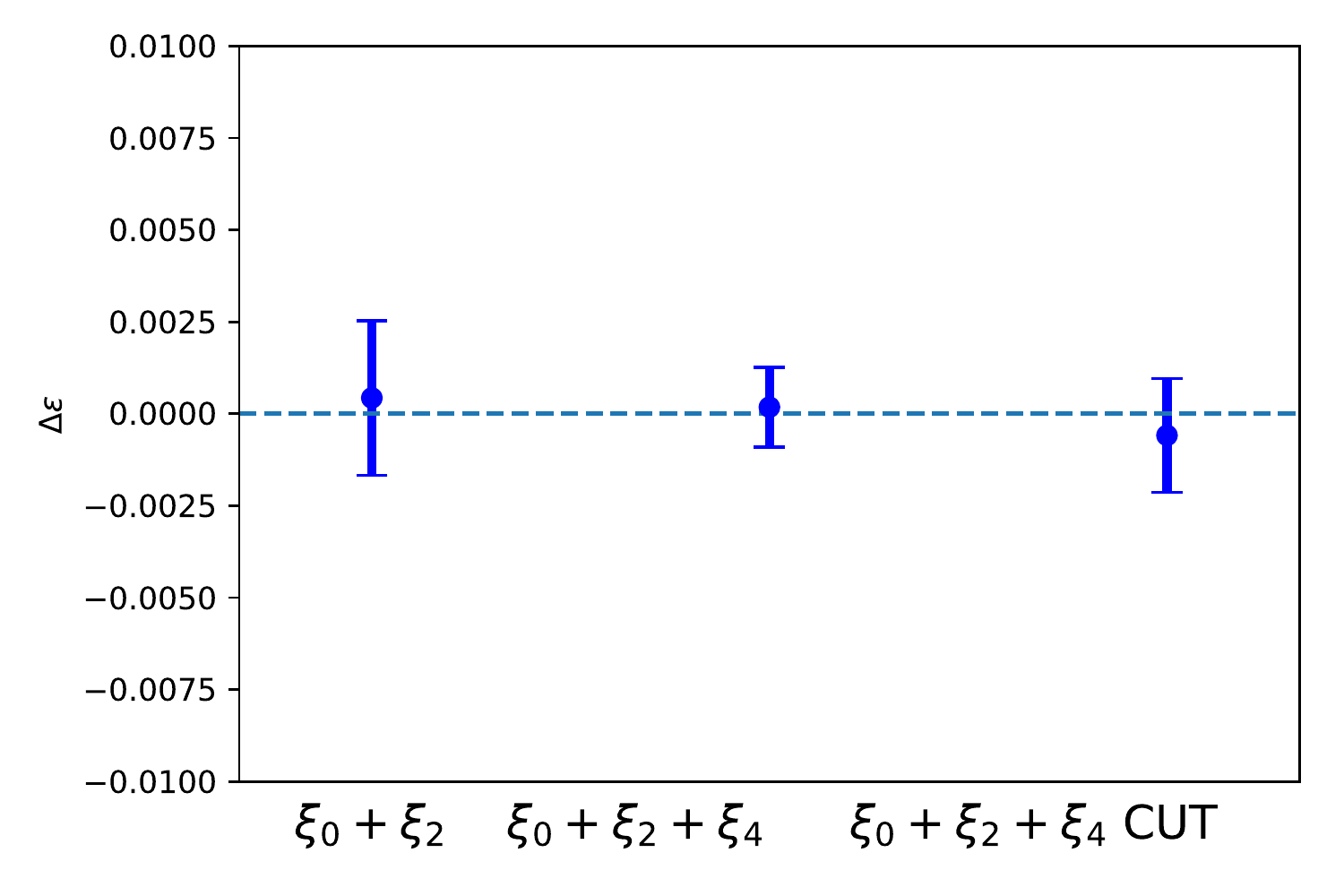}}
\caption{Systematic errors in RSD and AP parameters from using different multipole combinations in the fit. From left to right: mean of $\Delta f= f-f_{\rm exp}$,  $\Delta \alpha= \alpha -\alpha_{\rm exp}$, and $\Delta \epsilon= \epsilon -\epsilon_{\rm exp}$. These measurements were obtained from fitting N-Series sky mocks using two configurations: 1) multipoles up to order $\ell=2$ and 2) multipoles up to order $\ell=4$. The left panel includes the result from the previous work \citep{Alam2017}. The less significant biases are obtained by the monopole + quadrupole fits. Including the hexadecapole reduces the bias and variance producing more significant bias in f and larger biases in $\alpha$ and $\epsilon$.}
\label{fig:nseries}
\end{figure*}

Figure \ref{fig:nseries} shows the summary of the analysis for our three cases in the same format as the results reported in \cite{Alam2017}: the points correspond to the mean of the results obtained from fitting our 84 SkyCut with the BOSS mask mocks, the three quantities shown are (from left to right) the mean of $\Delta f= f-f_{\rm exp}$,  $\Delta \alpha= \alpha -\alpha_{\rm exp}$, and $\Delta \epsilon= \epsilon -\epsilon_{\rm exp}$, and the error indicated is the standard deviation of our fits.  The panels contain the results from: 1) the fits with monopole and quadrupole (left), 2) the fits also including the hexadecapole (middle), and 3) the fits using multipoles up to $\ell=4$ and using a constrained range on the hexadecapole (right). We also include the result for the growth factor obtained by BOSS and reported in \cite{Alam2017} (far right value of the left panel).

These results suggest that the most accurate results (smaller parameter biases in all parameters normalized by the dispersion) are obtained using the multipoles up to $\ell=4$ in the full range.

However, we would like to highlight that we noticed that the best model  of the hexadecapole does not accurately match the mean of the mocks (the value of the residuals is close to 50\% of the value of the model compared to 10\% and 4\% for the monopole and quadrupole respectively). For the individual fits, given the large error bars on the hexadecapole, this mismatch does not bias our individual measurements, but produces small bias in the best fit of the mean in $\epsilon$. Bearing all of this in mind we choose to analyze both cases (with and without hexadecapole), but we will take a conservative approach and  report the monopole and quadrupole only analysis as our final result of this work. Based on the results from N-Series we adopt $\sigma^{SYS}_{f\sigma_8}$ = 0.004, $\sigma^{SYS}_{\alpha}$ = 0.001, and $\sigma^{SYS}_{\epsilon}$ = 5e-4 as an estimate of the potential bias of $f,\alpha$ and $\epsilon$.

\begin{table*}
\caption{Results from fitting the 84 N-Series sky mocks with our fiducial methodology. The columns denoted by $\widetilde{x}$  are the mean, $S_x$ denotes the standard deviation, and the bias (defined by equation \ref{eq:biasf}) is denoted by $b_x$, with $x=f,\, \alpha, \,  \epsilon$. }
\label{tab:binning}

\begin{tabular}{@{}lcccccccccccc}
\hline
\multicolumn{11}{c}{Results for Fiducial Methodology with N-series Sky mocks}\\
\hline
\hline
Model&
$\widetilde{f\sigma_8}$&
$S_f$&
$b_f$&
$\widetilde{\alpha}$&
$S_{\alpha}$&
$b_{\alpha}$&
$\widetilde{\epsilon}$&
$S_\epsilon$&
$b_{\epsilon}$& $N_{mocks}$\\
\hline
\\[-1.5ex]
$\ell_{\max}=2$ [27.5,117.5] &
0.459&0.051&-0.003&
0.998&0.014&-0.002&
4e-4&0.019& 4e-4&81\\

$\ell_{\max}=4$ [27.5,117.5]&
0.471&0.037& -7e-5&
1.0&0.013&-8e-5&
1e-4&0.010&2e-4&83\\

$\ell_{\max}=4$ [47.5,117.5]&
0.471& 0.042&-5e-4&
0.999& 0.013&-0.001&
-5e-4& 0.014&-5e-4&84\\

\\[-1.5ex]
\hline
\label{tab:nseries}
\end{tabular}
\end{table*}

%% file: QPM_EZ_Accepted.tex
\subsection{Testing Systematics with eBOSS-Mocks}
\label{sec:QPMandEZ}
\begin{figure*}
\hspace*{-0.5cm}
\subfigure{\includegraphics[width=55mm,height=50mm,trim = 0cm 0cm 0cm 0cm, clip]{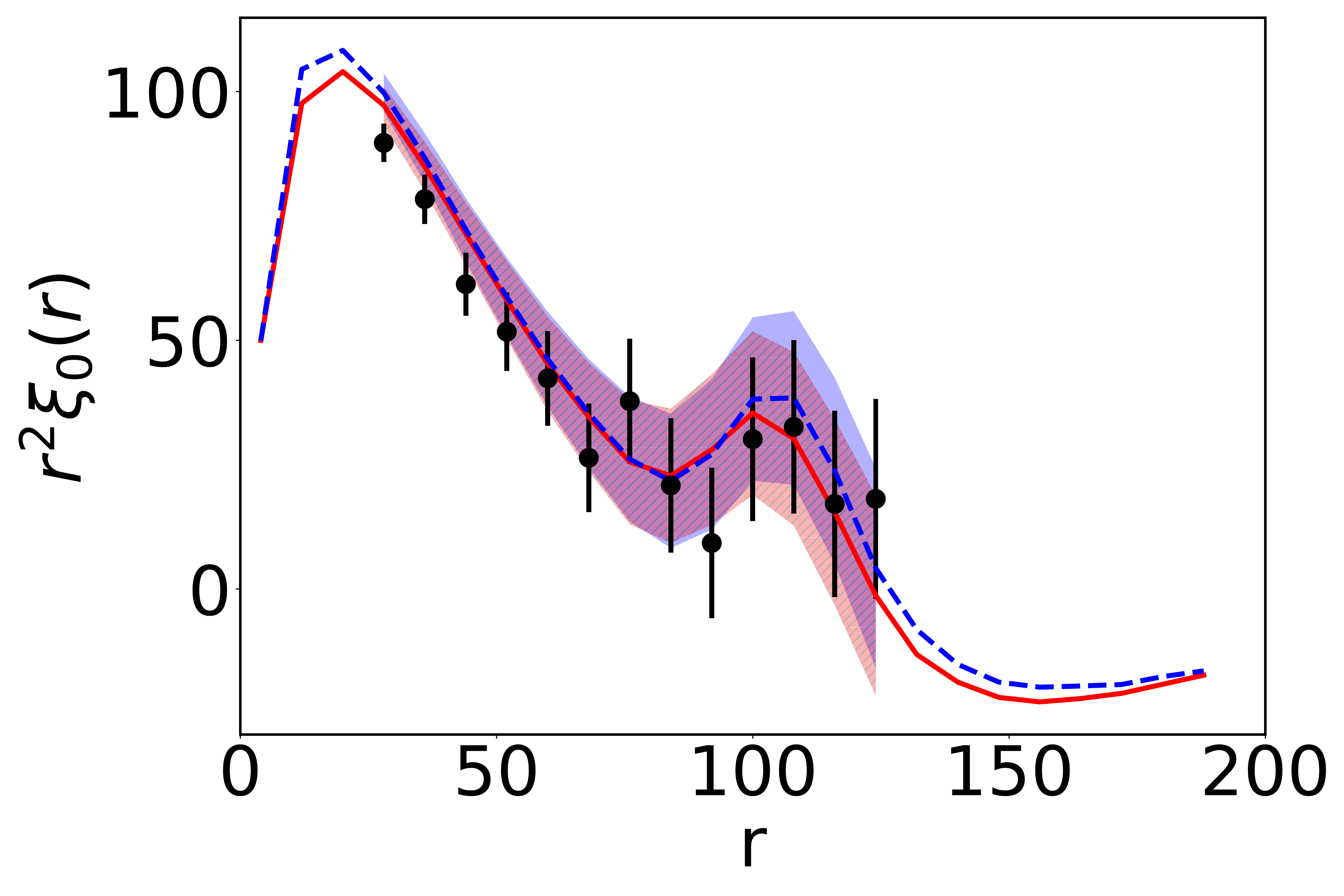}}
\subfigure{\includegraphics[width=55mm,height=50mm,trim = 0cm 0cm 0cm 0cm, clip]{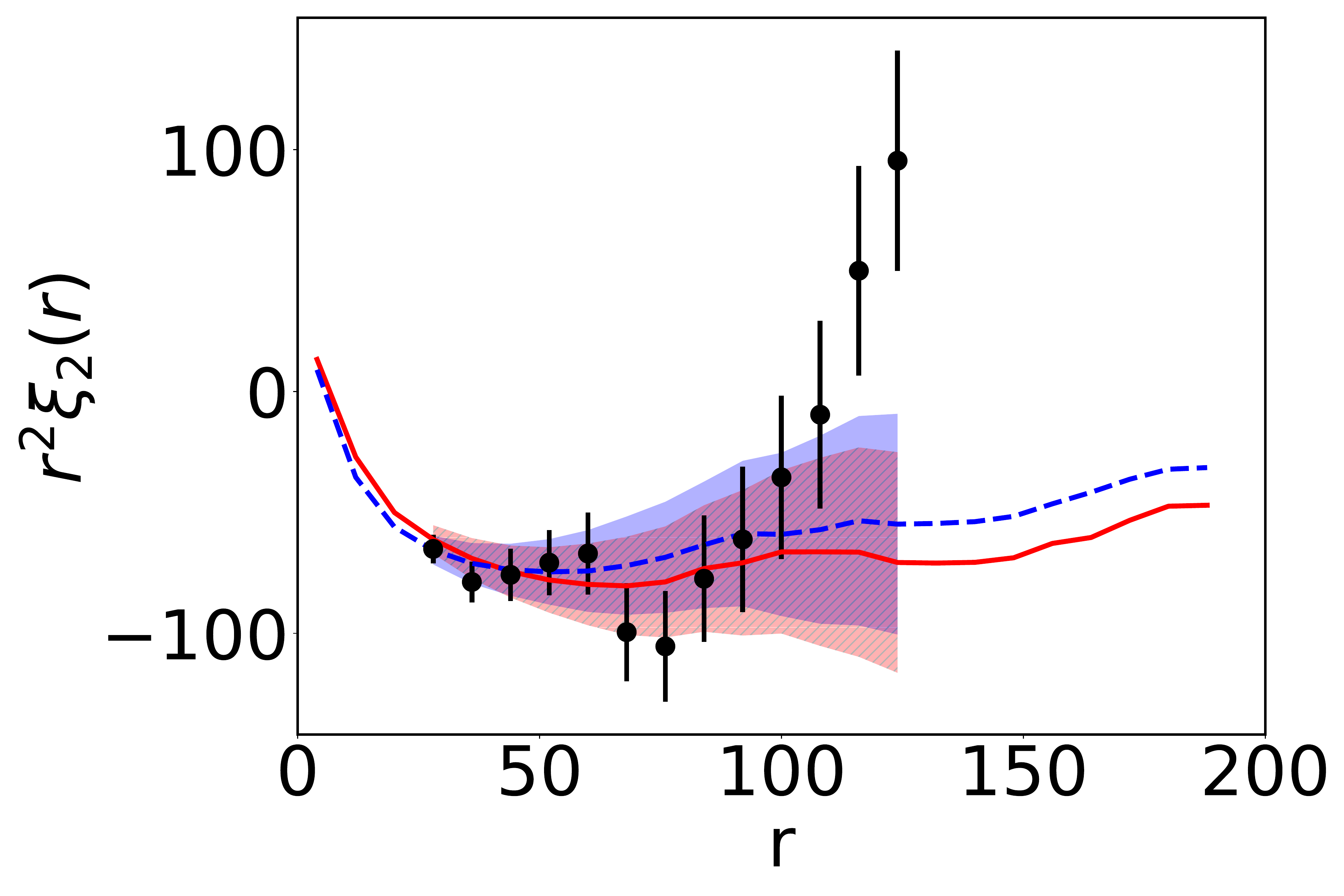}}
\subfigure{\includegraphics[width=55mm,height=50mm,trim = 0cm 0cm 0cm 0cm, clip]{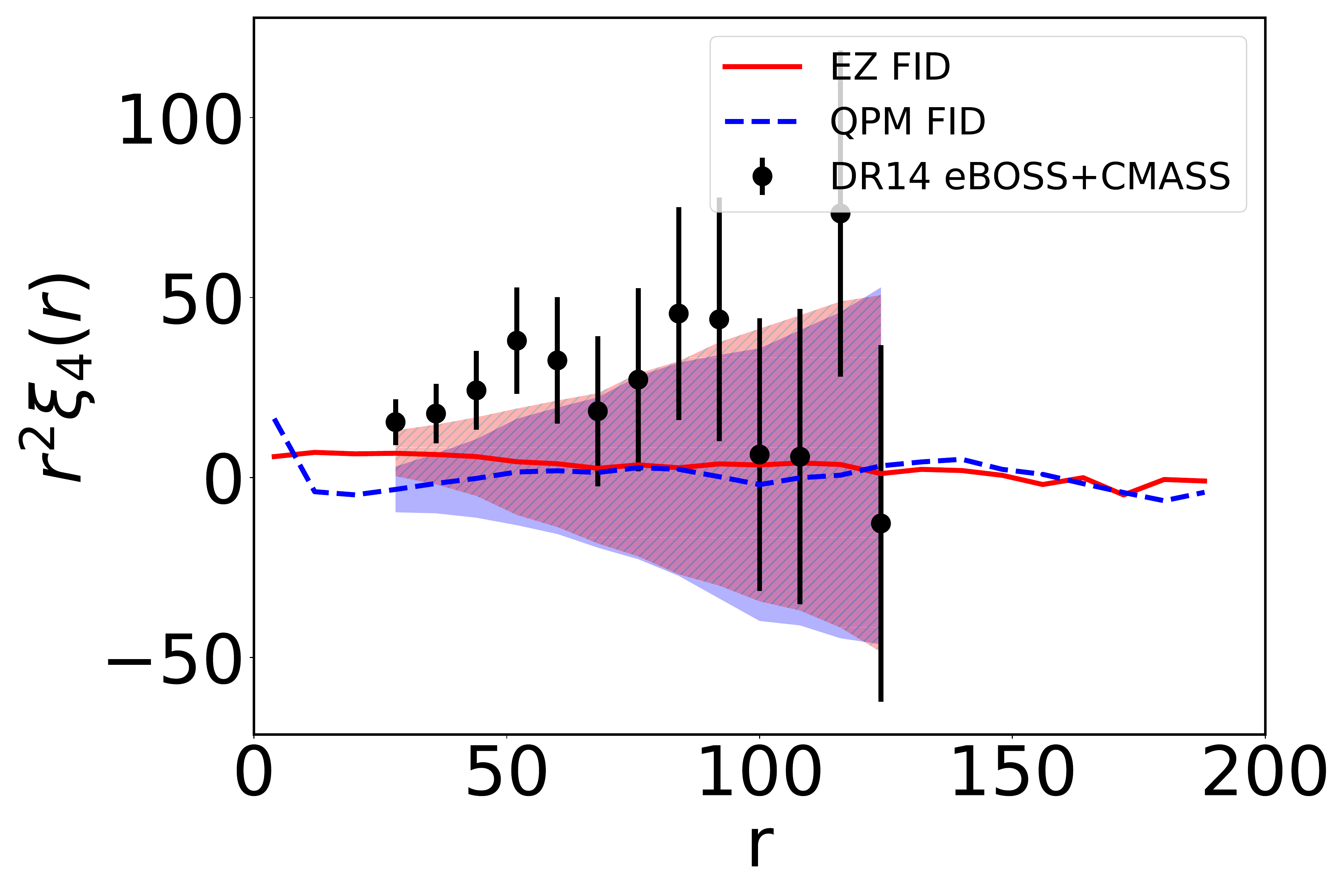}}
\caption{Mean QPM (dotted blue line) and EZ (solid red line) mocks and DR14 correlation function (black dots), all computed in the fiducial cosmology. QPM and EZ mock underestimate the hexadecapole. The monopole (left) shows small mismatches between the mocks and the data at small scales, the quadrupole data (center) presents a large correlation in the large scale quadrupole that lies outside the 1$\sigma$ variation observed in the mocks, the hexadecapole data (center) has a larger amplitude than the one predicted by the mocks.}
\hspace*{-0.5cm}
\label{fig:mean_data}
\end{figure*}

We will dedicate this section to test the variance of the methodology developed in Section \ref{section:model}. This analysis will be done using two sets of approximative mocks, the QPM and EZ described in Section \ref{section:mocks}, both built with the same properties of our eBOSS sample. The mocks were calibrated to match the data, however, these approximative mocks lack the accuracy to study the biases of our methodology. As shown in Figure \ref{fig:mean_data}, the QPM and EZ mock have a small mismatch in the monopole at small scales. Additionally, both seem to systematically underestimate the hexadecapole. Bearing this in mind, our estimates of the bias will only come from the results of the N-Series Cut-sky mocks obtained in the last section\footnote{The N-series mocks provided an estimate of the biases on a sample that is similar to that of BOSS-LRG; the mean redshift was slightly lower than the one from the eBOSS LRG sample considered in this work but it had similar clustering properties, i.e. the galaxy bias.}, we proceed to quantify the dispersion of the fitting methodology. Our specific goal is to estimate the dispersion expected around the  parameters of interest of our model. This will be done by applying the fitting methodology from section \ref{Fit_metho} to 100 of our individual QPM an EZ mocks, which will give us 100 estimates of the best fit values.

We test  two cases: 1) Considering only the multipoles up to $\ell=2$ (skipping the hexadecapole), and therefore following the methodology used in previous analysis performed with the LRG sample, which we will refer to as ``$\xi_0+\xi_2$".
We also consider the effect of extending the multipoles up to $\ell=4$ and using the full range for all multipoles, which we will refer to as ``$\xi_0+\xi_2+\xi_4$".

We used the cosmology use for the QPM mocks generation in order to compute their comoving coordinates, and the fiducial cosmology for computing those of the EZ mocks. Table \ref{tab:APtest} summarizes the results from our fits. The first block corresponds to the monopole + quadrupole fits using the QPM/EZ mocks; the second block describes the analysis adding the hexadecapole to our fits.
The dispersions obtained from our two sets of mocks when only using the  monopole and the quadrupole in the fits (first block of Table \ref{tab:APtest}) are fairly consistent for all of the parameters of interest: $S_{f\sigma_8}^{\rm QPM}= 0.113$ and $S_{f\sigma_8}^{EZ}= 0.122$, $S_{\alpha}^{\rm QPM}= 0.039$ and $S_{\alpha}^{\rm EZ}= 0.043$, and $S_{\epsilon}^{QPM}= 0.053$ and $S_{\epsilon}^{\rm EZ}= 0.044$. The dispersion is also consistent with previous results found on the anisotropic LRG DR14 BAO analysis from \cite{Bautista2017} where:  $S_{\alpha}^{\rm BAO}= 0.048$ and $S_{\epsilon}^{\rm BAO}= 0.055$.

In order to compare with the previous results from BOSS reported in \cite{2017MNRAS.470.2617A}, we need to rescale the variance using the differences in volume between the two samples; the effective volume of our sample is 0.9  $\rm Gpc^3$ while BOSS-CMASS accounts for 4.1 $\rm Gpc^3$ in the [0.5,0.75] redshift slice. The CMASS sample reported the following standard deviations for the [0.5,0.75] redshift slice: $S_{f\sigma_8}^{\rm BOSS}=0.058$, $S_{\alpha}^{\rm BOSS}= 0.016$, and $S_{\epsilon}^{\rm BOSS}= 0.022$ (Table 6 of \citealt{2017MNRAS.470.2617A}), we can scale them roughly to the eBOSS volume using $S_X^{\rm eBOSS^2}=(S_X^{\rm BOSS^2} \times$ 4.1 $\rm Gpc^3)/$0.9 $\rm Gpc^3$, yielding the following scaled dispersions: $S_{f\sigma_8}^{\rm BOSS}=0.124$, $S_{\alpha}^{\rm BOSS}= 0.034$, and $S_{\epsilon}^{\rm BOSS}= 0.047$. These values are in  agreement with the dispersion obtained with our QPM/EZ mocks.

Now, let us examine the fits that include the hexadecapole. The dispersion obtained from the two sets of mocks is also consistent for the parameters $f$ and $\epsilon$: $S_{f\sigma_8}^{\rm QPM}= 0.090$ and $S_{f\sigma_8}^{\rm EZ}= 0.089$, and $S_{\epsilon}^{\rm QPM}=S_{\epsilon}^{\rm EZ} 0.050$ and $S_{\alpha}^{\rm QPM}= S_{\alpha}^{\rm EZ}=0.028$. Also we observe the dispersion in all parameters decreases when considering the hexadecapole as expected.

 Figure \ref{fig:RSD_FS_triangle} shows the distribution of the differences between the parameters of interest and their expected values on a mock-by-mock basis, i.e. $\Delta f\sigma_8 =\langle f\sigma_8 -{f\sigma_8}_{\rm exp}\rangle $,
$\Delta \alpha =\langle \alpha -\alpha_{\rm exp}\rangle $, $\Delta \epsilon = \langle \epsilon -\epsilon_{\rm exp}\rangle $, and for $b=1+F'$, $\Delta b =\langle b \sigma_8-{b\sigma_8}_{\rm exp}\rangle $, for both the analyses using multipoles up to $\ell=2$ and up to $\ell=4$. Reviewing the monopole + quadrupole  fits (in blue dots) reveals that both sets of mocks show a well-behaved distribution that is centered close to zero and is symmetric.
From the hexadecapole fits (red x's), we also observe symmetric distributions centered around zero, however, especially the 1D distributions for the $f\sigma_8$  and $\epsilon$ parameters are slightly shifted.

These shifts in the distributions when considering the hexadecapole  are related to the QPM/EZ mocks poor precision  and to the fact that the model and the mean multipoles present mismatches, the following paragraphs will briefly show these mismatches. As we will see, the biggest mismatch between mock and model occurs in the hexadecapole for the QPM mocks and in the quadrupole for the EZ mocks.

\begin{figure*}
\hspace*{-0.5cm}
\subfigure{\includegraphics[width=88mm,height=90mm,trim = 0cm 1cm 2cm 2cm, clip]{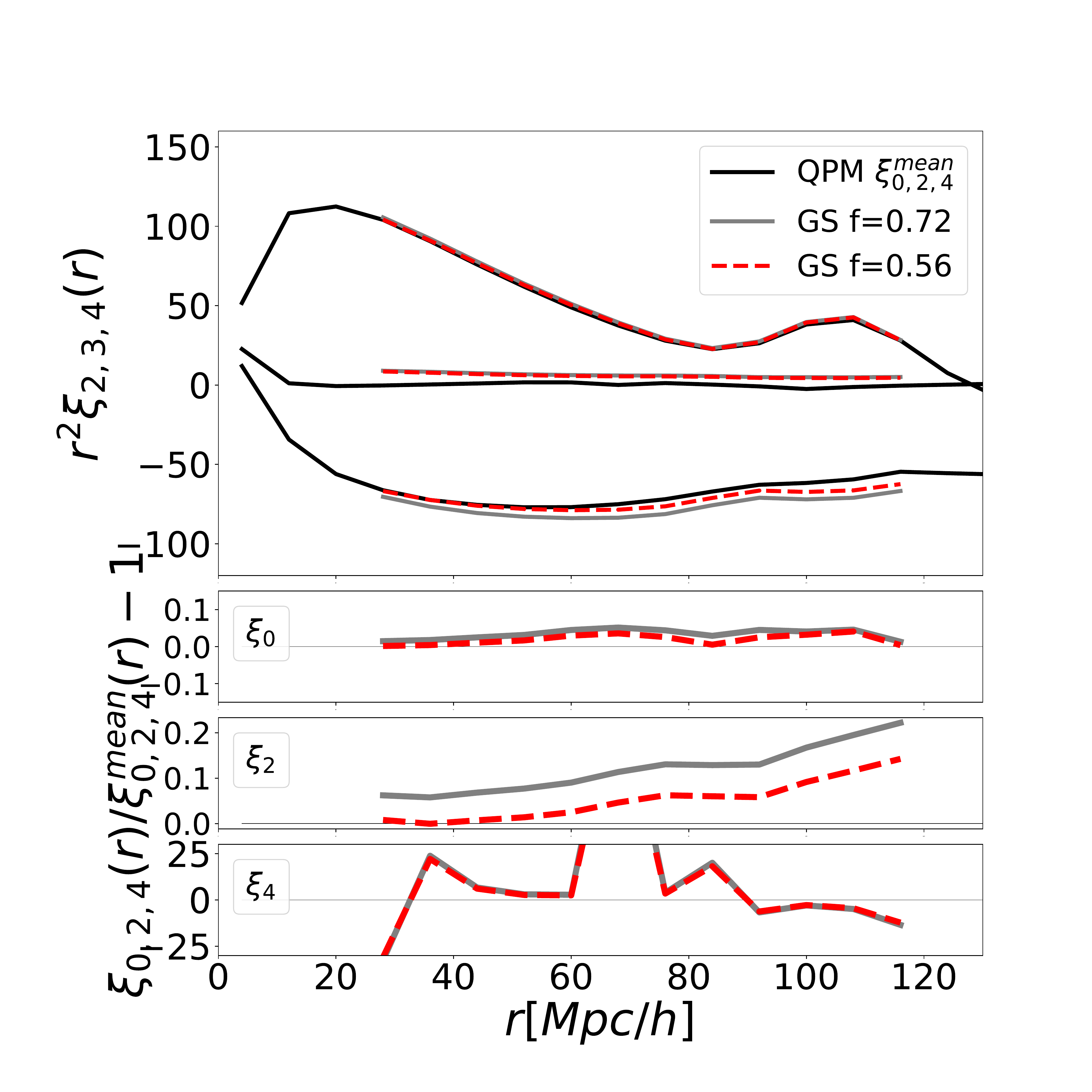}}
\subfigure{\includegraphics[width=88mm,height=90mm,trim = 0cm 1cm 2cm 2cm, clip]{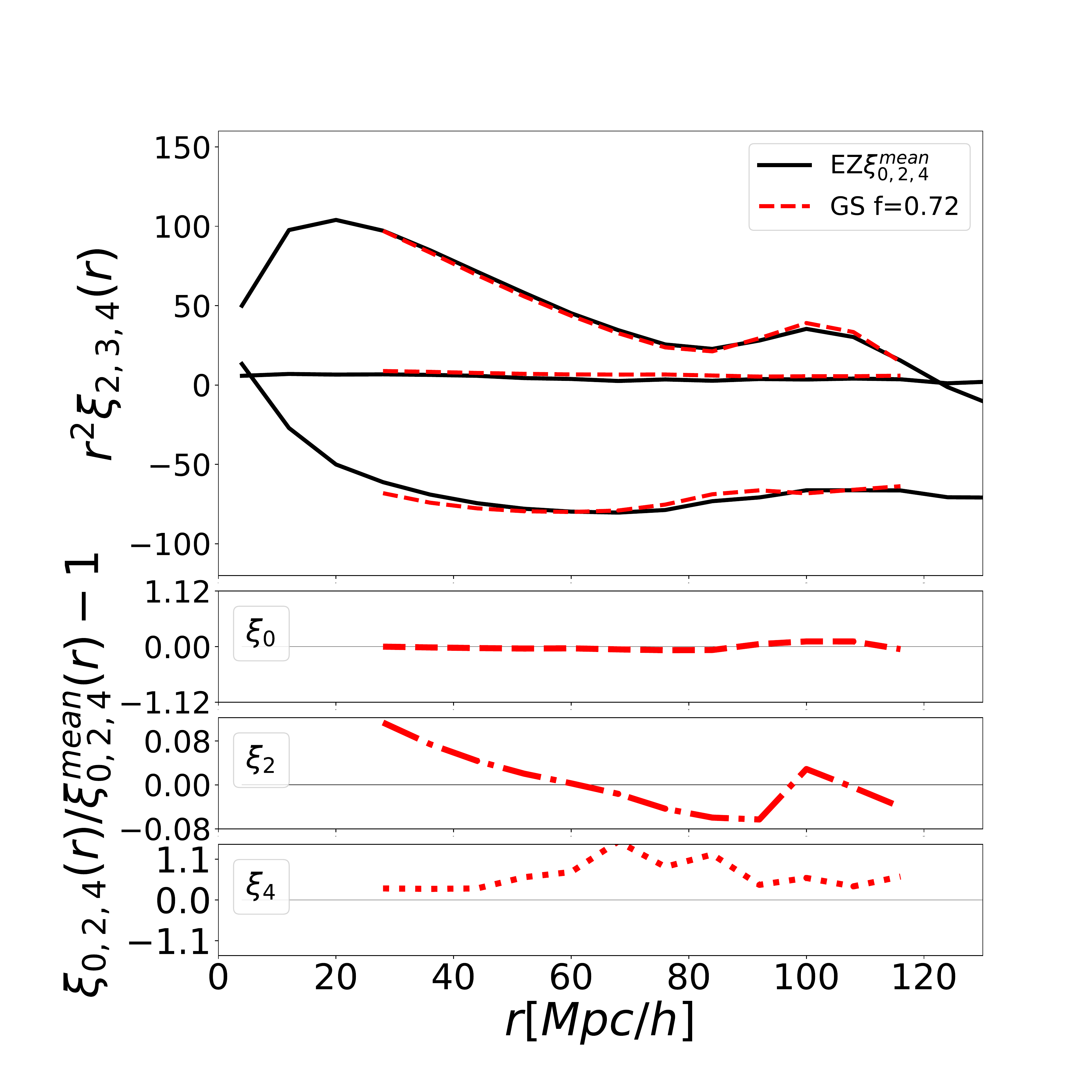}}
\caption{Mean QPM/EZ mocks vs. Template with Mock Cosmology. The error bars are smaller than the size of the points.  For QPM mocks, the template with the mocks cosmology  does not match the mean of the mocks (black line), this is evident in the quadrupole and hexadecapole residuals.  For the EZ mocks, the template with the mocks cosmology fits the mean better, but the template does not match all of the scales of the quadrupole and the hexadecapole simultaneously.}
\hspace*{-0.5cm}
\label{fig:mean_mocks_paper}
\end{figure*}

\begin{figure*}
\subfigure{\includegraphics[width=110mm,height=110mm,trim = 3.5cm 2cm 2cm 6.5cm, clip]{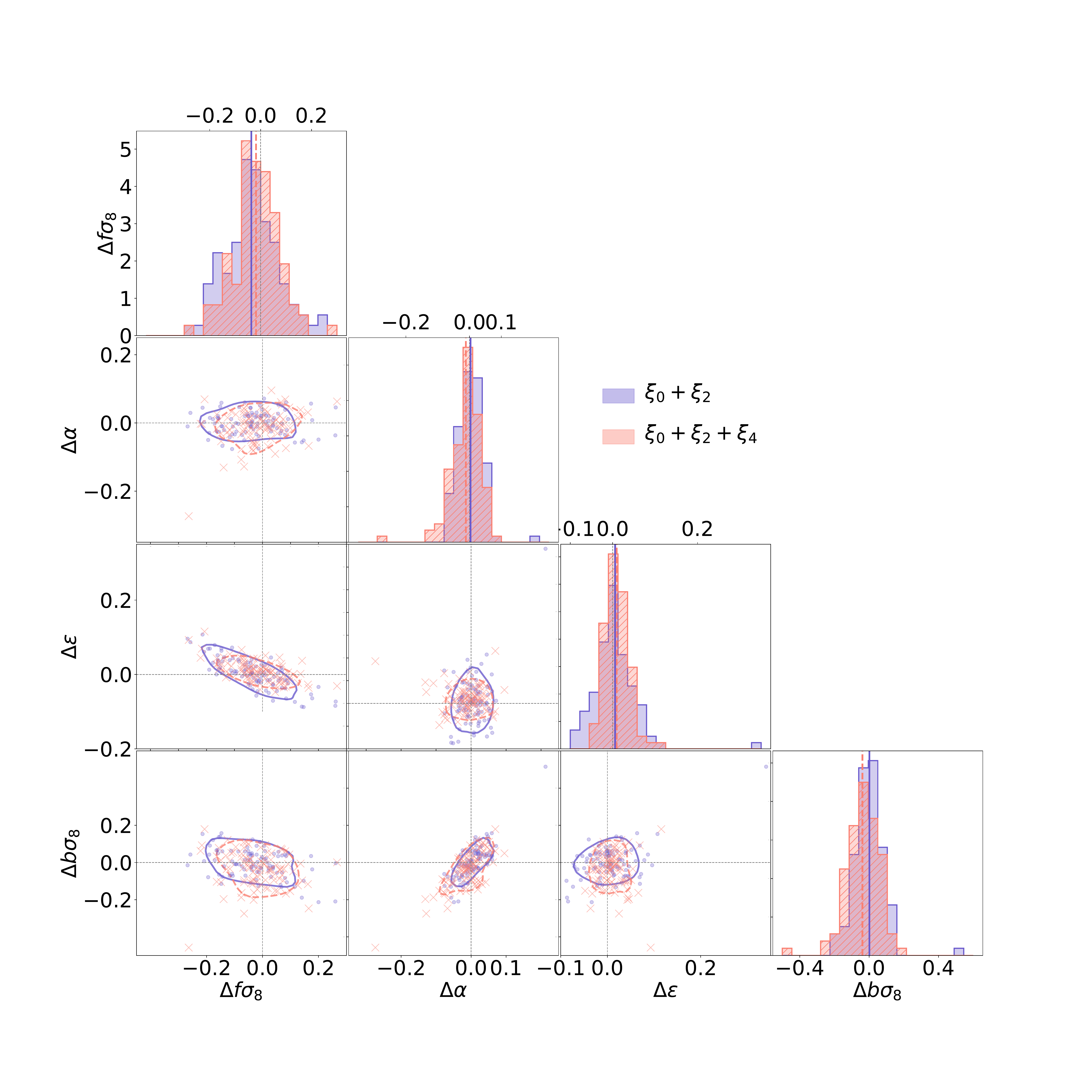}}
\subfigure{\includegraphics[width=110mm,height=110mm,trim = 3.5cm 2cm 2cm 6.0cm, clip]{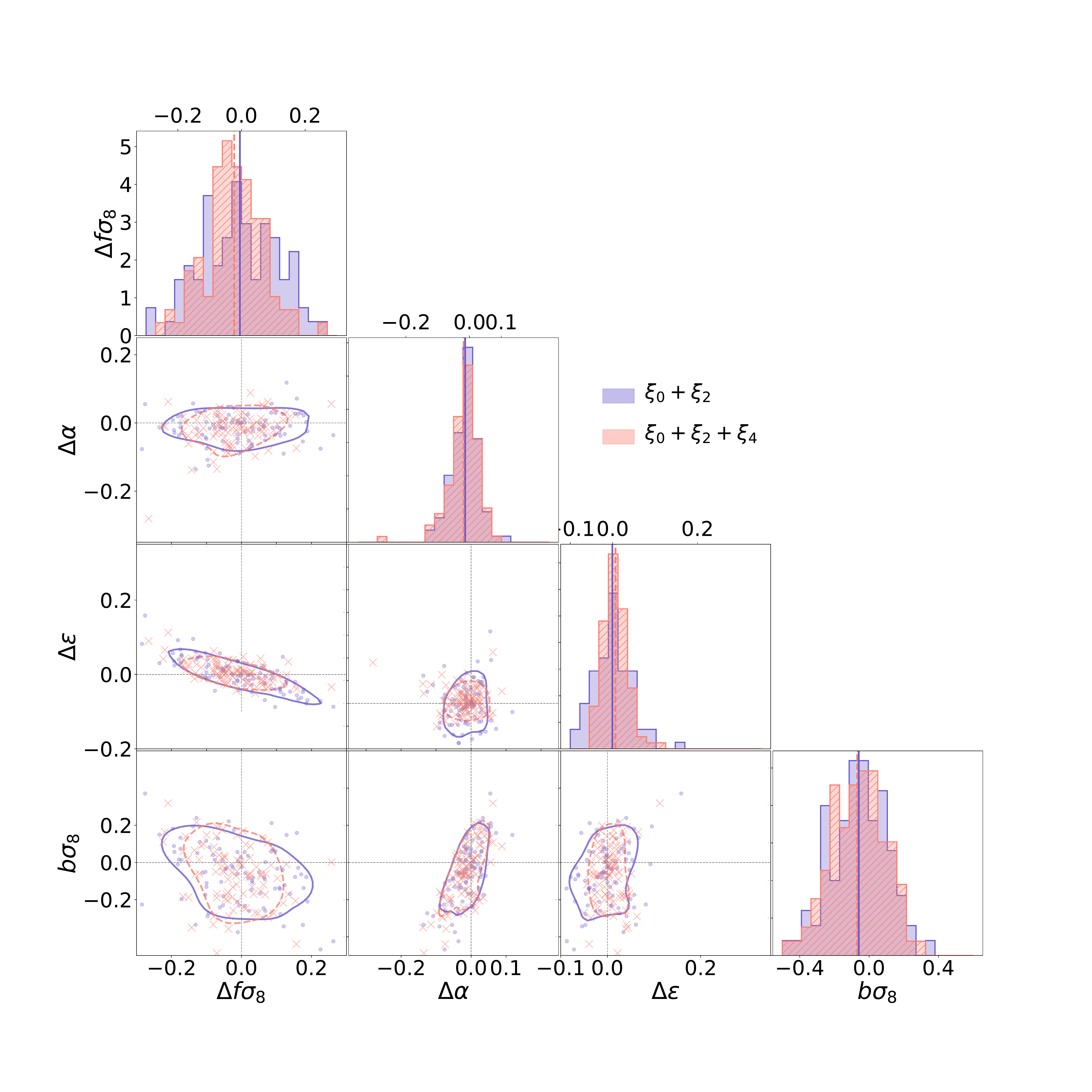}}

\caption{Scatter triangle plots comparing fits for full shape fits using $\xi_0+\xi_2$ (blue dots) and $\xi_0+\xi_2+\xi_4$ (red x's) for QPM (up) and EZ (down) mocks. We show the difference of the best fit values with respect to the expected values for each of the parameters of interest. The means are indicated as solid lines for the two cases explored. The dotted lines indicate the expected values, which are zero for all cases.}

\label{fig:RSD_FS_triangle}
\end{figure*}

Figure \ref{fig:mean_mocks_paper} shows a comparison between the mean of the mocks and the model templates built with the true cosmology of the mocks \footnote{the cosmology used for building the mocks} denoted by ``Model GS $f(z=0.72)$". The left panel shows the comparison between the mean of the QPM mocks and its model template  and the equivalent comparison for the EZ mocks is in the right panel. The growth factor used for building the model in the right panel is at the effective redshift of the mocks.

The figure reveals that the mean of the QPM mocks does not match the model with the cosmology used for their generation (gray solid line), which is evident in the quadrupole residuals. However, a template using a growth factor corresponding to a lower redshift ($z=0.56$) is a better match with the mean of the mocks (red dotted line); this model is denoted by ``Model GS $f(z=0.56)$" and is shown with red dotted lines.

From this analysis we can draw the following conclusions. First, the GSRSD model cannot match the multipoles of the QPM mocks, as they show a mismatch in the mean of the mocks and the model for the quadrupole, giving rise to a higher value than the input value of the simulations. Second, the model of the hexadecapole is systematically larger than the mean of the mocks, and in particular any conclusion about the bias of the hexadecapole cannot be extracted from the fits of the QPM mocks.

The right panel of Figure \ref{fig:mean_mocks_paper}  shows an equivalent comparison between mean and model template using the EZ mocks. As for the QPM mocks we also see a mismatch, but this time between the small scales of the quadrupole: the template with the cosmology and redshift of the EZ mocks (red dotted line) does not match with the mean quadrupole of the mocks (black solid line).

It is interesting to notice that the mean hexadecapole matches the template. EZ mocks describes better the hexadecapole than QPM  because the effective bias model encoded in EZ mocks accounts for both 2- and 3-point statistics (\cite{Chuang2015}, EZmock paper).
Indeed, the 3-point correlation function of EZmocks (Figure 8, \cite{Chuang2015}) is more consistent with N-body simulations, compared to QPM (Figure 9, \citep{2014MNRAS.437.2594W}, QPM paper). Since the high order statistics of EZmocks are better, the high order multipoles are also more reliable. Thus we can expect to get better fits when using the hexadecapole information.

We also notice that the mismatch in the quadrupole behaves different for different scales, the scales lower than 50 $h^{-1}$ Mpc are overestimated and the scales larger than 50  $h^{-1}$ Mpc are underestimated.
Thus, the EZ mocks seem to not be reproducible by the model. Apparently, the template with the mocks cosmology fits the mean better, but the template is not capable of fitting all of the scales of the quadrupole and the hexadecapole simultaneously.

We can summarize the results of this section as follow:  1) the dispersion obtained from both sets is consistent with each other and with previous results from BOSS \citep{2017MNRAS.470.2617A} and from the DR14 BAO group \citep{bautista2018}. 2) both sets of eBOSS mocks lack the accuracy to study the biases of our methodology: the QPM mocks seem to slightly overpredict the quadrupole expected by the GSRSD model and are not a good match to the hexadecapole. The EZ mocks have a better match to the hexadecapole, but can not match the quadrupole at small scales (lower than 50 $h^{-1}$ Mpc).

\begin{table*}
\caption{Results from fitting the 100 QPM/EZ mocks for FS analysis.
 We include the analysis for both cases using the hexadecapole in addition to the monopole and quadrupole. The columns denoted by $\widetilde{x}$  are the mean, and the $S_x$ denotes the standard deviation. The variables are the difference of the parameters of interest compared to their expected values on a mock-by-mock basis, i.e. $\Delta f\sigma_8 =\langle f\sigma_8 -{f\sigma_8}_{\rm exp}\rangle $,    $\Delta \alpha =\langle \alpha -\alpha_{\rm exp}\rangle $,  $\Delta \epsilon = \langle \epsilon -\epsilon_{\rm exp} \rangle $,
 for both the analysis using multipoles up to $\ell=2$ and using multipoles up to $\ell=4$.}
\label{tab:APtest}
\begin{tabular}{@{}lccccccccccc}
\hline
\multicolumn{8}{c}{Monopole-Quadrupole fits}\\
\hline
Model&
$\widetilde{\Delta f\sigma_8}$&
$S_{\Delta f\sigma_8}$&

$\widetilde{\Delta \alpha}$&
$S_{\Delta \alpha}$&

$\widetilde{\Delta\epsilon}$&
$S_{\Delta\epsilon}$&

$\chi^2/$d.o.f&
$N_{\rm mock}$\\
\hline
\\[-1.5ex]

FS-QPM MQ &
$-0.036$&$0.113$&
$0.003$&$0.039$&
$0.006$&$0.053$&
$ 1.0$&
$ 97$\\
\\[-1.5ex]

FZ-EZ MQ&
$-0.007$&$0.122$&
$0.009$&$0.043$&
$0.001$&$0.044$&
$ 1.0$&
$91$\\
\\[-1.5ex]
\hline
\multicolumn{8}{c}{Including Hexadecapole}\\
\hline
\\[-1.5ex]

FS-QPM &
$-0.018$&$0.090$&
$-0.011$&$0.050$&
$0.009$&$0.028$&
$ 1.1$&
$ 84$\\
\\[-1.5ex]

FS-EZ &
$-0.024$&$0.089$&
$0.005$&$0.050$&
$0.008$&$0.028$&
$ 1.0$&
$ 97$\\
\\[-1.5ex]

\\[-1.5ex]
\hline
\end{tabular}
\end{table*}

\subsection{Comparison of AP parameters results with BAO-only fits}\label{sec:bao}
\input{bao_FS_Accepted}
\subsection{Testing the Impact of Spectroscopic Incompleteness }
\label{sec:spec_incomp}
\input{spec_incomp_Accepted}

%% file: bao_FS_Accepted.tex
\begin{figure*}
\subfigure{\includegraphics[width=80mm,height=60mm,trim = 2cm 1cm 1cm 2cm]{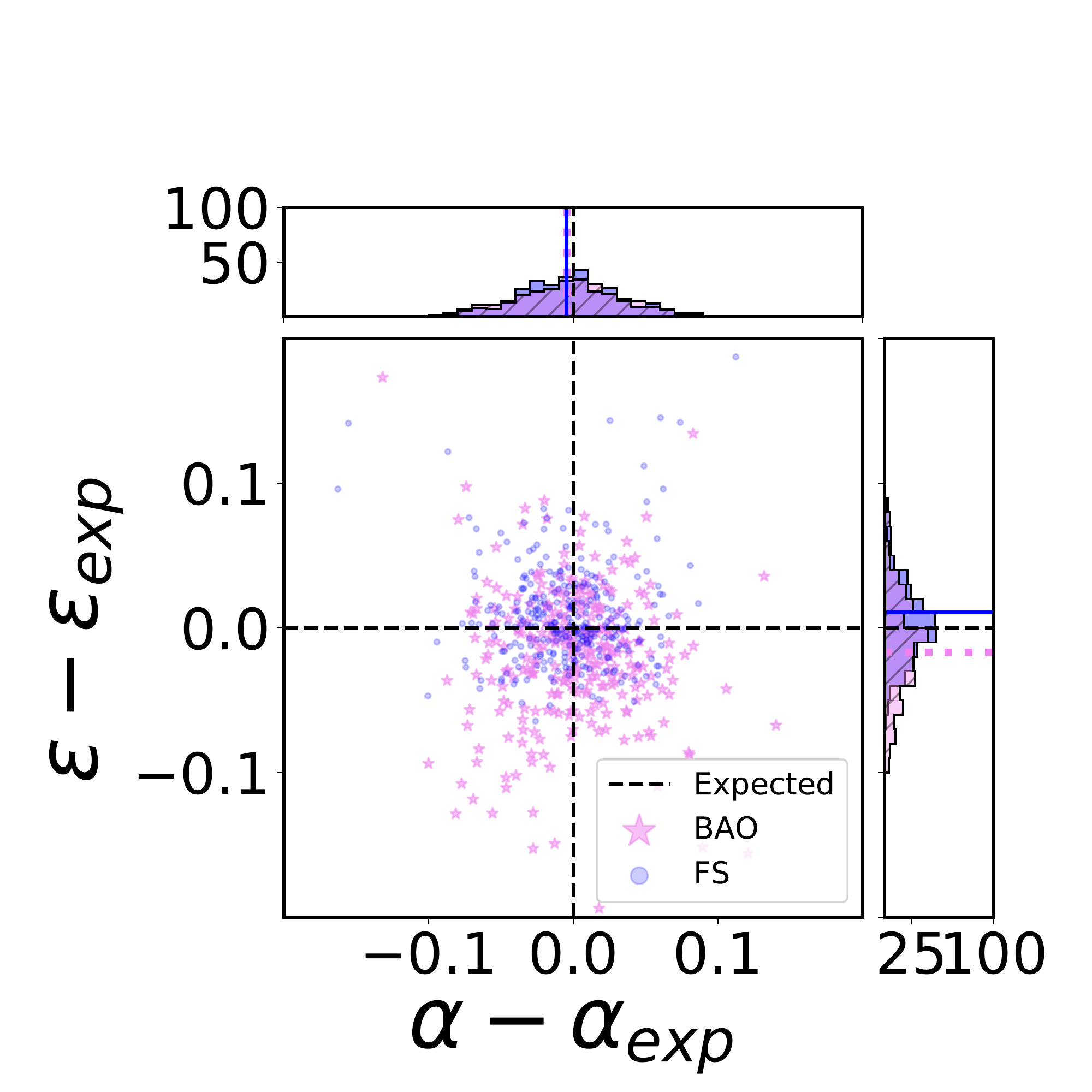}}
\subfigure{\includegraphics[width=80mm,height=60mm]{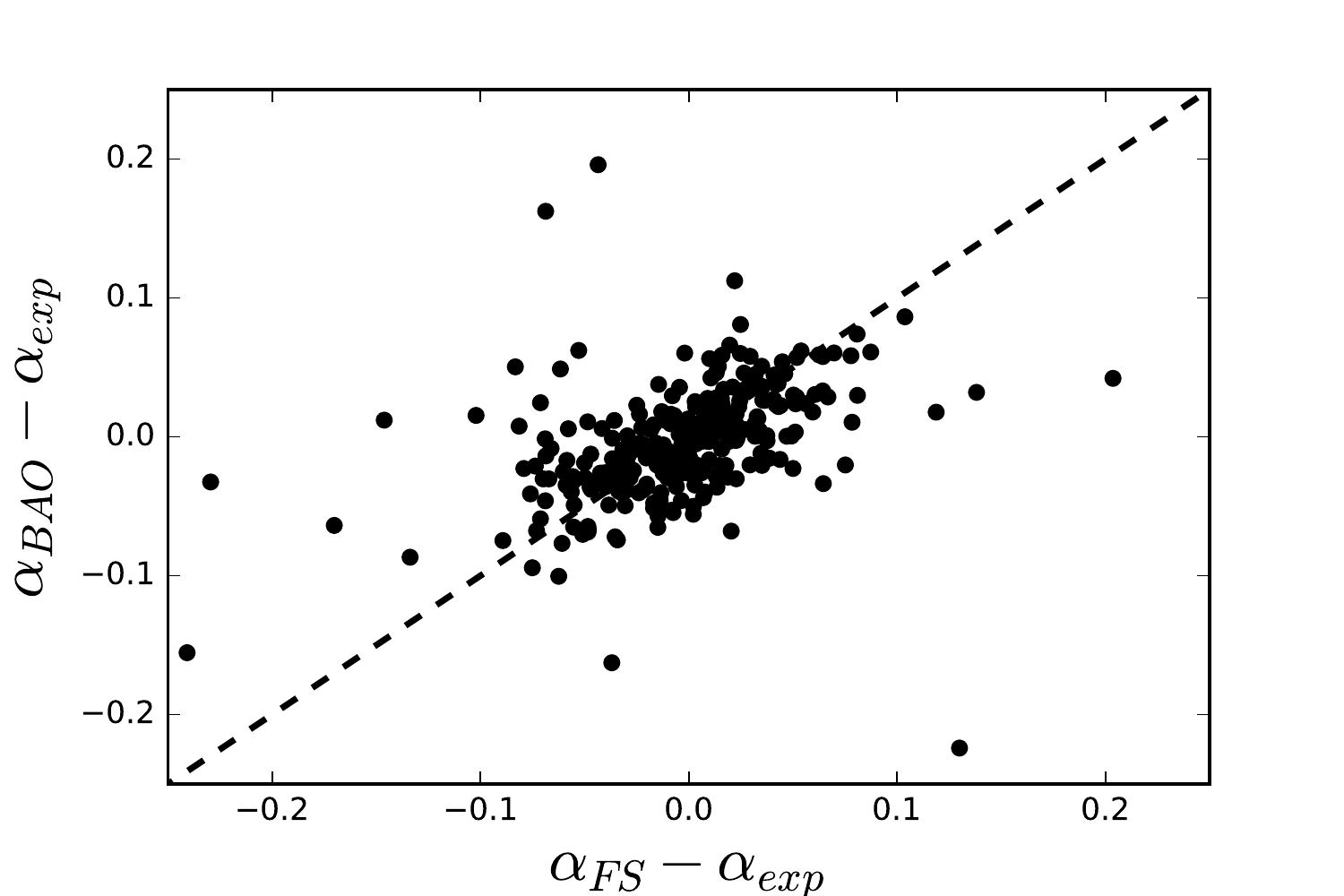}}
\caption{Left panel: A comparison of the BAO fits and full shape using $\xi_0+\xi_2$ for the mocks. Right panel: comparison of best fits in isotropic dilation parameter for FS and BAO for the mocks. The dispersion for the anisotropic warping, $\epsilon$,  from BAO fits is slightly larger compared to the FS best fits. FS analysis breaks some degeneracies in $\epsilon$ and reduces its dispersion. }
\label{fig:baovsfs}
\end{figure*}

In this section, we compare our results to those obtained in \cite{bautista2018}, which is a previous analysis using this same sample.
The left panel of Figure \ref{fig:baovsfs} shows the difference between our QPM FS fits to the combined sample and the expected value compared to those from the anisotropic BAO parameters, the later taken from \cite{Bautista2017}.
The dispersion for the anisotropic warping, $\epsilon$,  from BAO fits is slightly larger compared to the FS best fits.  In an RSD analysis other parameters that affect the quadrupole are included (most significantly the growth factor $f$), so it is not surprising that FS analysis breaks some degeneracies in $\epsilon$ and reduces its dispersion. There is also a small shift in the isotropic dilation parameter, $\alpha$, when comparing the FS analysis best fits to those coming from BAO. The left panel of Figure \ref{fig:baovsfs} shows the scatter plot for  $\alpha$, with a Pearson correlation factor of $r=0.5$. There are several differences in the fitting methodology between these two fits. Obviously the modeling of the signal is different in BAO and in our RSD+AP model, but in addition the fitting range used in BAO is wider in its $r$-range that is extended to 180 Mpc/$h$ while our FS analysis is constrained to $r$-values lower than 130 Mpc/$h$. Also, the binning used in BAO is 5 Mpc in width, while this work is using bins with a width of 8 Mpc.

%% file: spec_incomp_Accepted.tex
To test the effect of redshift incompleteness in our clustering, we consider three cases: the first is our mock catalogs with no redshift failures. Then, we study the effect of the two mitigations techniques described in Section \ref{sec:Corr_Spec_Comp}. The redshift failures are added to the mocks by associating a position in the plate to each galaxy, then the catalog of binned probabilities is used to mimic the effect of the redshift failures observed in our data. The second case explored is the up-weighting methodology, and finally, for the third case,  the forward-modeling technique.

\begin{figure}
\subfigure{\includegraphics[width=90mm, trim = 1.8cm 2.0cm 3cm 2cm, clip]{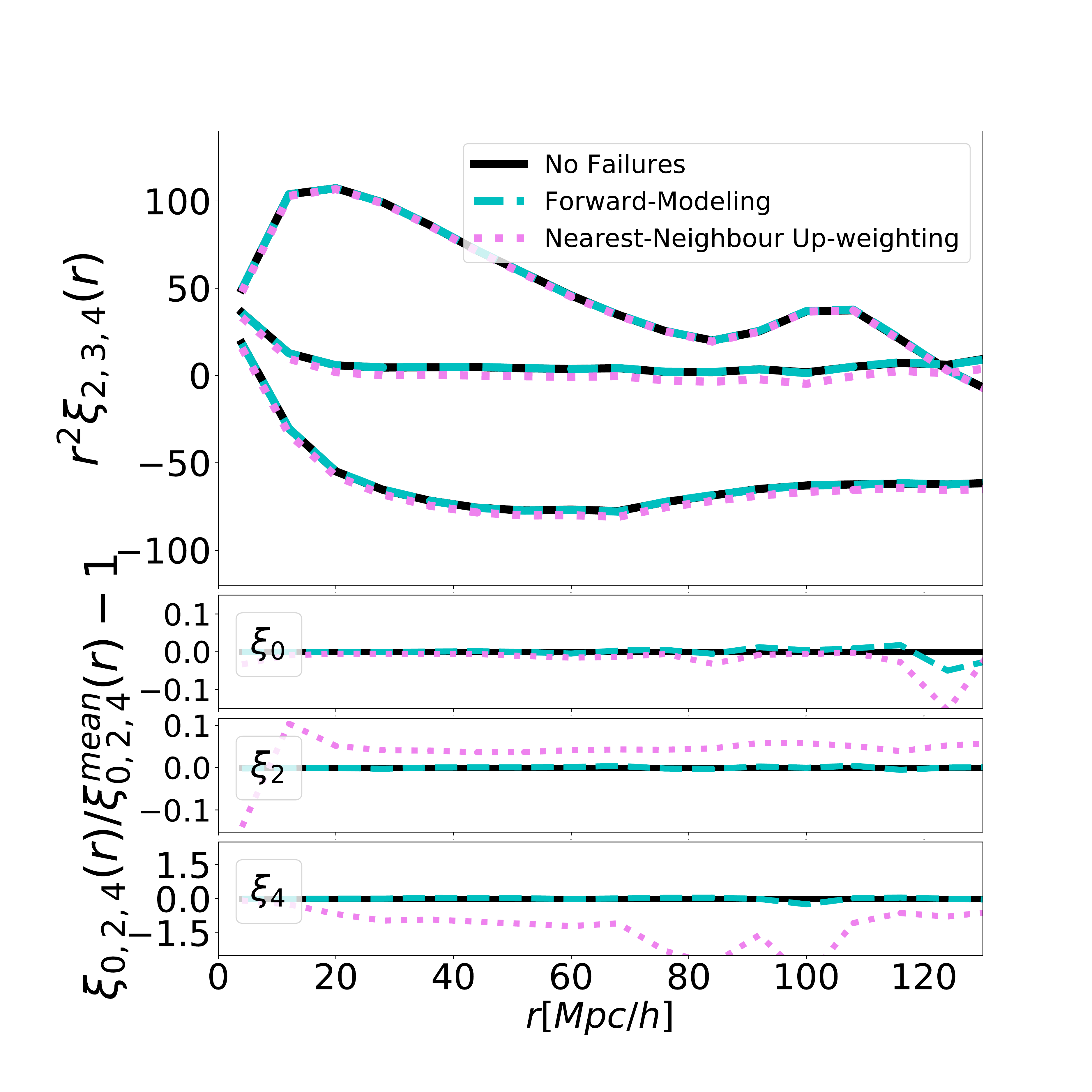}}
\caption{ Impact of the redshift completeness on the multipoles and the effect of the mitigation techniques for correcting potential biases. The monopole (top), the quadrupole (bottom) and hexadecapole (middle) are presented in three cases: without redshift failures, correcting by the up-weighting technique, and correcting using the Forward modeling technique. While the monopole is well recovered by the two correction techniques, the quadrupole/hexadecapole shows a clear shift (i.e., bias) at all scales when corrected with the up-weighting method. The forward-modeling recovers the expected values for scales smaller than $r=140$ Mpc/$h$.}
\label{fig:spec_com}
\end{figure}

Figure \ref{fig:spec_com} displays the impact of different mitigation methods on the average of all 1000 mock catalog correlation functions. The three lines represent the case without redshift failure corrections and the up-weighting and forward modeling corrections. While the monopole is equally well recovered in all three cases, the quadrupole shows a clear shift (i.e.,  bias) at all scales when using the up-weighting method. The forward-modeling corrections recover the expected values for scales smaller than $r=140$ Mpc/$h$, but show slight discrepancies at larger scales.

Table \ref{tab:failures} lists the results of the the best-fit parameters found by fitting all 100 QPM mocks using both correction schemes. We compared  the results of the mocks where redshift failures are applied and corrected by one of the two mitigation techniques with the case where no redshift failures are considered. We report the difference of the mean of the best fits as an indicator of the systematic bias related to the spectroscopic completeness denoted by $\Delta f, \,\Delta \alpha$,  and $\Delta \epsilon$; we also report the
the dispersion $ S_{x}$, where $x=f,\alpha, \,\epsilon$. We observe that the up-weighting technique differs from the case without redshift failures by  $|\Delta f |=0.016$  ($\Delta f/(S_f/\sqrt{N_{sim}}) \sim 0.7\sigma$ ), $|\Delta \alpha|=0.001$ ($0.1\sigma$), $|\Delta \epsilon|=0.003$ ($ 0.7\sigma$). When using the forward modeling, the systematic error reduces to $|\Delta f |=0.004$ ($0.1\sigma$), $|\Delta \alpha|=<0.001$ (<$0.1\sigma$) and $|\Delta \epsilon|=0.005$ ($0.8\sigma$). There is an increase of the dispersion for the case of the up-weighting technique in the parameters $f$ and $\epsilon$, which decreases for $f$ for the forward modeling scheme but is still larger when compared to the case without redshift failures, but increases the shift by $0.002$ on $\epsilon$. In any cases the biases are less than 1$\sigma$.
Given these results, we conclude that the forward modeling scheme performs slightly better than the up-weighting scheme. Therefore, in the rest of our analysis, we will adopt the forward modeling scheme for correcting the redshift failures.

\input{test_failures}

%% file: test_failures.tex
\begin{table*}
\caption{Testing for Redshift Failures. Fitting results from 100 QPM mocks using two different techniques for mitigating the redshift failures. We compared  the results of the the mocks where redshift failures are applied and corrected by one of the two mitigation techniques to the case where no redshift failures are considered. We report the difference of the mean of the best fits as an indicator of the bias related to the spectroscopic completeness denoted by $\Delta f$, $\Delta \alpha$, $\Delta \epsilon$ and we also report  the dispersion by $\Delta S_{x}$, where $x=f, \alpha, \epsilon$. }
\label{tab:failures}

\begin{tabular}{@{}lcccccccccc}
\hline
\multicolumn{10}{c}{Testing Impact of Mitigation Techniques for Redshift Failures.}\\
\hline
Mitigation Methodology&
$\Delta f$&
$S_f$&
$\Delta \alpha$&
$ S_\alpha$&
$\Delta \epsilon$&
$ S_\epsilon$&
$\Delta F'$&
$\Delta F''$&
$\Delta \sigma_{FOG}$
\\\\[-1.5ex]
\hline
No Fiber Collisions&-         & 0.232&-              &0.113 &-         &0.050&-              &-   &-\\\\[-1.5ex]
Forward Modeling&+0.003&0.252&$<$-0.001&0.116&+0.005&0.061&0.005      &0.394  &-0.271 \\\\[-1.5ex]
Fiber Weights       &-0.016&0.250 &      -0.001&0.112&+0.003&0.054&$<$0.001&-0.141&-0.234\\\\[-1.5ex]
\hline
\end{tabular}
\end{table*}

%% file: results_Accepted.tex
\label{section:results}

We performed the analysis on the eBOSS-CMASS sample combining the NGC and SGC (if not otherwise stated). The covariance matrices used in our fits were rescaled by a factor of 0.9753 in order to account for the slight mismatch between the footprint area of the data and of the mocks.

Before running a full Monte Carlo Markov Chains (MCMC) analysis, we will compute the Best fit parameters using the minimization methodology of the last chapter, which will help us understand how susceptible our models are to changes in the distance range of our analysis.

While these results were not expected to provide any information on the confidence contours of our parameters (as an MCMC would), they give an idea of the maximum likelihood values. The main reason for performing these tests is that our MCMC analysis in its current implementation is prohibitively time-consuming; we simply can not afford to run all the tests on our data using a full MCMC approach(as we will see in appendix \ref{Priors}, our models can be degenerated when using broad biases, how large the biases can be depends on the range of the bins and on the error sizes. These degeneracies make the convergence significantly slower). Further development needs to be done in order to reduce the time of convergence of our final analysis. These maximum likelihood tests can also be used as a check of the robustness of our MCMC results.

Our first test compares the robustness of the fit against variations in the maximum fitting range (the maximum distance in $h^{-1}$Mpc where the correlation function is measured). This test is particularly important in our analysis. Figure \ref{fig:meanmocks} shows that the quadrupole estimates made with the data show large correlations at scales larger than 100 $h^{-1}$Mpc, which are outside the variance observed in the mocks.
This anomalous correlation at large scales affects the capability of our model to fit the data multipoles. We suspect this behavior could be related to an unknown systematic or a statistical fluctuation. Given that we could not identify any systematic that affects the quadrupole, and that we can not exclude a large fluctuation, we also analyzed the behavior of the fits when those large scales are eliminated in all multipoles with $\ell \geq 2$. Our main result, however, is quoted with the complete range. If this behavior is repeated in the DR16 analysis, that will indicate a systematic error that needs to be analyzed properly to provide non-biased results. If the origin of this correlation is a statistical fluctuation, this feature will probably be diluted with an increase in volume.

Thus, before exploring the likelihood surface, we performed some maximum likelihood fits using a variety of ranges. The fiducial case uses the complete range between [28,124]$h^{-1}$Mpc  for the multipoles up to $\ell=0,2,4$. We also tested some variants of this range to investigate the impact of cutting the large scale of each multipole on the best fits.

\begin{figure}
\subfigure{\includegraphics[width=85mm]{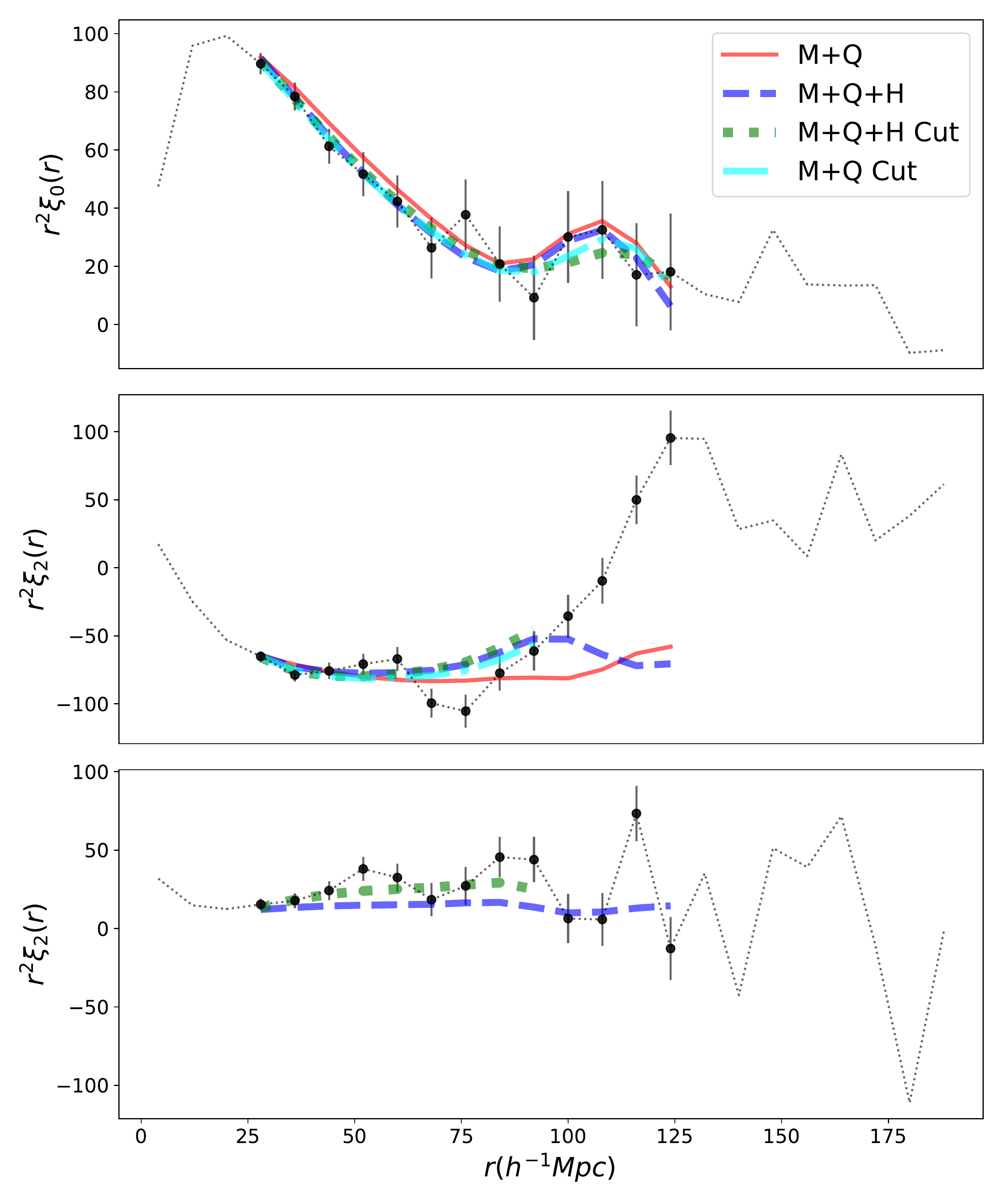}}
\caption{The maximum likelihood model for four cases: 1) using multipoles up to $\ell=2$ in the fiducial range, 2) using multipoles up to $\ell=4$ for the fiducial range, 3)  using multipoles up to $\ell=2$ but restricting the quadrupole range to [28,92] $h^{-1}$Mpc, and 4) using multipoles up to $\ell=4$ but restricting the range of the quadrupole and hexadecapole  to   [28,92] $h^{-1}$Mpc. }
\label{fig:data_bestfit_hexa}
\end{figure}

Table \ref{tab:data_bestfits_ML} lists the results of the best fits for the fiducial cases and several variants and figure \ref{fig:data_bestfit_hexa} shows how the best fit models compare to the data.
From table \ref{tab:data_bestfits_ML}, we conclude that reducing the range of the fit from 128 to 92 $h^{-1}$Mpc improves the goodness of the fit for both the monopole+quadrupole fit and the monopole+ quadrupole+hexadecapole fit.
The $\chi^2/\mathrm{d.o.f.}$, a measurement of the goodness of a fit, reduces from 2.1 to 1.35 for the $\ell_{max}=2$ case (However, there is no reduction when we reduce the range for both monopole and quadrupole where $\chi^2/\mathrm{d.o.f.}$ stays at 2.09), and from 1.81 to 1.16 if we eliminate large scales for all $\ell=0,2,4$ (It stays the same if we only limit hexadecapole, 1.14 if we restrict the range of the large scales for both $\ell=2,4$ but not for the monopole).
By using the complete range we increase the discrepancy between the fits using different order multipoles (i.e. $\ell_{max}=2$ versus $ \ell_{max}=4$). The difference in the best fit parameters for the growth factor $f$ is 0.211
 for the complete ranges (row one minus two), and it becomes 0.139
  when reducing the quadrupole and hexadecapole ranges to [28,92]$h^{-1}$Mpc (row three minus six). Similar trends occur  with $\epsilon$, where the differences in the best fit values range from  0.064 to 0.042. These trends indicate that the large correlation observed in the quadrupole is not properly modeled by our CLPT-GSRSD template.
When we exclude the large scales of the monopole, there is a significant shift in both $f$ and $\epsilon$, with $f$ shifting from 0.905 to 0.589 and $\epsilon$ from $-$0.026 to 0.088
 These shifts are expected when eliminating the large scales on the monopole. Finally, excluding the large scales on the hexadecapole affects the $f$ fits, and mildly affects the $\epsilon$ fits, as the quadrupole and hexadecapole capacity to break the degeneracy between $f$ and $\epsilon$ is derived from the BAO scales. The goodness of the fit, $\chi^2 /\mathrm{d.o.f.}$, improves when removing the large scales, due to the incapability of modeling the anomalous correlation; this approach loses all the information encoded in the BAO in the quadrupole and hexadecapole. Consequentially, the results for the AP parameters are degraded and potentially biased. We will perform the MCMC exploration for the same four cases for completeness, but we will quote the full range as our final result.

\begin{table*}
\caption{Best Fits from Maximum Likelihood Fits for different scenarios: using the fiducial ranges for the multipoles up to $\ell=2$ (first line), using multipoles up to $\ell=4$ (second line), and systematically excluding the large scales for the different multipoles considered in the fits (lines three to seven).} \label{tab:data_bestfits_ML}
\begin{tabular}{@{}lcccccccc}
\hline
\multicolumn{9}{c}{Best Fits from Maximum Likelihood for LRG sample DR14}\\
\hline
\multicolumn{9}{c}{Varying maximun range and $\ell$}\\
\hline
model& range ($h^{-1}$Mpc) &$F'$&$ F''$& $f$& $\alpha$& $\epsilon$& $\sigma_{FOG}$& $\chi^2$\\
\hline
$\xi_0+\xi_2$ &  [28,124][28,124]& 1.005& 0.74& 0.905& 0.947& -0.026& 0.009&42.4/20=2.1\\

$\xi_0+\xi_2+\xi_4$ & [28,124][28,124][28,124]&1.05& -2.7& 0.694& 0.965& 0.038& -1.51&59.81/33=1.81\\
\hline
$\xi_0+\xi_2$ & [28,124][28, 92] & 0.91 & -3.28& 0.710& 0.935&  0.050& 2.48&24.3/18=1.35\\

$\xi_0+\xi_2$ & [28,92][28, 92]&  0.753& -3.73& 0.589& 0.874&  0.088& 4.21& 25.13/12=2.09\\

\hline
$\xi_0+\xi_2+\xi_4$ & [28,124][28,124][28,92]&1.07& -2.58& 0.690& 0.969& 0.038& 0.96 &52.0/29=1.79\\
$\xi_0+\xi_2+\xi_4$ & [28,124][28,  92][28,92]&0.937& -2.96& 0.571& 0.92& 0.092& 5.07  &28.6/23=1.14\\
$\xi_0+\xi_2+\xi_4$ & [28,92][28,  92][28,92]&0.73& -3.79& 0.508& 0.858& 0.120& 6.49  &24.34/21=1.16\\
\hline
 \\[-1.5ex]

\end{tabular}
\end{table*}

As stated, we used an MCMC methodology for exploring the likelihood surface, which was done using the Monte Python public code \citep{Audren:2012wb}. We use flat priors for our parameters; the range of these priors is presented in Table \ref{tab:priors}. We run two different chains in the combined data set (NGC+SGC). The first is with the monopole and quadrupole only ($\xi_0+\xi_2$) and using the fiducial distance range.
 The second chain also runs with the monopole and quadrupole ($\xi_0+\xi_2$), but restricting the range in the quadrupole to [28,92] $h^{-1}$Mpc (the monopole stays in the same range of [28,124] $h^{-1}$Mpc).
Table \ref{tab:measurements} displays the results from the MCMC analysis. Our final  measurement was performed on the {\it combined} sample, which includes the NGC and the SGC, and was done using the fiducial methodology (i.e., 8 $h^{-1}$Mpc bins on the fiducial range). The first block reports the final result of this work, the monopole and quadrupole-only fits. The second block is for the $\xi_0+\xi_2$  fits when excluding the large scales of the quadrupole.
The third block lists the results for the Galactic hemispheres separately, this is shown for identification of any residual systematics in the data; we will discuss these results at the end of this subsection. The fourth block is quoted as a reference and it shows the fits of the BAO-only analysis done with this same sample in \cite{bautista2018}.
 MCMC chains using the hexadecapole are included in the final block for completeness and discussed in the Appendix \ref{sec:hexa_data}
as a robusteness test, but it is not part of our main results.

\begin{table}
\caption{ Flat priors ranges on the parameters of the model. }
\label{tab:priors}
\begin{tabular}{@{}ll}
\hline
\multicolumn{2}{c}{Measurements with LRG sample DR14.}\\
\hline
$f$&[0.0,2.0]\\
$F'$&[0.0,3.0]\\
$F''$&[-15,15]\\
$\sigma_{FOG}$&[0,40]\\
$\alpha$&[0.8,1.2]\\
$\epsilon$&[-0.2,0.2]\\
\hline
 \\[-1.5ex]
\hline
\end{tabular}
\end{table}

\begin{figure}
\includegraphics[width=85mm,trim = 3cm 3cm 5cm 2cm, clip]{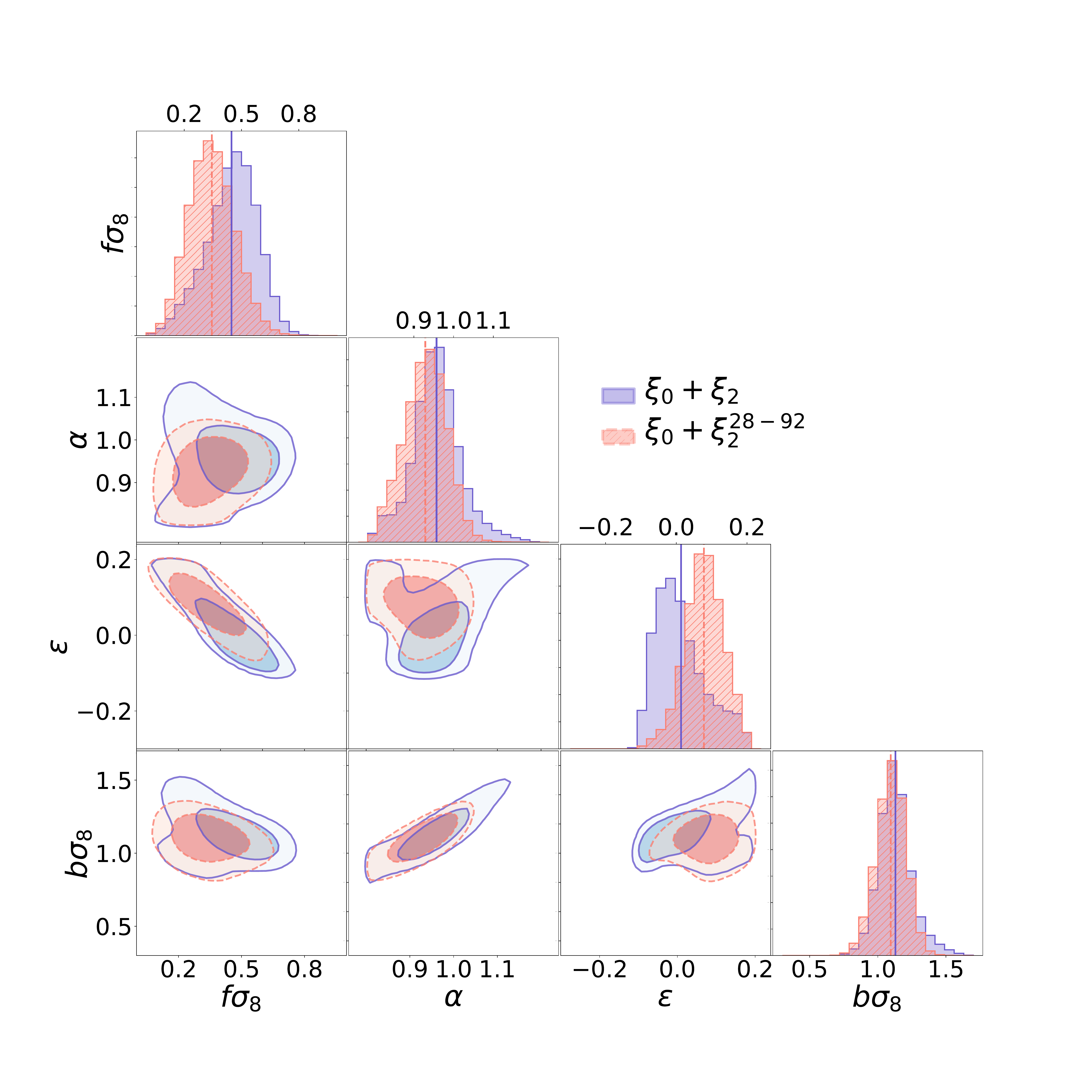}
\caption{ The shaded regions show the $1-2\sigma$ confidence surfaces found by our MCMC chains for the RSD-AP parameters for the $\xi_0+\xi_2$ space in the fiducial range (blue solid) and when excluding the large scales in quadrupole (red dashed). The confidence contours for the growth factor $f\sigma_8$, the linear bias $b\sigma_8$, the dilatation parameter $\alpha$, and the warping parameter $\epsilon$ are indicated, along with their 1D distributions. The dashed lines of each histogram are the mean values found by the MCMC chain.}
\label{fig:mcmccut1}
\end{figure}

Figure \ref{fig:mcmccut1} shows the likelihood surfaces for the two runs over the $\xi_0+\xi_2$, one chain is in the fiducial range and the other is eliminating the large scales for the quadrupole, i.e. [28,92] $h^{-1}$Mpc. The latter is added for completeness, but as stated before, our final result will be quoted using the full range.
The figure contains the $1-2\sigma$ confidence contours for the growth factor $f\sigma_8$, the linear bias $b\sigma_8$, the dilatation parameter $\alpha$, and the warping parameter $\epsilon$, together with their marginalized 1D distributions.
The 1-$\sigma$ regions are fully contained inside our priors for both cases. However, the 2-$\sigma$ regions are cut by our prior to large values of $\epsilon$ and small values of $\alpha$, our reasons for not using larger priors on the Alcock-Paczynski parameters will be discussed in the Appendix \ref{Priors}.

The results of $f\sigma_8$ and $\alpha$ are consistent within $1\sigma$, for both ranges. However, given the anti-correlation between the $\epsilon$ and $f$ parameters and the fact that the quadrupole is dominated by $\epsilon$ at the larger scales, when including the larger bins of the quadrupole the $\epsilon$ is driven from its expected value and as a consequence $f$ shifts as well. When the last three bins of the quadrupole are avoided we achieve a significant improvement in the goodness of the fit towards $\sim$ 1; the price paid for this approach is to eliminate the BAO information. This increases the degeneration of the parameters and biases the results, thus we lose information that constrains $\epsilon$, which in turn causes the contour areas to become larger, providing more freedom to the fitter to move $f$ to lower values (Figures \ref{fig:mcmccut1}).

Finally when comparing the results for the $\xi_0+\xi_2$  with the fit using  $\xi_0+\xi_2+\xi_4 $ we find agreement within $1-\sigma$ for $f\sigma_8$ and $\alpha$,  but the $\epsilon$ values have a  $1.3-\sigma$  difference. We should notice that tighter priors were used for hexadecapole because a bi-modality appears using the priors defined in table \ref{Priors}, more discussion about the results and the prior selection for the hexadecapole is provided in the Appendix \ref{sec:hexa_data}.

\begin{table*}
\caption{Results for the DR14 LRG sample. The first block is for our fiducial methodology, using the fiducial range for the $\xi_0+\xi_2$ fit. The second block is for
the $\xi_0+\xi_2$  fits when excluding the large scales of the quadrupole. The third block shows the fits separating the hemispheres NGC and SGC and using $\xi_0+\xi_2$ in the fiducial range. The fiducial value for the $\sigma_8(z_{eff=0.72})=0.55$. The eulerian bias is defined as $b=1+F'$.}

\label{tab:measurements}

\begin{tabular}{@{}lcccccc}
\hline
\multicolumn{7}{c}{Measurements with LRG sample DR14 Oficial Version.}\\
\hline
Case&
$f\sigma_8$ &
$b\sigma_8$&
<F''>&
$\sigma_{FOG}$&
$\alpha$&
$\epsilon$\\
\hline
\\[-1.5ex]
$\xi_0+\xi_2$ [28,124][28,124] &$0.454^{+0.119}_{-0.140}$ & $1.110^{+0.116}_{-0.100}$ & $2.245^{+3.849}_{-4.35}$ & $3.713^{+2.987}_{-2.31}$ & $0.955^{+0.055}_{-0.05}$ & $-4e-04^{+0.090}_{-0.050}$\\\\[-1.5ex]
\hline
$\xi_0+\xi_2 [28,124][28,92]$ &$0.337^{+0.121}_{-0.110}$ & $1.088^{+0.101}_{-0.100}$ & $-1.19^{+4.002}_{-2.900}$ & $5.027^{+2.721}_{-2.870}$ & $0.930^{+0.050}_{-0.050}$ & $0.083^{+0.059}_{-0.06}$\\\\[-1.5ex]
\hline
 $\xi_0+\xi_2$  NGC &$0.598^{+0.150}_{-0.190}$ & $1.262^{+0.121}_{-0.150}$ & $4.372^{+3.657}_{-5.810}$ & $3.008^{+2.740}_{-1.940}$ & $1.103^{+0.066}_{-0.100}$ & $-0.05^{+0.085}_{-0.040}$\\\\[-1.5ex]
 $\xi_0+\xi_2$ SGC &$0.359^{+0.168}_{-0.16}$ & $1.119^{+0.169}_{-0.12}$ & $0.328^{+1.725}_{-1.96}$ & $4.783^{+3.732}_{-3.00}$ & $0.929^{+0.087}_{-0.07}$ & $0.077^{+0.081}_{-0.07}$\\\\[-1.5ex]

\hline

 \hline
\multicolumn{7}{c}{Measurements BAO-only with LRG sample DR14  from Bautista et al 2018. Range of fits [32,182] and 8 Mpc/h bin size.}\\
\hline
Case&
range &
$\alpha_{\perp}$&
$\alpha_{||}$&
corr&
$\alpha$&
$\epsilon$\\
\hline
\\[-1.5ex]
Anisotropic&26-178&$1.01_{-0.05}^{+0.08}$&$0.82_{-0.08}^{+0.09}$&-0.39&$0.942_{-0.024}^{ +0.048}$&$-0.067_{-0.022}^{+0.033}$\\
\hline
\hline
\multicolumn{7}{c}{Measurements with LRG sample DR14 including hexadecapole.}\\
\hline
Case&
$f\sigma_8$ &
$b\sigma_8$&
<F''>&
$\sigma_{FOG}$&
$\alpha$&
$\epsilon$\\
\hline
\\[-1.5ex]
$\xi_0+\xi_2+\xi_4$  [28,124] [28,124] [28,124]&$0.31^{+0.09}_{-0.09}$ & $1.19^{+0.10}_{-0.10}$ & $-1.1^{+3.2}_{-3.3}$ & $5.8^{+3.3}_{-3.2}$ & $0.986^{+0.047}_{-0.046}$ & $0.091^{+0.046}_{-0.048}$\\ \\[-1.5ex]
\hline
\end{tabular}
\end{table*}

\begin{figure}
\subfigure{\includegraphics[width=85mm]{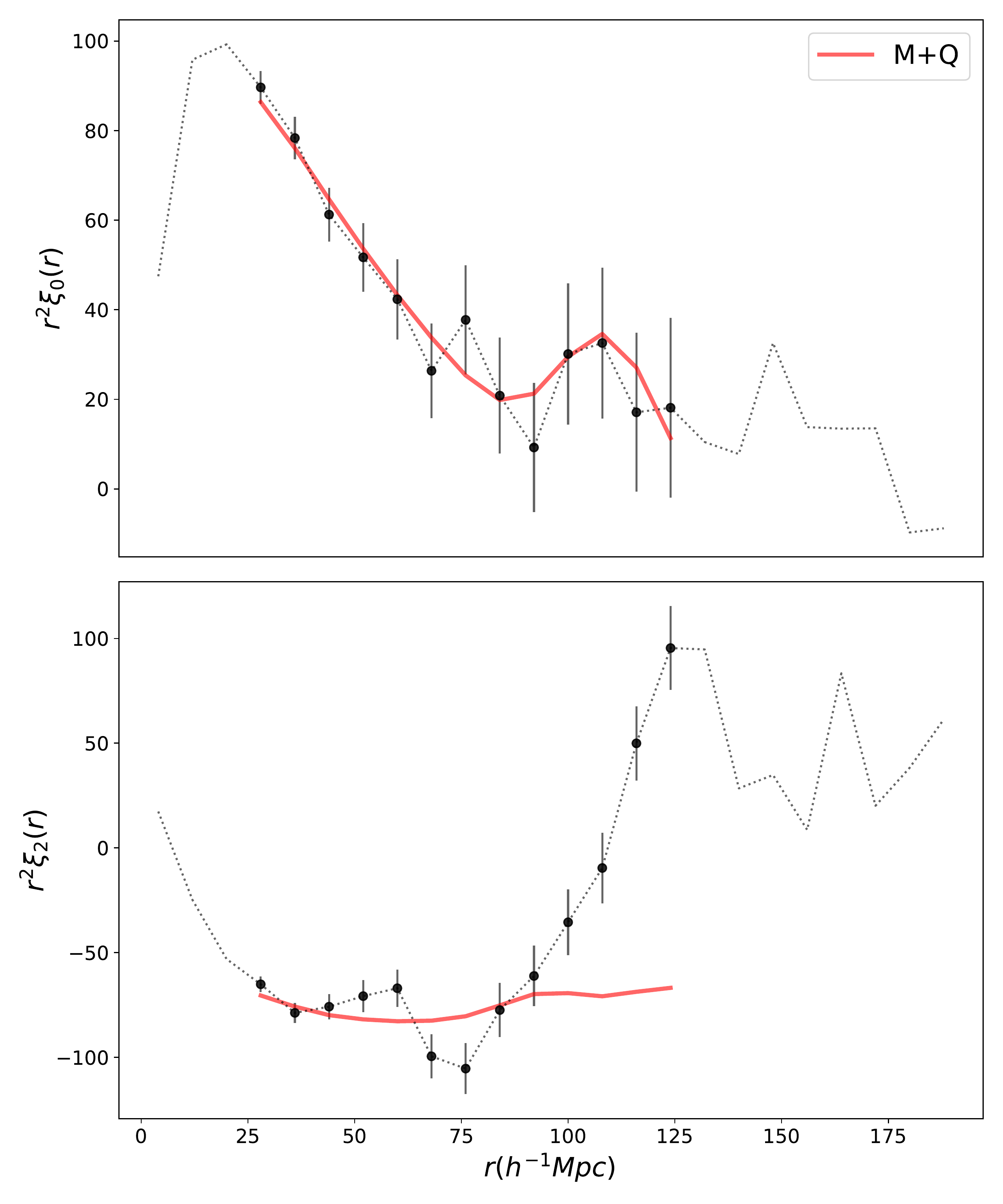}}
\caption{This plot shows how our data (black dots) compares to our model (red solid line). The model is built using the values from the first line of table \ref{tab:measurements}, that were computed using multipoles up to $\ell=2$ and our combined sample (NGC+SGC) in the fiducial range. The model is visually a good fit for the data, except for the large scales of the hexadecapole.}
\label{fig:MCMC_model}
\end{figure}

Figure \ref{fig:MCMC_model} displays the best-fit anisotropic models compared to the data for our fiducial choice of analysis. As expected, the monopole and quadrupole are visually good fits for the data, except for the large scales of the quadrupole where the correlation becomes strongly positive (scales larger than 90 $h^{-1}$Mpc).

\begin{figure}
\includegraphics[width=85mm,trim = 3cm 3cm 5cm 2cm, clip]{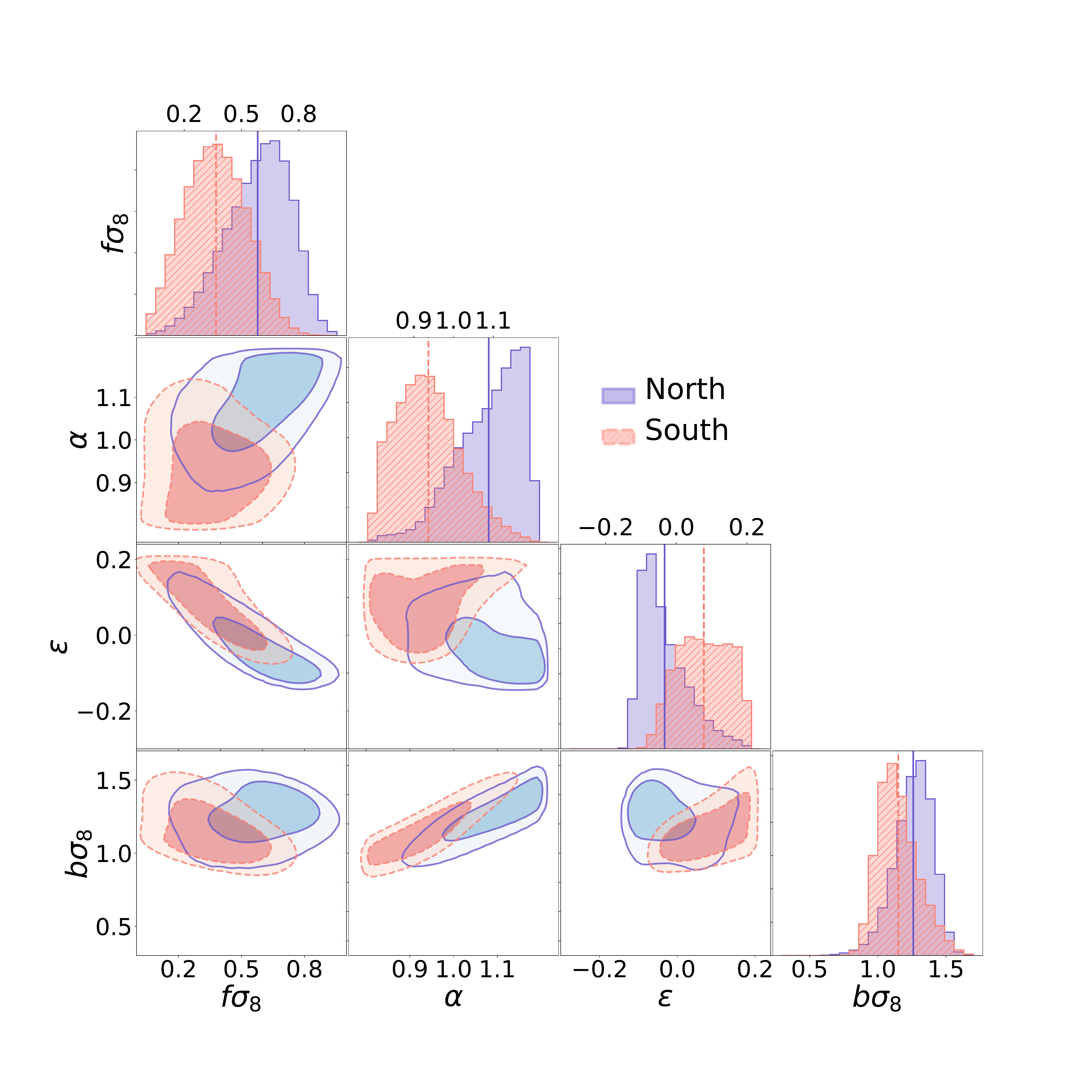}
\caption{Equivalent to Figure \ref{fig:mcmccut1} but presenting the MCMC chains for the RSD-AP parameters in the NGC (blue solid) and in the SGC (red dashed); both of them are in the fiducial range.}
\label{mcmcNGCSGC}
\end{figure}

To finalize this section, we analyze separately the North Galactic Cap (NGC) and the South Galactic Cap (SGC). Figure \ref{mcmcNGCSGC} displays the $1\sigma$ and $2\sigma$ confidence contours obtained from running an MCMC analysis separately on both hemispheres. They are computed using our standard priors quoted in Table \ref{tab:priors}. The contours in both galactic caps are poorly defined and the $1\sigma$ interval in $\epsilon$ and $\alpha$ are sharply cut by our imposed priors.

As discussed in appendix \ref{Priors} our methodology has difficulty on fixing the Alcock-Paczynski parameters to a unique value given the size of our errors compared to the the strength of the BAO signal, this leads to unphysical values of $\epsilon$ and $\alpha$ (and therefore $f$ due to their correlation with $\epsilon$) being accepted by the MCMC chain and affects all of the constraints of our parameters.

Given that the errors are larger for the North and the South separately that in the combined sample, the degeneration is stronger. This leads to several regions being accepted to within 1-$\sigma$ that would otherwise be rejected due to their inability to reproduce the BAO peak.
More data will tend to reduce this behavior and that is in fact what we see in the combined sample, where the errors are smaller.

\begin{figure}
\hspace*{-0.5cm}
\includegraphics[width=90mm]{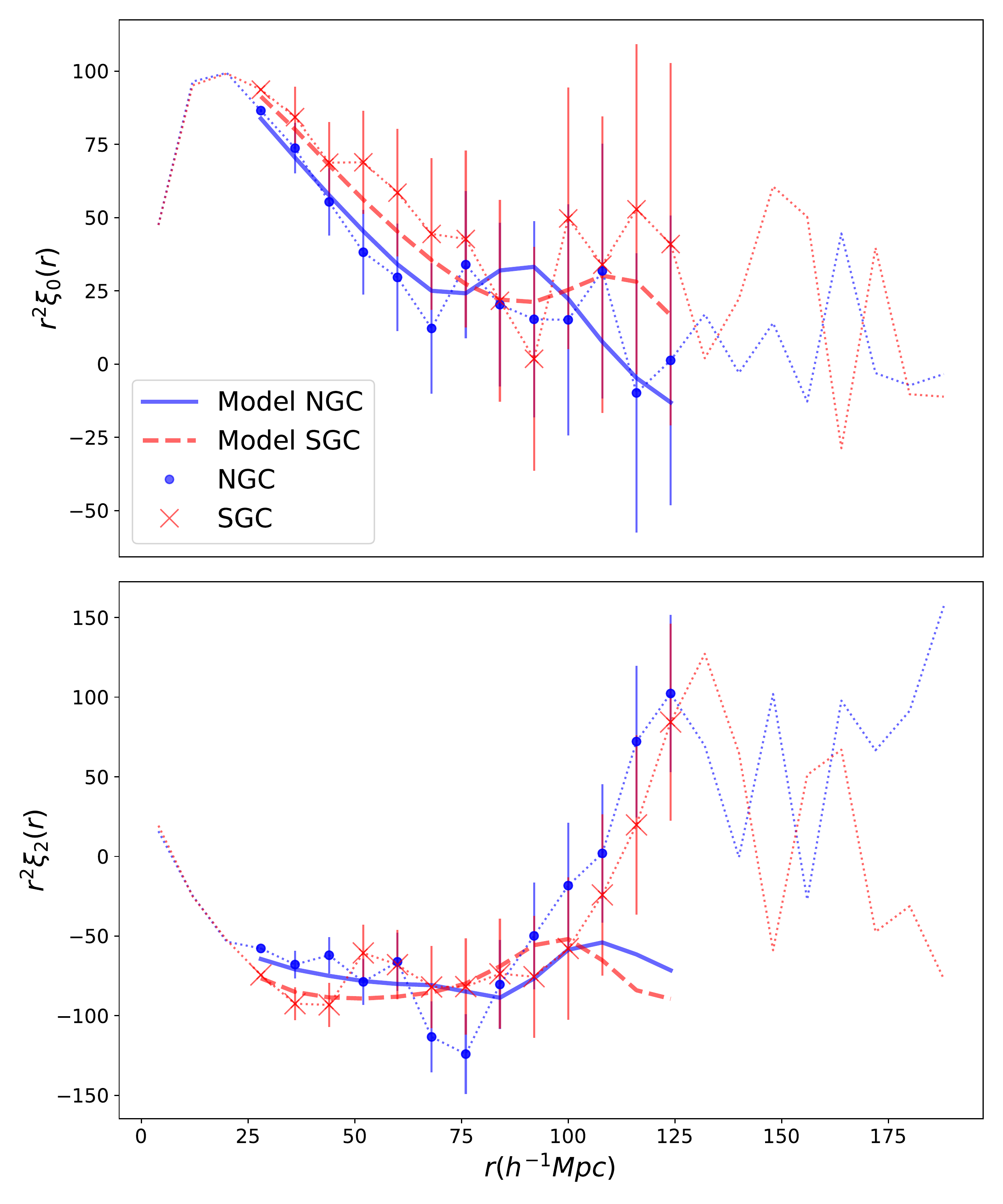}
\caption{The blue dots and red x's represent the NGC and the SGC measured data points for the Monopole (top) and the Quadrupole (bottom). The errors bars are the standard deviation computed using the QPM mocks. The solid lines indicates the best model found by the MCMC for the NGC/SGC.}
\hspace*{-0.5cm}
\label{fig:ModelvsDataNvsS_firstplot}
\end{figure}

Figure \ref{fig:ModelvsDataNvsS_firstplot} presents the data multipoles for the NGC (blue points) and SGC (red x's); the error bars correspond to their 1-$\sigma$ variance from the sample covariance matrix computed using the QPM mocks. The blue solid line represents the fits made by our MCMC analysis in the NGC, the red dashed line is the analog for the South Galactic Cap. These fits are done using the mean values obtained by our MCMC chains, which we use as our estimates of the best fits. The NGC and SGC have a significant difference in the clustering amplitude at small scales, and the peak is shifted in one hemisphere compared with the other (it is not well defined in either of the hemispheres).
Both models reasonably reproduce the multipoles; this is especially true in the smaller scales which are the ones with more weight in the fit (due to their smaller variance). There is a difference in the
multipole amplitude between both galactic caps, as a consequence the contours in Figure \ref{mcmcNGCSGC} are displaced among each other, the results for the combined sample surrounds the regions where both contours
intercept (see table \ref{tab:measurements}).

\section{Cosmological Implications}

\label{section:discussion}

Table \ref{tab:cosmo} presents our final constraints on the growth factor $f\sigma_8$, the angular diameter distance $D_A(z)$, and the Hubble parameter $H(z)$ including the statistical and the systematic error\footnote{The systematic error is based on the results from N-Series.}. Our fiducial cosmology was used to convert the best-fit dilation parameters $\alpha_{||}$ and $\alpha_\perp$, into distance measurements.
 The table includes the same variants of the methodology quoted in table \ref{tab:measurements} for the combined sample, and the values are in agreement with each other within $1\sigma$.

Our final constraint, the logarithmic growth of structure multiplied by the amplitude of dark matter density fluctuations, is  $f (z_{eff})\sigma_8(z_{eff})=0.454\pm 0.134 $.  Using the the Alcock-Paczynski dilation scales allowing us to constrain the angular diameter distance and the Hubble distance we arrive to:  $D_A(z_{eff})=1466.5  \pm 133.2 (r_s/r_s^{fid})$ and  $H(z_{eff})=105.8 \pm  15.7(r_s^{fid}/r_s) \mathrm{km.s^{-1}.Mpc^{-1}}$ where $r_s$ is the sound horizon at the end of the baryon drag epoch and $r_s^{fid}$ is its value in the fiducial cosmology at an effective redshift $z_{eff}=0.72$. These measurements correspond to relative errors of 29.4\%, 9.1\%, and 14.9\%, respectively considering the systematic error.

\cite{bautista2018}'s analysis with  the DR14 LRG sample reported a low statistical power of the current sample, and generated anisotropic BAO results yielded slightly worse results than isotropic fits. Further data releases from eBOSS should increase the statistical significance of our measurements.

\begin{figure}
   \includegraphics[width=85mm]{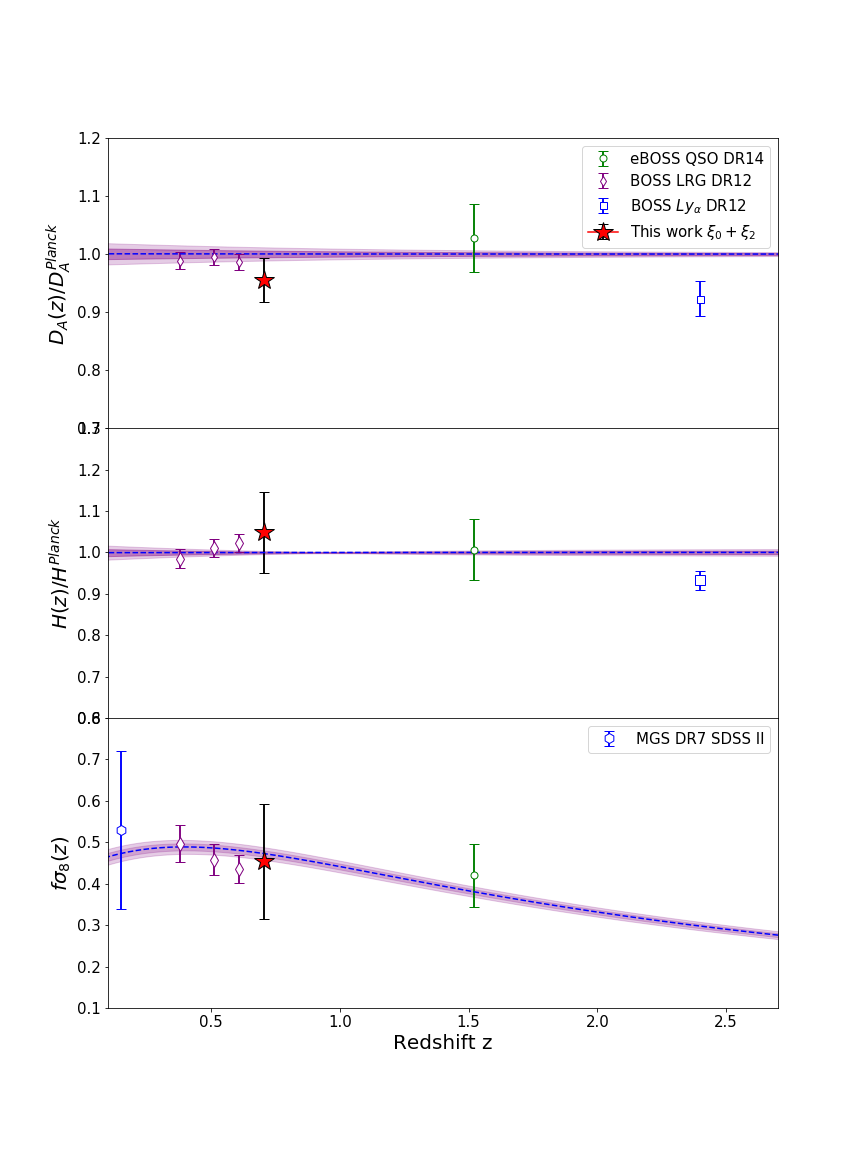}

 \caption{Measurements from DR14 eBOSS-CMASS sample using multipoles up to $\ell=2$ (red start) fitting in the range [24,128] $h^{-1}$Mpc. }
 \label{fig:finalresults}
\end{figure}

Figure \ref{fig:finalresults} presents our measurements compared with previous results from SDSS-III-BOSS DR12 from both galaxies \citep{Alam2017} and Lyman-alpha quasars \citep{Bautista2017,DuMas2017}, the eBOSS quasar measurements from \cite{Gil-Marin2018,Zarrouk2018,Hou2018}\footnote{Figure \ref{fig:finalresults} quotes the \citep{Gil-Marin2018} result; however, the three measurements from the different analyses were shown to be fully consistent.},  and the Main Galaxy Sample (MGS) from SDSS-II-DR7 \citep{Ross2015}. Our measurements are consistent with previous analyses and the $\Lambda$CDM model.

Our measurements with the CMASS-eBOSS sample are correlated with the CMASS measurements. The correlation coefficient between the two measurements was roughly estimated to be 0.16 \citep{bautista2018}; a proper measurement of this correlation will be achieved for the DR16 analysis.

\begin{table*}
\caption{ Cosmological constraints on DR14 LRG sample, using $D_A(z=0.72)^{fid}=1535, H(z=0.72)^{fid}=101$. Systematic error included.}
\label{tab:cosmo}
\begin{tabular}{@{}lcccccc}
\hline
\multicolumn{7}{c}{Measurements with LRG sample DR14.}\\
\hline
Model&range ($h^{-1}$Mpc) & $\alpha_{||}$ & $\alpha_\perp$&
$f\sigma_8$&
$D_A(z)(r_s^{fid}/r_s)$ &$H(z) (r_s/r_s^{fid})$\\
\hline

$\xi_0+\xi_2$&[28,124][28,124] & 0.954$\pm$0.149 &0.955$\pm$0.083&0.454$\pm$0.134&1466.5$\pm$133.2 &105.8$\pm$15.7\\

\hline
\multicolumn{7}{c}{Forecast.}\\
\hline
-&-&0.707&0.0382 &0.0271&0.0538\\
\hline
\multicolumn{7}{c}{BAO-only}\\
\hline
$\xi_0+\xi_2$&[32,182]&0.72&0.82 $\pm$  0.085 & 1.1 $\pm$  0.065 & -\\
\hline
\end{tabular}
\end{table*}

Finally, in order to validate our results, we
compute a forecast for the joint BAO and RSD parameters using LRG DR14 samples. The observed galaxy power spectrum is modeled as in \cite{Ballinger:1996cd, Simpson2010} using the following equation:
\begin{equation}
\begin{split}
P^{\rm obs}_{gg}(k, \mu) = \frac {1}{ \alpha_\parallel \alpha_\perp^2} \left[b \sigma_8(z)+ \frac{f \sigma_8(z)\mu^2}{A^2+(1-A^2)\mu^2}\right]^2 \\
\times \frac{P_{{mm},z=0}\left[\frac{k}{\alpha_\perp}\sqrt{1+(\frac{1}{A^2}-1)\mu^2} \right]}{\sigma^2_{8,z=0}}+ \frac{1}{n} \,,
\end{split}
\end{equation}
where $A=\alpha_\parallel/\alpha_\perp$ denotes the relative distortion in the radial and transverse directions due to the AP effect. $P_{mm}$ and $P_{gg}$  are, respectively, the matter and galaxy power spectrum; $n$ is the number density of the galaxies. The uncertainties on the BAO and RSD parameters are estimated by computing the Fisher matrix from the observed galaxy power spectrum, $F_{ij}$, following \cite{Tegmark:1997rp} with the parameter set $\mathbf{ p}\equiv \left\{\ln (\alpha_\parallel), \ln (\alpha_\perp), \ln (b \sigma_8), \ln(f \sigma_8) \right\}$. In order to account for the error induced by redshift uncertainties, we multiply the integrand of $F_{ij}$ by a damping factor of exp$[-(k\mu \sigma_r)^2]$, where $\sigma_r= \partial r / \partial z \sigma_z$ and $r$ is the comoving distance.

The second block of Table \ref{tab:cosmo}  lists the forecast when including the AP test. The measurements of $H(z)$, $D_A$, and $f\sigma_8$ are respectively 3.9, 3.1, and 2.6 times larger than the predictions.
 While this result might be caused by the non-uniform footprint of the sample at its current state, the BAO measurements also reported errors larger than the predictions.

%% file: Conclusions_Accepted.tex
The RSD effect generates an artificial anisotropy on the clustering of galaxies which can be used to constrain the growth factor, $f(z)\sigma_8$, and the radial and angular distances to the sample ( i.e., the $H(z)$ and $D_A(z)$ parameters). We used the LRG sample from the first two years of the eBOSS, denoted as DR14, to measure these parameters at the mean redshift of the survey ($z=0.72$). We presented the first full-shape analysis of this sample (i.e. modeling Redshift-Space Distortions (RSD) simultaneously with an Alcock-Paczynski (AP) parametrization), and that should be followed up on and improved on once the full observational time of the eBOSS survey is completed for the final DR16 sample. The measured correlation function was decomposed into the first three non-zero multipoles of its Lagrange expansion, and compared with theoretical predictions made with a Convolution Lagrangian Perturbation Theory (CLPT) model combined with a Gaussian Streaming model (GS). We considered six free parameters, four RSD-parameters ${ [f, F', F'', \sigma_{FoG} ]}$ and  two AP parameters $[\alpha,\epsilon]$.

We tested our methodology using a set of 84 high-precision N-Series CutSky mocks built with BOSS-CMASS properties. We fitted all individual mocks using two different methodologies: using only multipoles up to $ \ell=2$, and using all multipoles up to $\ell=4$.The fits using all the multipoles were computed in two different distance ranges, first using the complete [28,124] $h^{-1}$Mpc range for all of them, then removing the smaller scales of the hexadecapole. This extends on previous works that performed this exploration using only the monopole and the quadrupole. From the individual fits the most accurate results (smaller parameter biases in all parameters normalized by the dispersion) are obtained using the multipoles up to $\ell=4$ in the full range. Besides the fact we do not find significant biases in the distributions, when fitting the mean we noticed that the model hexadecapole does not accurately match the mean of the mocks, this generates small biases in the fits of the mean $\epsilon$ parameter. The reason  why this mismatch does not bias our measurements in the individual realizations is because of the larger errors bars we have on the hexadecapole. This behavior is related to the fact that the fits are driven by $\epsilon$ when we include the lower bins of the hexadecapole. The error bars for those lower scales are smaller, and therefore their constraining power is larger. This makes the accuracy of the model at small scales critical.

In order to characterize the statistical properties of the sample, especially its variance, we run our fitting methodology on two different sets of low-precision mocks with eBOSS properties: the QPM and EZ mocks. All of the mocks in both sets are fitted twice, the first considering only the multipoles up to $\ell=2$, and the second with multipoles up to $\ell=4$. The dispersion obtained from the two sets of low-precision mocks was fairly consistent in all cases and for all of the parameters of interest. However, the biases and distributions were not consistent with those obtained using high-precision mocks. The discrepancy arises because the GSRSD model can not match the multipoles of QPM/EZ mocks, thus no conclusion about the bias could be extracted from these fits. They were only used as a reference for the variance of the best fits for eBOSS-like mocks.

The tests performed with mocks (high and low precision), demonstrated that the constraining power of the lower bins of the hexadecapole is large due to the smaller error bars of those points. We concluded that including the hexadecapole is desirable; however, it becomes critical to  have accurate models, particularly of the small scales of the quadrupole and the hexadecapole. In this work, we adopted the conservative approach of reporting the $\xi_0$+$\xi_2$ as our final result and used the hexadecapole  results only a as a consistency test.

We considered that even if the results with high-precision mocks validated fitting the hexadecapole with our model, the biases observed when fitting the mean and the mismatch in the model hexadecapole for the mean needs further exploration. Additionally, we did not have high-precision mocks with the properties of the eBOSS sample available (higher redshift and lower mean density) and we could not properly study the statistical properties of the fitting methodology with low-precision mocks.

Our final  measurement was performed  on the ``combined" sample, using the fiducial methodology considering only the monopole and quadrupole. We constrained the logarithmic growth of structure $f\sigma_8=0.454\pm$ 0.134, $\alpha_{||}=0.954\pm$ 0.149 and $\alpha_{\perp}=0.955\pm$ 0.083.

The eBOSS DR14 LRG sample presents a large correlation in the large scale quadrupole that lies outside the 1$\sigma$ variation observed in the mocks. This feature could be related to an unknown systematic effect or just a large statistical fluctuation. Given that we could not find any systematic that affects the quadrupole, and that a large fluctuation cannot be excluded, we analyzed the behavior of the fits when we eliminated those large scales in multipoles with $\ell >= 2$ as a robustness test for our main result. Avoiding the latest three bins of the quadrupole achieves a significant improvement in the goodness of the fit to $\sim$1; however the price paid is to eliminate the BAO information, which increases the degeneration of the parameters and biases the results, thus we lose information that constrains $\epsilon$, and the contour regions become larger, giving more freedom to the fitter  to move $f$ to lower values.

We quote as our final cosmological constraint the logarithmic growth of structure multiplied by the amplitude of dark matter density fluctuations, $f (z_{eff})\sigma_8(z_{eff})=0.454 \pm 0.134 $, and the Alcock-Paczynski dilation scales which allow constraints to be placed on the angular diameter distance $D_A(z_{eff})=1466.5  \pm 133.2  (r_s/r_s^{fid})$ and the Hubble distance $H(z_{eff})=105.8 \pm  15.7  (r_s^{fid}/r_s)  \mathrm{km s^{-1} Mpc^{-1}}$, where $r_s$ is the sound horizon at the end of the baryon drag epoch and $r_s^{fid}$ is its value in the fiducial cosmology at an effective redshift $z_{eff}=0.72$. These measurements correspond to relative errors of 29.4\%, 9.1\%, and 14.9\%, respectively considering the systematic error.

Our results are consistent with previous measurements and with a $\Lambda$CDM model using Planck 2018 cosmology. Comparing our result with the forecasted ones, the measurements of $H(z)$, $D_A$ and $f\sigma_8$ are respectively about 3.9, 3.1, and 2.6 larger than the predictions. This result might be caused by the non-uniform footprint in the current state of the survey, although the BAO DR14 LRG measurements also reported errors larger than the predictions \citep{bautista2018}.
We expect a reduction on the statistical error of a factor two by the end of the experiment for DR16 final analysis.

%% file: Appendices_Accepted.tex
\section{Selecting priors}
\label{Priors}

 As discussed in chapter \ref{section:results}, figure \ref{fig:mcmccut1} shows that the 1-$\sigma$ regions are fully contained inside our priors. However, the 2-$\sigma$ regions are clearly cut by our prior to large values of $\epsilon$. In this section we will discuss our reasons for not using larger $\epsilon$ priors in our analysis, as we will see prior selection was challenging giving the size of our errors.

 Figure \ref{fig:MCMC_model} shows a comparison between our final model and the multipoles of the data set, it is clear that the detection of the BAO signal is week: the error-bars of the monopole have a similar size to the power of the BAO peak. This is problematic as the BAO peak locks the Alcock Paczynski parameters around a specific value. Given the limited capability of our methodology to fix the cosmology our model is vulnerable to being degenerated. As a consequence, we have to be very careful when choosing our priors as a large prior in the Alcock Paczynski parameters will result in degenerated regions contributing significantly to our statistics.

\begin{figure}
 \includegraphics[width=85mm,trim = 3cm 3cm 5cm 2cm, clip]{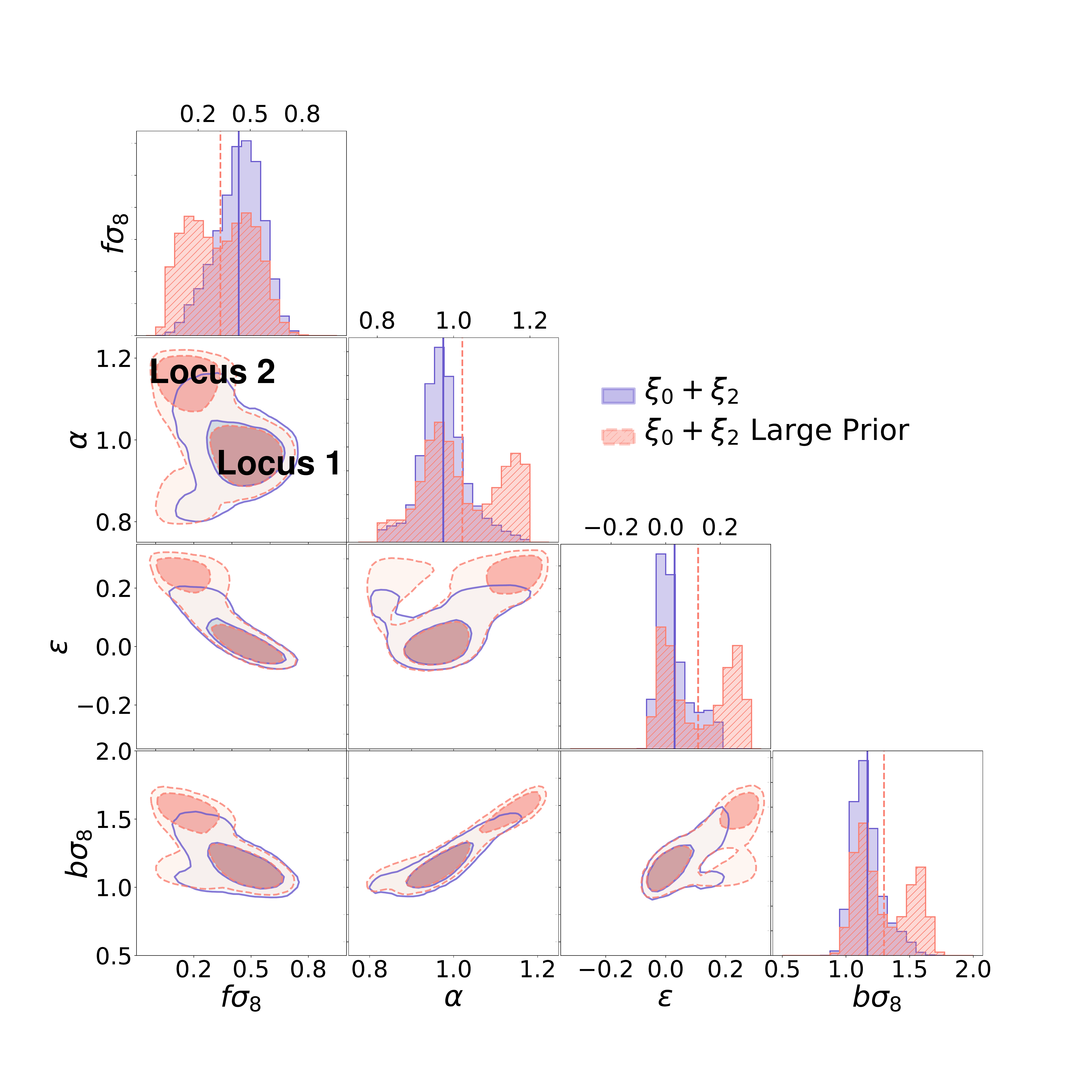}
 \caption{This plot is equivalent to Figure \ref{fig:mcmccut1}, here we are presenting the MCMC chains of two  fits to the RSD-AP parameters in the $\xi_0+\xi_2$ space done with diferent priors in $\epsilon$. The blue solid-line contours use the priors quoted in table \ref{tab:priors} for all parameters but $F2$ that is set to zero. The red dashed-line contours have larger priors on $\epsilon$ which are expanded to $[-3,3]$ and also set $F2$ to zero.}
 \label{fig:2peak}
 \end{figure}

 This is shown in figure \ref{fig:2peak} where we have run a second MCMC chain of our fiducial methodology but extending the priors of $\epsilon$ to $[-0.3,0.3]$. These chains were done with  a fixed value of $F2=0.0$ to save computational time as the goal of is not to obtain precise statistics but to show the effect of larger $\epsilon$ priors (F2 contributions to our model corresponds to second-order corrections on small scales, primarily broadening the parameter contours). The priors for the other parameters (i.e. neither $\epsilon$ nor F2) stay at the value quoted in table \ref{tab:priors}.

 In figure \ref{fig:2peak}, the solid-line blue contours show our default results, while the dashed-line red ones show those with the enlarged priors on $\epsilon$.

  A second locus is present for large values of $\epsilon$ and small values of $f\sigma_8$. This second locus is centered somewhere around $f\approx0.3$ assuming a nominal $\sigma_8$ value consistent with Planck ($\sigma_8(z_{eff})=0.55$), which would result in the Alcock-Paczynski parameters switching the cosmology to $\Omega_M(z=0)\approx0.03$ (for $\sigma_8(z=0)=0.8$ and a flat universe, assuming that $f(z)\approx\Omega_M(z)^{0.6}$). This strongly disagrees with previous constraints made by Planck, that predicts a value of $\Omega_M(z=0)=0.315\pm0.007$ \citep{2018arXiv180706209P}. Hence, the DR14 data does not allow us to broaden the priors too much, as the accuracy is not yet there in the data to rule out cosmological parameters already strongly rejected by Planck measurements.

\begin{figure}
\includegraphics[width=85mm,trim = 0cm 0cm 0cm 0cm, clip]{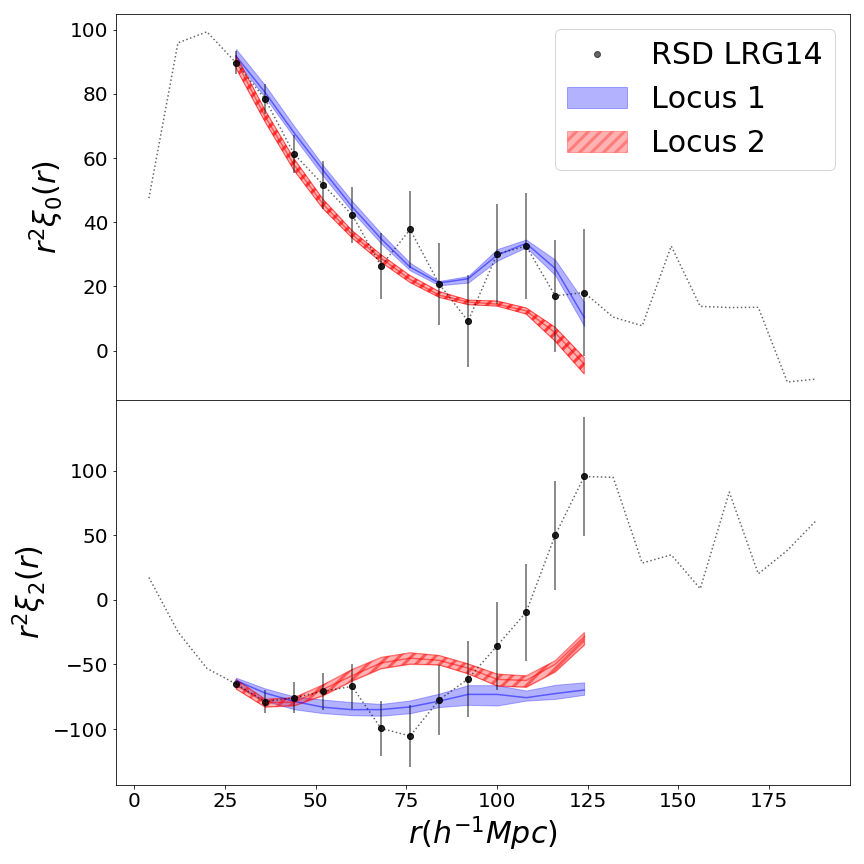}
\caption{This figure shows the median and 1-sigma percentiles for 2 sets of 100 models built with 100 points chosen randomly from the subset of those explored by our MCMC. The blue shaded regions correspond to points inside the peak centered around the expected cosmology ($\alpha \approx 1$ and $\epsilon \approx 0$). The red line and line-shaded region contours are computed with points inside the second peak that appears for large values of $\epsilon$.}
\label{fig:2Peack}
\end{figure}

Figure \ref{fig:2Peack} shows why this secondary locus is chosen by our MCMC analysis to be an acceptable fit. The blue line is the median of the models of 100 points chosen randomly from the subset of MCMC points within the locus centered around $\alpha \approx 1$ and $\epsilon \approx 0$ ({\it locus 1} in figure \ref{fig:2peak}). The blue shaded regions indicate the $18^{th}$ and $84^{th}$ percentile confidence range. The red line and line-shaded region correspond to models randomly selected from points of our MCMC chain inside locus 2 (top panel of \ref{fig:2peak}).

From figure \ref{fig:2Peack} we observe that the best fit model within locus 2 do not show a well defined BAO peak. However, statistically, both sets of models are equally good and indistinguishable in terms of their likelihood. DR16 should have smaller errors around the BAO signal which could in principle discard this second solution (locus 2). As we have stated, this second locus is discarded using Planck CMB constraints, therefore we consider reasonable to choose priors on the Alcock Paczynski parameters that  keep it out of our statistics. Considering mild Planck CMB constraints, it is reasonable to assume priors on $\alpha$ and epsilon of $\pm 0.2$ around their nominal value, as Planck strongly rejects cosmologies that are beyond that alpha and epsilon range to several sigmas.

\begin{figure}
\includegraphics[width=85mm,trim = 3cm 3cm 5cm 2cm, clip]{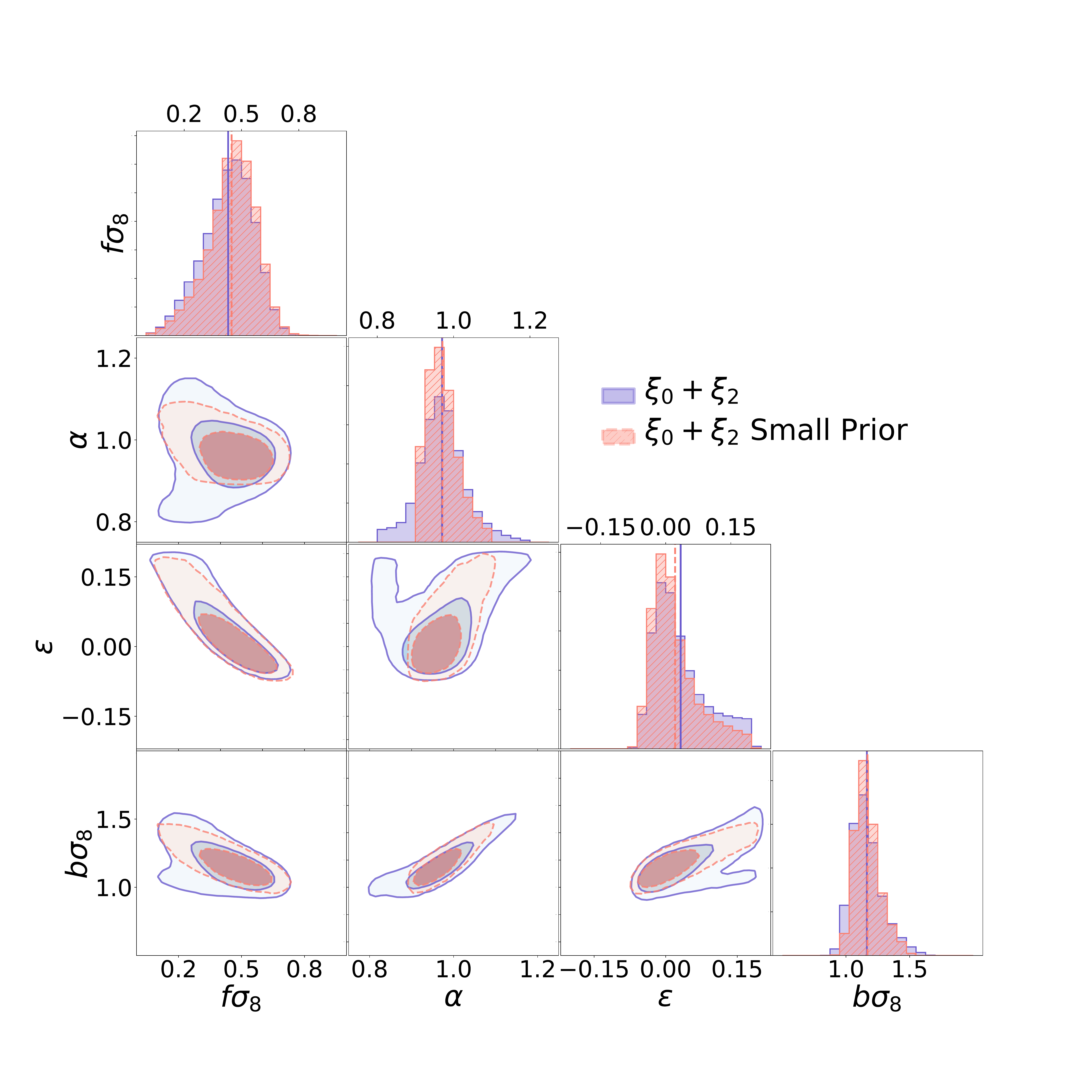}
\caption{This plot is equivalent to Figure \ref{fig:mcmccut1}, here we presenting the MCMC chains of two  fits to the RSD-AP parameters in the $\xi_0+\xi_2$ space done with diferent priors in $\alpha$. The solid-line contours use the priors quoted in table \ref{tab:priors} for all parameters but $F2$ that is set to zero. The red dashed-line contours have smaller priors on $\alpha$ which are reduced to $[0.9,1.1]$ and also set $F2$ to zero.}
\label{fig:smallalpha}
\end{figure}

Figure \ref{fig:smallalpha} is included as a robustness test of our methodology, here our fiducial result is compared with a new MCMC chain computed reducing the priors of  $\alpha$ to $[0.9,1.1]$. As in \ref{fig:2peak} $F2$ is set to zero for both chains to save computational time. The plot shows that the $\alpha$ contours are cut by the new priors, nevertheless, the 1-$\sigma$ contours of both chains are centered around the same values and have a similar shape, the main difference being marginaly reduced size of the contours, which is expected when reducing the priors.

\section{Likelihoods for eBOSS sample using hexadecapole.}
\label{sec:hexa_data}
In section \ref{High_resolution_sys} we applied our methodology to find the maximum likelihood fits of 84 Nseries high-resolution simulations. We have shown that our methodology provides consistent results with and without hexadecapole information.

 As stated in section \ref{sec:results}, we adopted the conservative approach of reporting the $\xi_0$+$\xi_2$ as our final result and using the hexadecapole  results just as a consistency test. In this appendix, we show results including the hexadecapole.

\begin{figure}
\includegraphics[width=85mm,trim = 3cm 3cm 4cm 2cm,clip]{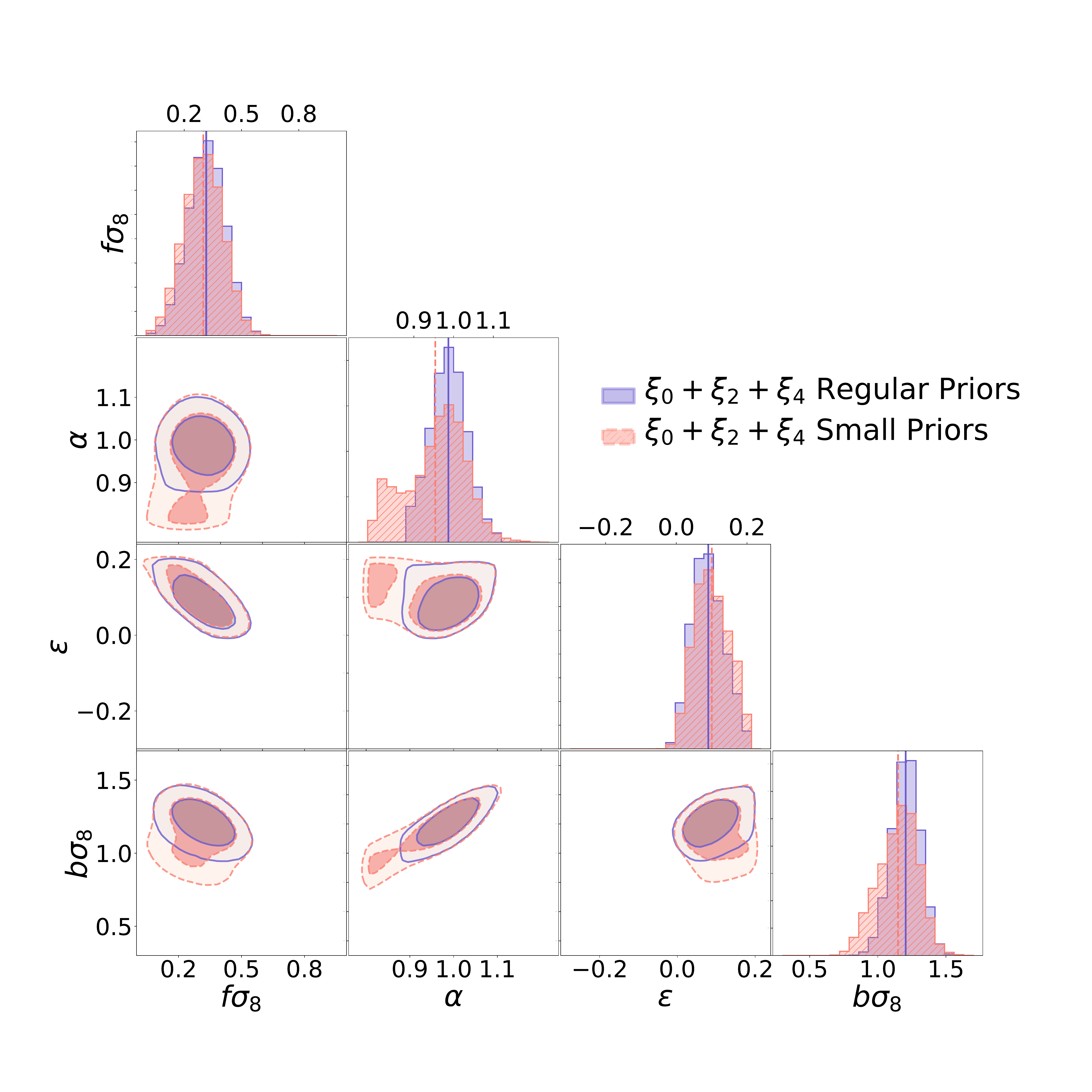}
\caption{ The shaded regions show the $1-2\sigma$ confidence surfaces found by our MCMC chains for the RSD-AP parameters using $\xi_0+\xi_2+\xi_4$ in the [28,124] $h^{-1}$Mpc range. The red dashed-line contours represent a chain with the priors of table \ref{tab:priors}, in the blue solid-line contours the priors in $\alpha$ have been reduced to the interval [0.88, 1.12]. The confidence contours for the growth factor $f\sigma_8$, the linear bias $b\sigma_8$, the dilatation parameter $\alpha$, and the warping parameter $\epsilon$ are indicated, along with their 1D distributions. The dashed lines of each histogram are the mean values found by the MCMC chain .}
\label{fig:Hexa_doublepeack}
\end{figure}

We run two different chains that include $\xi_4$  using the combined data set (NGC+SGC). One using the priors shown in table \ref{tab:priors}, and a second chain with more constraining priors. The main reason for this choice is that when considering the priors quoted in table \ref{tab:priors} we find a double peak when fitting the full range, which is shown in Figure \ref{fig:Hexa_doublepeack}.  The Figure shows the $1-2\sigma$ confidence contours for the growth factor $f\sigma_8$, the linear bias $b\sigma_8$, the dilatation parameter $\alpha$, and the warping parameter $\epsilon$, together with their 1D distributions. The only difference between both plots are the priors; The red dashed-line represent a chain with the priors of table \ref{tab:priors}, in the blue  solid-line contours the priors in $\alpha$ have been reduced to the interval [0.88, 1.12].

 As discussed in appendix \ref{Priors}, this double peaked distribution is a consequence of degenerated solutions not being rejected due to the size of our errors. Following the same procedure done in appendix \ref{Priors} we will only analyse the solution that is not in disagreement with mild Planck CMB constraints. In order to try to avoid this second degenerate solution we will reduce the size of our priors in the $\alpha$ parameter to the interval [0.88, 1.12], while the rest of the parameters are fixed to the values of table \ref{tab:priors}, these priors were chosen arbitrarily so that they contain the 1-$\sigma$ region of the main peak and completely exclude the second. We acknowledge that it is possible for the statistics obtained from this chain to still be slightly distorted by the presence of this second peak or by the position of the more constrained prior, the reduced error bars of DR16 should make the second peak less significant which could allow us to use larger priors.

\begin{figure}
\includegraphics[width=85mm,trim = 3cm 3cm 5cm 2cm,clip]{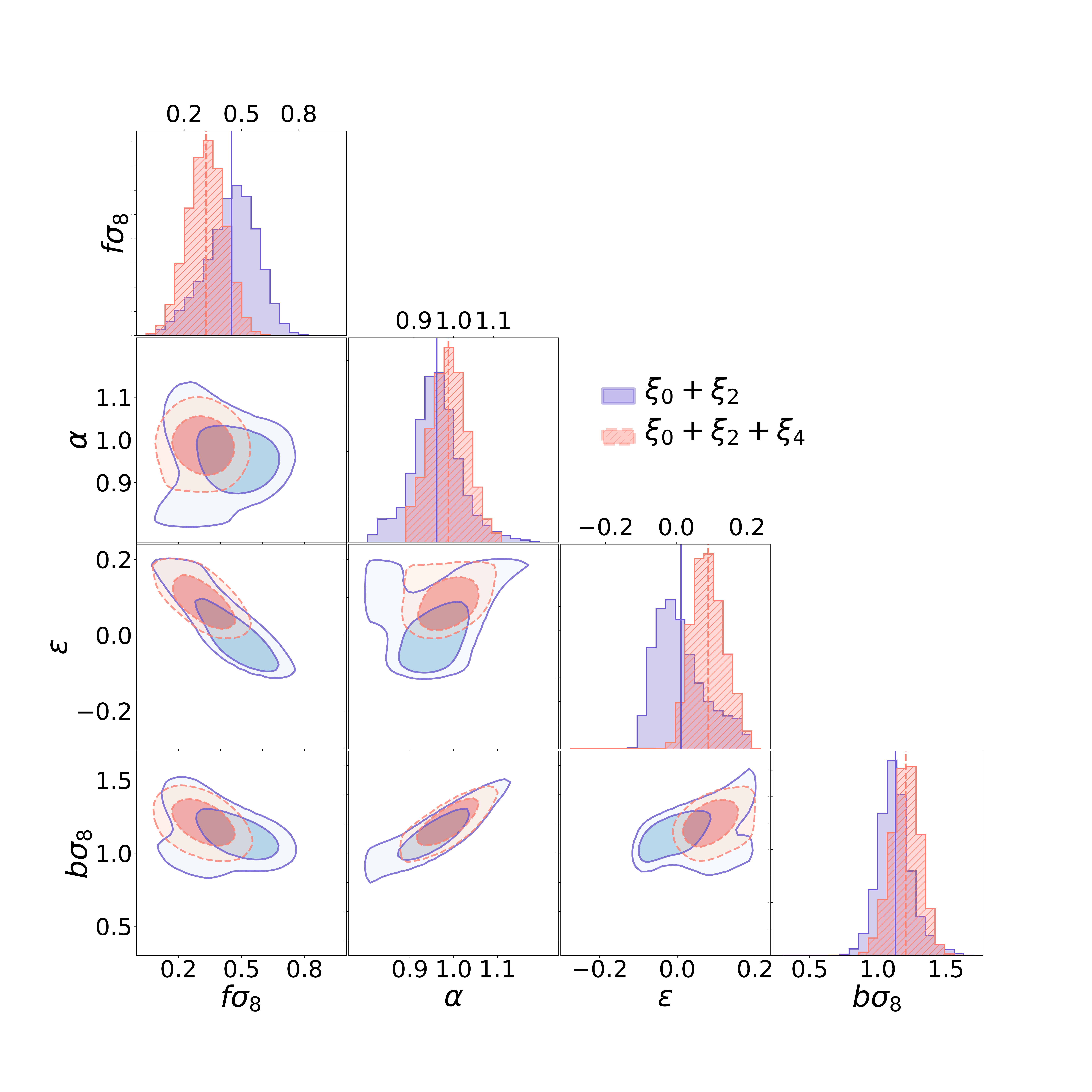}
\caption{ The shaded regions show the $1-2\sigma$ confidence surfaces found by our MCMC chains for the RSD-AP parameters for two cases: $\xi_0+\xi_2$ (red dashed-line contours) and $\xi_0+\xi_2+\xi_4$ (blue solid-line contours), all multipoles in both models are in the [24,128] $h^{-1}$Mpc range. }
\label{fig:mcmc_fullrange}
\end{figure}

 The statistical results of our parameters are quoted in table \ref{tab:measurements_hexa}: the first line repeats for comparison purposes the results for monopole and quadrupole only ($\xi_0+\xi_2$). The rest of the table includes the results using monopole, quadrupole, and hexadecapole ($\xi_0+\xi_2+\xi_4$). The second line uses the full [28,124] $h^{-1}$Mpc range in all multipoles.
In the last line, the monopole is in the [28,124] $h^{-1}$Mpc range, and the quadrupole  and hexadecapole are in the [28,92] $h^{-1}$Mpc range, we cut out the large scales for the quadrupole and hexadecapole where potential systematic errors could be present. The results of $f\sigma_8$ and $\alpha$ are consistent in the fiducial ranges within $1-\sigma$, for the two cases, $\xi_0+\xi_2$  and $\xi_0+\xi_2+\xi_4 $, but the $\epsilon$ values have a  $1.3-\sigma$  difference.
Figure \ref{fig:mcmc_fullrange} shows the likelihood surfaces of the $\xi_0+\xi_2+\xi_4$ compared with our fiducial methodology ($\xi_0+\xi_2$), they are both in the fiducial [28,124]$h^{-1}$Mpc range for all multipoles and both chains and they use the priors from table \ref{tab:priors}.

Figure \ref{fig:mean_data}, show that our hexadecapole data have a stronger amplitude on small scales that both sets of mocks. This mismatch in amplitude could be a problem of the mocks or  could be due to a real signal in the data, or could  be due to either an undetected systematic error in our data or a statistical fluctuation. If it is the latter then the increase in data with DR16 should reduce this shift. If it is a real cosmological signature it should become more significant in DR16. Regardless of the origin of this larger amplitud, the MCMC fitter prefers a large value of $\epsilon$ and a small value of $f\sigma_8$  to fit the amplitud of the data hexadecapole (see Figure \ref{fig:mcmc_fullrange}.).

\begin{figure}
\includegraphics[width=85mm,trim = 3cm 3cm 5cm 2cm,clip]{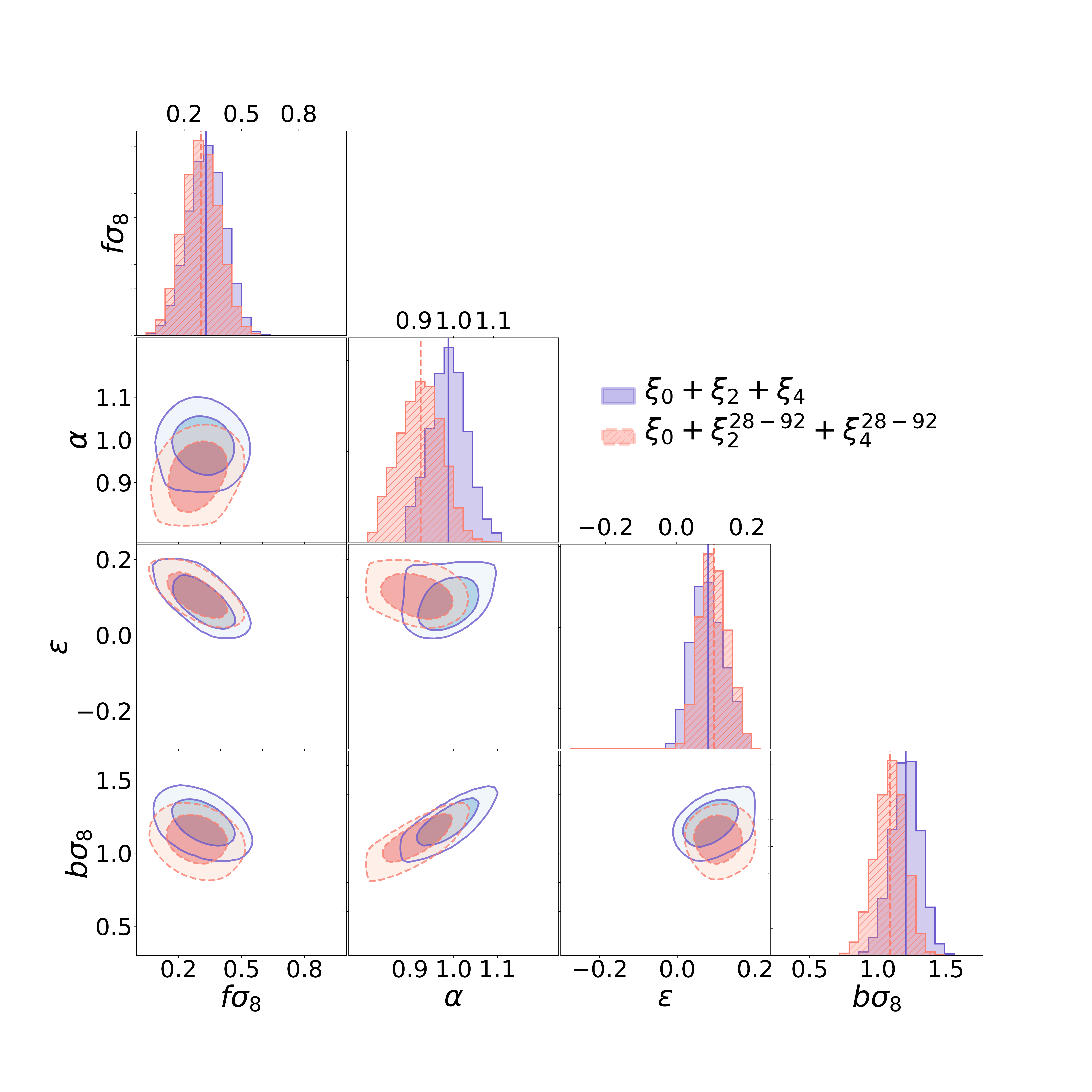}
\caption{Similar to figure \ref{fig:mcmc_fullrange}, here the red dashed-line contours represent a fit where the quadrupole and hexadecapole are reduced to the [28,92] $h^{-1}$Mpc range while the monopole stays in the full range ([24,128] $h^{-1}$Mpc), the blue solid-line contours show the fit where monopole quadrupole and hexadecapole are in the full range.}
\label{fig:mcmccut2}
\end{figure}

Figure \ref{fig:mcmc_fullrange} shows the results of cutting the large scales for the quadrupole and hexadecapole ($\ell=2,4$), the red  dashed-line contours come from a chain where the monopole is in the [28,124] $h^{-1}$Mpc range, and the quadrupole and hexadecapole in the constrained range of [28,92] $h^{-1}$Mpc. When the last three bins of the quadrupole and hexadecapole are avoided we achieve a significant improvement in the goodness of the fit (we saw this same behavior in section \ref{section:results} when removing the large scales of the quadrupole), we also lose the secondary locus that was present in the full approach whithout having to reduce our priors; however, the price paid is to eliminate the BAO information. This increases the degeneration of the parameters and biases the results, we lose information that constrains $\epsilon$ and $\alpha$, which in turn causes the contour areas to become larger, providing more freedom to the fitter to move $f$ and $\alpha$ to lower values (Figure \ref{fig:mcmccut2}).

The main impact of removing the large scales in the hexadecapole fits is in parameters that require the BAO peak to be constrained, as expected. When excluding the large scales, the BAO information is lost, and $\alpha$ is shifted in consequence.

\begin{table*}
\caption{Results for the DR14 LRG sample. The first block is for our fiducial methodology, using the fiducial range for the $\xi_0+\xi_2$ fit. The second block is for the $\xi_0+\xi_2+\xi_4$ fits in the ranges [28,124], [28,124], and [44,124] $h^{-1}$Mpc for their multipoles $\ell=0,2,4$. The third block is for the $\xi_0+\xi_2$ and $\xi_0+\xi_2+\xi_4$ fits when excluding the large scales for quadrupole and quadrupole/hexadecapole, respectively. The fiducial value for the $\sigma_8(z_{eff=0.72})=0.55$ (0.5495932). The eulerian bias is defined by $b=1+F'$.}
\label{tab:measurements_hexa}

\begin{tabular}{@{}lcccccc}
\hline
\multicolumn{7}{c}{Measurements with LRG sample DR14 Oficial Version.}\\
\hline
Case&
$f\sigma_8$ &
$b\sigma_8$&
<F''>&
$\sigma_{FOG}$&
$\alpha$&
$\epsilon$ \\\\[-1.5ex]
\hline
$\xi_0+\xi_2$ [28,124][28,124] &$0.454^{+0.119}_{-0.140}$ & $1.110^{+0.116}_{-0.100}$ & $2.2^{+3.8}_{-4.4}$ & $3.7^{+3.0}_{-2.3}$ & $0.955^{+0.055}_{-0.05}$ & $0.000^{+0.090}_{-0.050}$\\\\[-1.5ex]
\hline

$\xi_0+\xi_2+\xi_4$  [28,124] [28,124] [28,124]&$0.31^{+0.09}_{-0.09}$ & $1.19^{+0.10}_{-0.10}$ & $-1.1^{+3.2}_{-3.3}$ & $5.8^{+3.3}_{-3.2}$ & $0.986^{+0.047}_{-0.046}$ & $0.091^{+0.046}_{-0.048}$\\ \\[-1.5ex]

$\xi_0+\xi_2+\xi_4 [28,124][28,92][28,92]$&$0.285^{+0.093}_{-0.094}$ & $1.079^{+0.108}_{-0.110}$ & $-1.5^{+3.3}_{-3.0}$ & $5.5^{+2.6}_{-2.8}$ & $0.917^{+0.054}_{-0.056}$ & $0.107^{+0.041}_{-0.039}$\\ \\[-1.5ex]
\hline
\end{tabular}
\end{table*}

%% file: RSD_Accepted.bbl
\begin{thebibliography}{}
\makeatletter
\relax
\def\mn@urlcharsother{\let\do\@makeother \do\$\do\&\do\#\do\^\do\_\do\%\do\~}
\def\mn@doi{\begingroup\mn@urlcharsother \@ifnextchar [ {\mn@doi@}
  {\mn@doi@[]}}
\def\mn@doi@[#1]#2{\def\@tempa{#1}\ifx\@tempa\@empty \href
  {http://dx.doi.org/#2} {doi:#2}\else \href {http://dx.doi.org/#2} {#1}\fi
  \endgroup}
\def\mn@eprint#1#2{\mn@eprint@#1:#2::\@nil}
\def\mn@eprint@arXiv#1{\href {http://arxiv.org/abs/#1} {{\tt arXiv:#1}}}
\def\mn@eprint@dblp#1{\href {http://dblp.uni-trier.de/rec/bibtex/#1.xml}
  {dblp:#1}}
\def\mn@eprint@#1:#2:#3:#4\@nil{\def\@tempa {#1}\def\@tempb {#2}\def\@tempc
  {#3}\ifx \@tempc \@empty \let \@tempc \@tempb \let \@tempb \@tempa \fi \ifx
  \@tempb \@empty \def\@tempb {arXiv}\fi \@ifundefined
  {mn@eprint@\@tempb}{\@tempb:\@tempc}{\expandafter \expandafter \csname
  mn@eprint@\@tempb\endcsname \expandafter{\@tempc}}}

\bibitem[\protect\citeauthoryear{{Abolfathi} et~al.,}{{Abolfathi}
  et~al.}{2017}]{Abolfathi}
{Abolfathi} B.,  et~al., 2017, preprint, \href
  {http://adsabs.harvard.edu/abs/2017arXiv170709322A} {} (\mn@eprint {arXiv}
  {1707.09322})

\bibitem[\protect\citeauthoryear{{Alam}, {Ho}, {Vargas-Maga{\~n}a}  \&
  {Schneider}}{{Alam} et~al.}{2015}]{Alam2015}
{Alam} S.,  {Ho} S.,  {Vargas-Maga{\~n}a} M.,   {Schneider} D.~P.,  2015,
  \mn@doi [\mnras] {10.1093/mnras/stv1737}, \href
  {https://ui.adsabs.harvard.edu/abs/2015MNRAS.453.1754A} {453, 1754}

\bibitem[\protect\citeauthoryear{{Alam} et~al.,}{{Alam}
  et~al.}{2017a}]{Alam2017}
{Alam} S.,  et~al., 2017a, \mn@doi [\mnras] {10.1093/mnras/stx721}, \href
  {http://adsabs.harvard.edu/abs/2017MNRAS.470.2617A} {470, 2617}

\bibitem[\protect\citeauthoryear{{Alam} et~al.,}{{Alam}
  et~al.}{2017b}]{2017MNRAS.470.2617A}
{Alam} S.,  et~al., 2017b, \mn@doi [\mnras] {10.1093/mnras/stx721}, \href
  {https://ui.adsabs.harvard.edu/abs/2017MNRAS.470.2617A} {470, 2617}

\bibitem[\protect\citeauthoryear{{Alcock} \& {Paczynski}}{{Alcock} \&
  {Paczynski}}{1979}]{AP}
{Alcock} C.,  {Paczynski} B.,  1979, \mn@doi [\nat] {10.1038/281358a0}, \href
  {http://adsabs.harvard.edu/abs/1979Natur.281..358A} {281, 358}

\bibitem[\protect\citeauthoryear{{Alonso}}{{Alonso}}{2012}]{2012arXiv1210.1833A}
{Alonso} D.,  2012, preprint, \href
  {http://adsabs.harvard.edu/abs/2012arXiv1210.1833A} {} (\mn@eprint {arXiv}
  {1210.1833})

\bibitem[\protect\citeauthoryear{{Anderson} et~al.,}{{Anderson}
  et~al.}{2014}]{Anderson2014}
{Anderson} L.,  et~al., 2014, \mn@doi [\mnras] {10.1093/mnras/stu523}, \href
  {http://adsabs.harvard.edu/abs/2014MNRAS.441...24A} {441, 24}

\bibitem[\protect\citeauthoryear{Audren, Lesgourgues, Benabed  \&
  Prunet}{Audren et~al.}{2013}]{Audren:2012wb}
Audren B.,  Lesgourgues J.,  Benabed K.,   Prunet S.,  2013, \mn@doi [JCAP]
  {10.1088/1475-7516/2013/02/001}, 1302, 001

\bibitem[\protect\citeauthoryear{{Ballinger}, {Peacock}  \&
  {Heavens}}{{Ballinger} et~al.}{1996}]{Ballinger:1996cd}
{Ballinger} W.~E.,  {Peacock} J.~A.,   {Heavens} A.~F.,  1996, \mn@doi [\mnras]
  {10.1093/mnras/282.3.877}, \href
  {https://ui.adsabs.harvard.edu/abs/1996MNRAS.282..877B} {282, 877}

\bibitem[\protect\citeauthoryear{{Bautista} et~al.,}{{Bautista}
  et~al.}{2017}]{Bautista2017}
{Bautista} J.~E.,  et~al., 2017, preprint, \href
  {http://adsabs.harvard.edu/abs/2017arXiv171208064B} {} (\mn@eprint {arXiv}
  {1712.08064})

\bibitem[\protect\citeauthoryear{{Bautista} et~al.,}{{Bautista}
  et~al.}{2018}]{bautista2018}
{Bautista} J.~E.,  et~al., 2018, \mn@doi [\apj] {10.3847/1538-4357/aacea5},
  \href {https://ui.adsabs.harvard.edu/abs/2018ApJ...863..110B} {863, 110}

\bibitem[\protect\citeauthoryear{{Blanton} et~al.,}{{Blanton}
  et~al.}{2017}]{Blanton}
{Blanton} M.~R.,  et~al., 2017, \mn@doi [\aj] {10.3847/1538-3881/aa7567}, \href
  {http://adsabs.harvard.edu/abs/2017AJ....154...28B} {154, 28}

\bibitem[\protect\citeauthoryear{{Cabr{\'e}} \& {Gazta{\~n}aga}}{{Cabr{\'e}} \&
  {Gazta{\~n}aga}}{2009}]{Cabre2009}
{Cabr{\'e}} A.,  {Gazta{\~n}aga} E.,  2009, \mn@doi [\mnras]
  {10.1111/j.1365-2966.2008.14281.x}, \href
  {https://ui.adsabs.harvard.edu/abs/2009MNRAS.393.1183C} {393, 1183}

\bibitem[\protect\citeauthoryear{{Carlson}, {Reid}  \& {White}}{{Carlson}
  et~al.}{2013}]{Carlson2013}
{Carlson} J.,  {Reid} B.,   {White} M.,  2013, \mn@doi [\mnras]
  {10.1093/mnras/sts457}, \href
  {http://adsabs.harvard.edu/abs/2013MNRAS.429.1674C} {429, 1674}

\bibitem[\protect\citeauthoryear{{Chuang}, {Kitaura}, {Prada}, {Zhao}  \&
  {Yepes}}{{Chuang} et~al.}{2015}]{Chuang2015}
{Chuang} C.-H.,  {Kitaura} F.-S.,  {Prada} F.,  {Zhao} C.,   {Yepes} G.,  2015,
  \mn@doi [\mnras] {10.1093/mnras/stu2301}, \href
  {https://ui.adsabs.harvard.edu/abs/2015MNRAS.446.2621C} {446, 2621}

\bibitem[\protect\citeauthoryear{{Cole}, {Fisher}  \& {Weinberg}}{{Cole}
  et~al.}{1995}]{Cole1995}
{Cole} S.,  {Fisher} K.~B.,   {Weinberg} D.~H.,  1995, \mn@doi [\mnras]
  {10.1093/mnras/275.2.515}, \href
  {https://ui.adsabs.harvard.edu/abs/1995MNRAS.275..515C} {275, 515}

\bibitem[\protect\citeauthoryear{{Dawson} et~al.,}{{Dawson}
  et~al.}{2016}]{Dawson}
{Dawson} K.~S.,  et~al., 2016, \mn@doi [\aj] {10.3847/0004-6256/151/2/44},
  \href {http://adsabs.harvard.edu/abs/2016AJ....151...44D} {151, 44}

\bibitem[\protect\citeauthoryear{{Eisenstein} et~al.,}{{Eisenstein}
  et~al.}{2005}]{Eisenstein2005}
{Eisenstein} D.~J.,  et~al., 2005, \mn@doi [\apj] {10.1086/466512}, \href
  {http://cdsads.u-strasbg.fr/abs/2005ApJ...633..560E} {633, 560}

\bibitem[\protect\citeauthoryear{{Fang}, {Wang}, {Hu}, {Haiman}, {Hui}  \&
  {May}}{{Fang} et~al.}{2008}]{Fang2008}
{Fang} W.,  {Wang} S.,  {Hu} W.,  {Haiman} Z.,  {Hui} L.,   {May} M.,  2008,
  \mn@doi [\prd] {10.1103/PhysRevD.78.103509}, \href
  {http://adsabs.harvard.edu/abs/2008PhRvD..78j3509F} {78, 103509}

\bibitem[\protect\citeauthoryear{{Feldman}, {Kaiser}  \& {Peacock}}{{Feldman}
  et~al.}{1994}]{Feldman1994}
{Feldman} H.~A.,  {Kaiser} N.,   {Peacock} J.~A.,  1994, \mn@doi [\apj]
  {10.1086/174036}, \href
  {https://ui.adsabs.harvard.edu/abs/1994ApJ...426...23F} {426, 23}

\bibitem[\protect\citeauthoryear{{Fukugita}, {Ichikawa}, {Gunn}, {Doi},
  {Shimasaku}  \& {Schneider}}{{Fukugita} et~al.}{1996}]{Fukugita}
{Fukugita} M.,  {Ichikawa} T.,  {Gunn} J.~E.,  {Doi} M.,  {Shimasaku} K.,
  {Schneider} D.~P.,  1996, \mn@doi [\aj] {10.1086/117915}, \href
  {http://adsabs.harvard.edu/abs/1996AJ....111.1748F} {111, 1748}

\bibitem[\protect\citeauthoryear{{Gil-Mar{\'\i}n} et~al.,}{{Gil-Mar{\'\i}n}
  et~al.}{2018}]{Gil-Marin2018}
{Gil-Mar{\'\i}n} H.,  et~al., 2018, \mn@doi [\mnras] {10.1093/mnras/sty453},
  \href {https://ui.adsabs.harvard.edu/abs/2018MNRAS.477.1604G} {477, 1604}

\bibitem[\protect\citeauthoryear{{Gunn} et~al.,}{{Gunn}
  et~al.}{1998}]{Gunn1998}
{Gunn} J.~E.,  et~al., 1998, \mn@doi [\aj] {10.1086/300645}, \href
  {https://ui.adsabs.harvard.edu/abs/1998AJ....116.3040G} {116, 3040}

\bibitem[\protect\citeauthoryear{{Gunn} et~al.,}{{Gunn} et~al.}{2006}]{Gunn}
{Gunn} J.~E.,  et~al., 2006, \mn@doi [\aj] {10.1086/500975}, \href
  {http://adsabs.harvard.edu/abs/2006AJ....131.2332G} {131, 2332}

\bibitem[\protect\citeauthoryear{{Hamilton}}{{Hamilton}}{1992}]{Hamilton1992}
{Hamilton} A.~J.~S.,  1992, \mn@doi [\apjl] {10.1086/186264}, \href
  {https://ui.adsabs.harvard.edu/abs/1992ApJ...385L...5H} {385, L5}

\bibitem[\protect\citeauthoryear{{Hartlap}, {Simon}  \& {Schneider}}{{Hartlap}
  et~al.}{2007}]{2007A&A...464..399H}
{Hartlap} J.,  {Simon} P.,   {Schneider} P.,  2007, \mn@doi [\aap]
  {10.1051/0004-6361:20066170}, \href
  {http://adsabs.harvard.edu/abs/2007A%26A...464..399H} {464, 399}

\bibitem[\protect\citeauthoryear{{H{\o}g} et~al.,}{{H{\o}g}
  et~al.}{2000}]{Hog2000}
{H{\o}g} E.,  et~al., 2000, \aap, \href
  {http://adsabs.harvard.edu/abs/2000A%26A...355L..27H} {355, L27}

\bibitem[\protect\citeauthoryear{{Hou} et~al.,}{{Hou} et~al.}{2018}]{Hou2018}
{Hou} J.,  et~al., 2018, \mn@doi [\mnras] {10.1093/mnras/sty1984}, \href
  {https://ui.adsabs.harvard.edu/abs/2018MNRAS.480.2521H} {480, 2521}

\bibitem[\protect\citeauthoryear{{Joyce}, {Jain}, {Khoury}  \&
  {Trodden}}{{Joyce} et~al.}{2015}]{2015_galileon_rev}
{Joyce} A.,  {Jain} B.,  {Khoury} J.,   {Trodden} M.,  2015, \mn@doi [\physrep]
  {10.1016/j.physrep.2014.12.002}, \href
  {http://adsabs.harvard.edu/abs/2015PhR...568....1J} {568, 1}

\bibitem[\protect\citeauthoryear{{Kaiser}}{{Kaiser}}{1987}]{Kaiser1987}
{Kaiser} N.,  1987, \mn@doi [\mnras] {10.1093/mnras/227.1.1}, \href
  {http://adsabs.harvard.edu/abs/1987MNRAS.227....1K} {227, 1}

\bibitem[\protect\citeauthoryear{{Landy} \& {Szalay}}{{Landy} \&
  {Szalay}}{1993}]{LZ}
{Landy} S.~D.,  {Szalay} A.~S.,  1993, \mn@doi [\apj] {10.1086/172900}, \href
  {http://adsabs.harvard.edu/abs/1993ApJ...412...64L} {412, 64}

\bibitem[\protect\citeauthoryear{{Lewis}, {Challinor}  \& {Lasenby}}{{Lewis}
  et~al.}{2000}]{2000ApJ...538..473L}
{Lewis} A.,  {Challinor} A.,   {Lasenby} A.,  2000, \mn@doi [\apj]
  {10.1086/309179}, \href {http://adsabs.harvard.edu/abs/2000ApJ...538..473L}
  {538, 473}

\bibitem[\protect\citeauthoryear{{Linder}}{{Linder}}{2005}]{Linder2005}
{Linder} E.~V.,  2005, \mn@doi [\prd] {10.1103/PhysRevD.72.043529}, \href
  {http://adsabs.harvard.edu/abs/2005PhRvD..72d3529L} {72, 043529}

\bibitem[\protect\citeauthoryear{{Matsubara}}{{Matsubara}}{2008}]{2008PhRvD..77f3530M}
{Matsubara} T.,  2008, \mn@doi [\prd] {10.1103/PhysRevD.77.063530}, \href
  {http://adsabs.harvard.edu/abs/2008PhRvD..77f3530M} {77, 063530}

\bibitem[\protect\citeauthoryear{{Mueller}, {de Bernardis}, {Bean}  \&
  {Niemack}}{{Mueller} et~al.}{2014}]{Mueller2014}
{Mueller} E.-M.,  {de Bernardis} F.,  {Bean} R.,   {Niemack} M.,  2014,
  preprint, \href {http://adsabs.harvard.edu/abs/2014arXiv1408.6248M} {}
  (\mn@eprint {arXiv} {1408.6248})

\bibitem[\protect\citeauthoryear{{Peacock} et~al.,}{{Peacock}
  et~al.}{2001}]{Peacock2001}
{Peacock} J.~A.,  et~al., 2001, \nat, \href
  {https://ui.adsabs.harvard.edu/abs/2001Natur.410..169P} {410, 169}

\bibitem[\protect\citeauthoryear{{Perlmutter} et~al.,}{{Perlmutter}
  et~al.}{1999}]{Perlmutter1999}
{Perlmutter} S.,  et~al., 1999, \mn@doi [\apj] {10.1086/307221}, \href
  {http://adsabs.harvard.edu/abs/1999ApJ...517..565P} {517, 565}

\bibitem[\protect\citeauthoryear{{Planck Collaboration} et~al.,}{{Planck
  Collaboration} et~al.}{2016}]{Planck2015_Mod_Grav}
{Planck Collaboration} et~al., 2016, \mn@doi [\aap]
  {10.1051/0004-6361/201526681}, \href
  {http://adsabs.harvard.edu/abs/2016A%26A...594A..16P} {594, A16}

\bibitem[\protect\citeauthoryear{{Planck Collaboration} et~al.,}{{Planck
  Collaboration} et~al.}{2018}]{2018arXiv180706209P}
{Planck Collaboration} et~al., 2018, arXiv e-prints, \href
  {https://ui.adsabs.harvard.edu/abs/2018arXiv180706209P} {p. arXiv:1807.06209}

\bibitem[\protect\citeauthoryear{{Prakash} et~al.,}{{Prakash}
  et~al.}{2016}]{Prakash}
{Prakash} A.,  et~al., 2016, \mn@doi [\apjs] {10.3847/0067-0049/224/2/34},
  \href {http://adsabs.harvard.edu/abs/2016ApJS..224...34P} {224, 34}

\bibitem[\protect\citeauthoryear{{Press}, {Teukolsky}, {Vetterling}  \&
  {Flannery}}{{Press} et~al.}{2002}]{2002nrca.book.....P}
{Press} W.~H.,  {Teukolsky} S.~A.,  {Vetterling} W.~T.,   {Flannery} B.~P.,
  2002, {Numerical recipes in C++ : the art of scientific computing}

\bibitem[\protect\citeauthoryear{{Reid} \& {White}}{{Reid} \&
  {White}}{2011a}]{Reid2011}
{Reid} B.~A.,  {White} M.,  2011a, \mn@doi [\mnras]
  {10.1111/j.1365-2966.2011.19379.x}, \href
  {http://adsabs.harvard.edu/abs/2011MNRAS.417.1913R} {417, 1913}

\bibitem[\protect\citeauthoryear{{Reid} \& {White}}{{Reid} \&
  {White}}{2011b}]{2011MNRAS.417.1913R}
{Reid} B.~A.,  {White} M.,  2011b, \mn@doi [\mnras]
  {10.1111/j.1365-2966.2011.19379.x}, \href
  {http://adsabs.harvard.edu/abs/2011MNRAS.417.1913R} {417, 1913}

\bibitem[\protect\citeauthoryear{{Reid} et~al.,}{{Reid}
  et~al.}{2012}]{Reid2012}
{Reid} B.~A.,  et~al., 2012, \mn@doi [\mnras]
  {10.1111/j.1365-2966.2012.21779.x}, \href
  {http://adsabs.harvard.edu/abs/2012MNRAS.426.2719R} {426, 2719}

\bibitem[\protect\citeauthoryear{{Riess} et~al.,}{{Riess}
  et~al.}{1998}]{Riess1998}
{Riess} A.~G.,  et~al., 1998, \mn@doi [\aj] {10.1086/300499}, \href
  {http://adsabs.harvard.edu/abs/1998AJ....116.1009R} {116, 1009}

\bibitem[\protect\citeauthoryear{{Ross}, {Percival}  \& {Manera}}{{Ross}
  et~al.}{2015}]{Ross2015}
{Ross} A.~J.,  {Percival} W.~J.,   {Manera} M.,  2015, \mn@doi [\mnras]
  {10.1093/mnras/stv966}, \href
  {https://ui.adsabs.harvard.edu/abs/2015MNRAS.451.1331R} {451, 1331}

\bibitem[\protect\citeauthoryear{{Ross} et~al.,}{{Ross}
  et~al.}{2017}]{Ross2017}
{Ross} A.~J.,  et~al., 2017, \mn@doi [\mnras] {10.1093/mnras/stw2372}, \href
  {http://adsabs.harvard.edu/abs/2017MNRAS.464.1168R} {464, 1168}

\bibitem[\protect\citeauthoryear{{Rykoff} et~al.,}{{Rykoff}
  et~al.}{2014}]{Rykoff2014}
{Rykoff} E.~S.,  et~al., 2014, \mn@doi [\apj] {10.1088/0004-637X/785/2/104},
  \href {http://adsabs.harvard.edu/abs/2014ApJ...785..104R} {785, 104}

\bibitem[\protect\citeauthoryear{{Satpathy} et~al.,}{{Satpathy}
  et~al.}{2017}]{Satpathy2017}
{Satpathy} S.,  et~al., 2017, \mn@doi [\mnras] {10.1093/mnras/stx883}, \href
  {https://ui.adsabs.harvard.edu/abs/2017MNRAS.469.1369S} {469, 1369}

\bibitem[\protect\citeauthoryear{{Schlafly} \& {Finkbeiner}}{{Schlafly} \&
  {Finkbeiner}}{2011}]{Schlafly}
{Schlafly} E.~F.,  {Finkbeiner} D.~P.,  2011, \mn@doi [\apj]
  {10.1088/0004-637X/737/2/103}, \href
  {http://adsabs.harvard.edu/abs/2011ApJ...737..103S} {737, 103}

\bibitem[\protect\citeauthoryear{{Sheldon} et~al.,}{{Sheldon}
  et~al.}{2004}]{Sheldon2004}
{Sheldon} E.~S.,  et~al., 2004, \mn@doi [\aj] {10.1086/383293}, \href
  {http://cdsads.u-strasbg.fr/abs/2004AJ....127.2544S} {127, 2544}

\bibitem[\protect\citeauthoryear{{Simpson} \& {Peacock}}{{Simpson} \&
  {Peacock}}{2010}]{Simpson2010}
{Simpson} F.,  {Peacock} J.~A.,  2010, \mn@doi [\prd]
  {10.1103/PhysRevD.81.043512}, \href
  {https://ui.adsabs.harvard.edu/abs/2010PhRvD..81d3512S} {81, 043512}

\bibitem[\protect\citeauthoryear{{Smee} et~al.,}{{Smee} et~al.}{2013}]{Smee}
{Smee} S.~A.,  et~al., 2013, \mn@doi [\aj] {10.1088/0004-6256/146/2/32}, \href
  {http://adsabs.harvard.edu/abs/2013AJ....146...32S} {146, 32}

\bibitem[\protect\citeauthoryear{{Sotiriou} \& {Faraoni}}{{Sotiriou} \&
  {Faraoni}}{2010}]{Sotiriou2010}
{Sotiriou} T.~P.,  {Faraoni} V.,  2010, \mn@doi [Reviews of Modern Physics]
  {10.1103/RevModPhys.82.451}, \href
  {http://adsabs.harvard.edu/abs/2010RvMP...82..451S} {82, 451}

\bibitem[\protect\citeauthoryear{{Springel}}{{Springel}}{2005}]{2005MNRAS.364.1105S}
{Springel} V.,  2005, \mn@doi [\mnras] {10.1111/j.1365-2966.2005.09655.x},
  \href {http://adsabs.harvard.edu/abs/2005MNRAS.364.1105S} {364, 1105}

\bibitem[\protect\citeauthoryear{{Tegmark}}{{Tegmark}}{1997}]{Tegmark:1997rp}
{Tegmark} M.,  1997, \mn@doi [Physical Review Letters]
  {10.1103/PhysRevLett.79.3806}, \href
  {https://ui.adsabs.harvard.edu/abs/1997PhRvL..79.3806T} {79, 3806}

\bibitem[\protect\citeauthoryear{{Tinker} et~al.,}{{Tinker}
  et~al.}{2012}]{2012ApJ...745...16T}
{Tinker} J.~L.,  et~al., 2012, \mn@doi [\apj] {10.1088/0004-637X/745/1/16},
  \href {http://adsabs.harvard.edu/abs/2012ApJ...745...16T} {745, 16}

\bibitem[\protect\citeauthoryear{{Vargas-Maga{\~n}a}
  et~al.,}{{Vargas-Maga{\~n}a} et~al.}{2014}]{Vargas2014}
{Vargas-Maga{\~n}a} M.,  et~al., 2014, \mn@doi [\mnras]
  {10.1093/mnras/stu1681}, \href
  {http://adsabs.harvard.edu/abs/2014MNRAS.445....2V} {445, 2}

\bibitem[\protect\citeauthoryear{{Wang}, {Reid}  \& {White}}{{Wang}
  et~al.}{2014}]{Wang2014}
{Wang} L.,  {Reid} B.,   {White} M.,  2014, \mn@doi [\mnras]
  {10.1093/mnras/stt1916}, \href
  {http://adsabs.harvard.edu/abs/2014MNRAS.437..588W} {437, 588}

\bibitem[\protect\citeauthoryear{{White}, {Tinker}  \& {McBride}}{{White}
  et~al.}{2014}]{2014MNRAS.437.2594W}
{White} M.,  {Tinker} J.~L.,   {McBride} C.~K.,  2014, \mn@doi [\mnras]
  {10.1093/mnras/stt2071}, \href
  {http://adsabs.harvard.edu/abs/2014MNRAS.437.2594W} {437, 2594}

\bibitem[\protect\citeauthoryear{{Wright} et~al.,}{{Wright}
  et~al.}{2010}]{Wright}
{Wright} E.~L.,  et~al., 2010, \mn@doi [\aj] {10.1088/0004-6256/140/6/1868},
  \href {http://adsabs.harvard.edu/abs/2010AJ....140.1868W} {140, 1868}

\bibitem[\protect\citeauthoryear{{Xu}, {Padmanabhan}, {Eisenstein}, {Mehta}  \&
  {Cuesta}}{{Xu} et~al.}{2012}]{Xu2012}
{Xu} X.,  {Padmanabhan} N.,  {Eisenstein} D.~J.,  {Mehta} K.~T.,   {Cuesta}
  A.~J.,  2012, \mn@doi [\mnras] {10.1111/j.1365-2966.2012.21573.x}, \href
  {http://adsabs.harvard.edu/abs/2012MNRAS.427.2146X} {427, 2146}

\bibitem[\protect\citeauthoryear{{Zarrouk} et~al.,}{{Zarrouk}
  et~al.}{2018}]{Zarrouk2018}
{Zarrouk} P.,  et~al., 2018, \mn@doi [\mnras] {10.1093/mnras/sty506}, \href
  {https://ui.adsabs.harvard.edu/abs/2018MNRAS.477.1639Z} {477, 1639}

\bibitem[\protect\citeauthoryear{{Zhai} et~al.,}{{Zhai}
  et~al.}{2017}]{2017ApJ...848...76Z}
{Zhai} Z.,  et~al., 2017, \mn@doi [\apj] {10.3847/1538-4357/aa8eee}, \href
  {http://adsabs.harvard.edu/abs/2017ApJ...848...76Z} {848, 76}

\bibitem[\protect\citeauthoryear{{du Mas des Bourboux} et~al.,}{{du Mas des
  Bourboux} et~al.}{2017}]{DuMas2017}
{du Mas des Bourboux} H.,  et~al., 2017, \mn@doi [\aap]
  {10.1051/0004-6361/201731731}, \href
  {https://ui.adsabs.harvard.edu/abs/2017A&A...608A.130D} {608, A130}

\makeatother
\end{thebibliography}
